\documentclass[12pt]{article}

\input epsf
\usepackage{amssymb}
\usepackage[dvips]{graphicx}
\usepackage{amsmath,amscd}

\newcommand{\be}{\begin{equation}}
\newcommand{\ee}{\end{equation}}

\def\dj{\hbox{d\kern-0.347em \vrule width 0.3em height 1.252ex depth
-1.21ex \kern 0.051em}}

\def\ee{{\rm e}\,}

\numberwithin{equation}{section}

\begin{document}

\setlength{\oddsidemargin}{0cm}
\setlength{\baselineskip}{7mm}


\thispagestyle{empty}
\setcounter{page}{0}

\begin{flushright}
CERN-PH-TH/2008-031  \\
IFT-UAM/CSIC-08-20    \\
\end{flushright}

\vspace*{0.5cm}

\begin{center}
{\bf \Large Critical gravitational collapse: towards a holographic understanding of the Regge region}

\vspace*{0.7cm}

Luis \'Alvarez-Gaum\'e,$^{\,\rm a,}$\footnote{E-mail: 
{\tt Luis.Alvarez-Gaume@cern.ch}}
C\'esar G\'omez,$^{\rm\, b,}$\footnote{E-mail: 
{\tt Cesar.Gomez@uam.es}} 
Agust\'{\i}n Sabio Vera,$^{\rm\, a,}$\footnote{E-mail: 
{\tt sabio@cern.ch}} 
Alireza Tavanfar$^{\rm\, a,c,}$\footnote{E-mail: 
{\tt Alireza.Tavanfar@cern.ch}} 

and 
Miguel A. V\'azquez-Mozo$^{\rm\, d,}$\footnote{E-mail: 
{\tt Miguel.Vazquez-Mozo@cern.ch}}

\vspace*{0.25cm}

\begin{quote}

$^{\rm a}$\,\,{\sl Theory Group, Physics Department, CERN, CH-1211
Geneva 23, Switzerland}

$^{\rm b}$\,\,{\sl Instituto de F\'{\i}sica Te\'orica UAM/CSIC, 
Universidad Aut\'onoma de Madrid, E-28049 Madrid, Spain}

$^{\rm c}$\,\,{\sl Institute for Studies in Theoretical Physics and Mathematics (IPM),
P.O. Box 19395-5531, Tehran, Iran}

$^{\rm d}$\,\, {\sl Departamento de F\'{\i}sica Fundamental, Universidad de 
Salamanca, Plaza de la Merced s/n, E-37008 Salamanca, Spain}

\end{quote}

{\bf \large Abstract}
\end{center}

\noindent
We study the possible holographic connection between the Regge limit in 
QCD and critical gravitational collapse of a perfect fluid in higher dimensions. 
We begin by analyzing the problem of critical gravitational collapse of a perfect fluid
in any number of dimensions and numerically compute the associated Choptuik exponent 
in $d=5$, $6$ and $7$ for a range of values of the speed of sound of the fluid.
Using continuous self-similarity as guiding principle, a holographic correspondence 
between this process and the phenomenon of parton saturation in high-energy scattering in 
QCD is proposed. This holographic connection relates strong gravitational physics in the bulk
with (nonsupersymmetric) QCD at weak coupling in four dimensions.


\newpage

\tableofcontents

\section{Introduction}

Recently it has been proposed that critical black hole formation 
might provide the gravitational dual
description of diffractive scattering in QCD \cite{us} 
(see also \cite{brower_et_al,hatta_iancu_mueller,BPST} for other attempts at a 
holographic description of Pomeron physics). More precisely, 
the proposal identifies the onset of
strong curvature effects in the five-dimensional scalar field gravitational 
collapse with the corresponding
phenomenon in the BFKL evolution, leading 
to unitarity saturation in QCD in the Regge limit. The computation 
of the corresponding exponents on
the gravity and the QCD side render numerical values that agree quite well.

There are still a number of open questions to be clarified in 
this conjectured duality. One of them
is the reason to consider a scalar field instead of any other 
kind of collapsing matter. On physical grounds 
it seems that a perfect fluid would be more appropriate to 
describe the gravitational dual of hadron 
scattering.  In the center of mass, in the  Regge limit,
the colliding protons are composed of a large collection
of partons,  in particular gluons. Hence, a hydrodynamic 
description, specially when we are interested in collective
properties of the scattering process seems appropriate.
In this respect it is important to notice that in 
four-dimensional gravitational collapse one finds that 
the critical exponents for a collapsing massless scalar field, 
and for a radiation perfect fluid are numerically very close.
The best current value for the  Choptuik exponent for a
massless scalar field in four dimensions is $\gamma=0.374$
\cite{scalar_hD},
while for the conformal perfect fluid one has $\gamma=0.3558$ \cite{fluid,hara_koike_adachi}.
In five dimensions, the best value for a massless scalar is
$\gamma=0.431\pm 0.0001$ \cite{scalar_hD},
while in this paper we compute that for a perfect conformal
fluid in $d=5$ the  corresponding value is $\gamma=0.388$.
The value to  be compared with in the Regge limit of four-dimensional QCD according
to \cite{us} is $\gamma=0.409$. All these numbers are 
reasonably ``close", but not close enough that further
analysis are not required to establish that they
are indeed related by some underlying physics.
This implies a double task. On one hand it is necessary to draw
a clear enough holographic map between the gravity and the gauge theory
descriptions. On the other hand a more precise understanding of the QCD exponents
is required.

This paper is divided in two clearly distinct parts.
The first part studies a well-posed problem: the determination
of the critical exponents in the gravitational collapse of a perfect fluid
in dimensions other than four, in particular in five
dimensions.  The second is dedicated to outline
the possible holographic map of this critical collapse with the Regge
region.  Needless to say, this latter part is wanting
a more concrete and straightforward formulation.


The discovery of critical black hole formation has been one of the 
most exciting developments 
in numerical general relativity (see \cite{reviews} for a review). 
This phenomenon was discovered by 
Choptuik \cite{choptuikPRL} when 
studying numerically the spherically symmetric collapse of a 
massless scalar field. 
Considering a set of initial conditions parametrized by a 
real number $p$, such that for ``large" $p$ a 
black hole is formed, whereas for ``small" $p$ the initial scalar 
field configuration disperses to infinity, it
was found numerically that there is a critical value 
$p^{*}$ marking the threshold of
black hole formation. Surprisingly, for $p\gtrsim p^{*}$ 
the size of the black hole follows a scaling relation,
\begin{eqnarray}
r_{\rm BH} \sim (p-p^{*})^{\gamma}.
\label{scaling_law_r}
\end{eqnarray}
The critical exponent $\gamma$ is independent of the 
particular family of initial conditions
chosen. It changes, however, when the scalar field is 
replaced by other types of matter. 

Notice that the
critical  solution corresponds to the formation of
a ``zero mass" black hole. If
the scaling behavior (\ref{scaling_law_r}) were observed only in
a very restricted value of the parameter $p$ close
to $p^{*}$, the very scaling law would be quite
questionable.  One could argue that  higher curvature
corrections to the Einstein equations might wash
out this behavior. Indeed, the curvature at the black hole
horizon behaves roughly like $1/m^2$ hence, as
the mass goes to zero, the horizon size vanishes
and the curvature grows quadratically.  Therefore we should
expect not only higher curvature effects, but
also quantum gravitational effects to become
relevant.  However, the beauty of Choptuik's result lies in 
that the scaling law is observed in a long range
of the parameter $p$ before the black hole forms.  It
sets in before the curvature becomes overwhelmingly
big.  This means that we can expect that there is a region
in parameter space where the conjectured holographic
map could be defined.  Trying to delineate this
region is one of the things we attempt in the
second part of  this paper, although more work
remains to be done.

A very interesting feature of the numerical solution found 
by Choptuik \cite{choptuikPRL} is that when $p\sim p^{*}$
the metric near $r=0$ is approximately discretely 
self-similar (DSS).  This means that the metric functions
satisfy the property $Z_{*}(t,r)\approx Z_{*}(e^{\Delta}t,e^{\Delta}r)$, 
with $\Delta\approx 3.44$ in the case of the spherical collapse
of a massless scalar field in four dimensions. This is
known as the ``echoing" phenomenon.
A similar analysis for the case of a perfect fluid collapse \cite{fluid} 
shows that the critical solution
exhibits continuous self-similarity (CSS) near the center. 
This implies that there exists a  vector
field $v$ such that
\begin{eqnarray}
\mathcal{L}_{v}g_{\mu\nu}\equiv \nabla_{\mu}v_{\nu}+\nabla_{\nu} v_{\mu}=2g_{\mu\nu},
\label{conformal}
\end{eqnarray}
where $\mathcal{L}_{v}$ represents the Lie derivative 
with respect to the vector field $v$. 

Space-times with CSS are very interesting from various 
points of view (see \cite{reviews_CSS}
for a comprehensive rewiew). Although DSS is a remarkable
phenomenon in gravity, it seems to be a disadvantage when trying to 
establish a holographic correspondence with
the Regge region in QCD, where no echo behavior is to be found. 
However, the so-called
leading $(\log s)$-behavior of the amplitudes does indeed show scale invariance
(see Section \ref{sec_holo}).  This is a fundamental reason to abandon
the construction of the holographic map using collapsing
massless scalar fields and to use instead a system where the critical 
solution exhibits scale invariance.  The archetypical system of this
kind is the spherical collapse of a perfect fluid.

\begin{figure}[t]
 \begin{center}
 \includegraphics[width=4.0in]{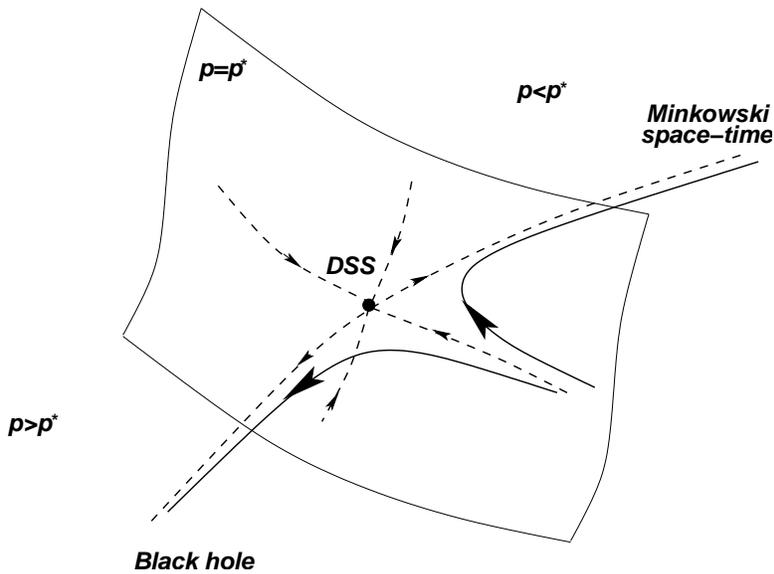}
 \caption{Phase space picture of the critical gravitational collapse.}
  \end{center}
  \label{fig1}
\end{figure}

One of the main (technical) difficulties
in the original computation of the 
Choptuik exponent \cite{choptuikPRL} is that
it requires a very involved numerical solution of the 
Einstein equations.  In \cite{fluid,hara_koike_adachi} an alternative  
procedure to compute $\gamma$ was 
proposed based on a renormalization group analysis of 
critical gravitational collapse. In this
picture, the surface $p=p^{*}$ represents a critical surface 
in the space of solutions separating
the basins of attraction of two fixed points, corresponding 
respectively to Minkowski and the black hole space-times (see Fig. \ref{fig1}). 
The critical solution with DSS or CSS has a single 
unstable direction normal to the critical surface. 

In this approach, the critical solution is characterized by 
having a single growing mode for perturbations
around it.  We can characterize it by the corresponding Lyapunov exponent.
If $\lambda_{1}>0$ represents the Lyapunov exponent associated to the 
repulsive direction, the critical exponent $\gamma$ is given in terms of it by 
\cite{hara_koike_adachi,reviews}
\begin{eqnarray}
\gamma={1\over \lambda_{1}}.
\end{eqnarray}
Consequently, in order to compute the Choptuik exponent associated with
a collapsing system,
one should identify the critical solutions
with either CSS or DSS and look for their single unstable mode. 

For systems in which the critical solution has 
DSS, this renormalization-group approach provides a nice geometrical
picture to critical  collapse. However, from a practical
point of view, the computation of the DSS critical
solution still remains a fairly formidable numerical task.  If, on the 
other hand, we deal with systems exhibiting CSS 
the problem of finding the Choptuik exponent is reduced to numerically
solving a system of ordinary differential equations 
and studying their linear perturbations. For this it suffices
to use standard numerical routines to solve the corresponding
system of ordinary differential equations, for example Runge-Kutta methods.
This is a substantial simplification with respect to simulating the whole system
using dynamical triangulations. This is the approach we follow in this paper.

This paper is organized as follows:
in Sections
\ref{sec_back} and \ref{sec_pert} we present the gravitational
part of the problem. 
We start with the study of self-similar, spherically 
symmetric, solutions to the Einstein equations 
in $d$ dimensions, sourced by a perfect fluid with equation
of state $p=k\rho$. These can be interpreted as critical 
solutions for the spherical gravitational collapse of a perfect fluid.
We write the corresponding equations of motion and study 
in detail the regularity conditions to be
imposed on the solutions. Next, we integrate these equations 
numerically and study the properties of the solutions. 
This part has a value on its own. To our knowledge 
this is the first time such an analysis is carried
out.  We also present a number of technical
clarifications that should be useful to those
who would like to engage in similar computations. 

Before proceeding to numerically integrate the 
differential equations 
there are a number of non-trivial issues
that need to be resolved.  The Einstein equations
with CSS have three singular points in terms of the
scaling variable $z$.  The solutions near
two of them, $z=0$ and $z=\infty$, can be dealt with using 
standard Frobenius analysis.  The third,
the so-called sonic point, is far more subtle.  The generic
solution present sonic booms where part of the collapsing
matter is thrown out to infinity and part forms
a compact object near the origin. For arbitrary 
initial conditions it is not possible 
to analytically join the solution below and above
the sonic surface.  The critical
solution is precisely the one that can cross analytically
the sonic point (surface). This analyticity constraint
implies that there is only a discrete number of possible
solutions.  The absence of ``moduli" in the critical
solution adds to  the difficulty in its determination.
The gravitational part of this paper is essentially
the story of how to cross the sonic point, both
for the critical solutions as well as their
perturbations. This turns out to be a long and winding
road we have tried to make as practicable as possible for the
reader.

Once the self-similar solution is 
obtained, we study in Section \ref{sec_pert} its perturbations
and compute the corresponding repulsive Lyapunov exponent for
diverse values of $k$ and $d$. From them we obtain 
the Choptuik exponents in each case, in particular the
ones quoted above.

After the detailed discussion of the critical gravitational collapse
of a perfect fluid in higher dimensions, in Section~\ref{sec_holo} we
discuss
its holographic interpretation in terms of four dimensional high energy
scattering in QCD in the perturbative Regge limit.

In Subsection \ref{subsec_holo1} we
provide a general introduction to the concept of the hard Pomeron as a bound
state of two Reggeized gluons. The Reggeization of the gluon at high
energies
calls for an interpretation in terms of an effective string theory. We are
always working in the small 't Hooft coupling limit in the gauge theory
side.
However, in deep inelastic scattering (DIS), the gluon distribution
function
is sensitive to nonperturbative effects not related to confinement: the
so-called saturation effects. They
are responsible for the transition from a dilute regime, where
parton-parton
correlations are weak, towards a dense limit where these correlations 
are very strong.
We argue that this dilute-dense transition corresponds to the formation
of a higher dimensional tiny black in the gravity side of our holographic
proposal.

In Subsection \ref{subsec_holo2} a brief introduction to the generalities of
evolution equations in DIS is provided. We then focus on the small-$x$
limit
and describe in some detail the building blocks of the BFKL (Balitsky-Fadin-Kuraev-Lipatov)
formalism \cite{pomeron}.
The BFKL equation describes a linear evolution of the parton densities
with respect to $x$ with no ordering in the virtualities of the internal
propagators. This generates a UV/IR symmetric random walk in transverse
momentum space which can be described by a simple diffusion equation. We
identify the variable $1/x$, proportional to the center of mass energy,
as the holographic extra dimension.

The saturated region, corresponding to a dense system of packed
partons,
is dominated
by nonlinearities required to respect the Froissard-Martin bound on total cross
sections for very large energies. In Subsection \ref{subsec_holo3} the simplest
generalization of the BFKL equation containing a nonlinear term, the BK (Balitsky-Kovchegov)
equation \cite{BK}, is introduced. However, we do not use the BK equation in this
work since it is posible to study the transition to saturation using an
absorptive barrier in the BFKL equation in the limit of asymptotically
large center of mass energies. This barrier introduces an asymmetry
in transverse momentum space suppressing the propagation towards IR
virtualities. These correspond to gluons with a large transverse size and
a larger probability of destructive interference. In Subsection \ref{subsec_hol4} we discuss the
holographic relation between saturation and critical gravitational collapse and argue that the
suppression of IR modes in BFKL evolution corresponds to the large
fraction of initial matter which evolves into flat space-time in the critical
formation of the tiny black hole.

To close the paper, in Section \ref{sec_final}
we do not present conclusions, but rather a collection
of unanswered questions and work in progress.
For the reader's convenience some technical details have been
deferred to the Appendices.

\section{Self-similar perfect fluid solutions in $\boldsymbol{d}$ dimensions}
\label{sec_back}

This is a very long Section where we try to collect in a self-contained
way our results on the gravitation collapse of perfect fluids in diverse
dimensions.  Although we are interested in the case of a conformal fluid
in five dimensions, the complexity of the problem is similar in any dimension.
Hence we have analyzed the problem including the space-time dimension
$d$ as a parameter.  The basic results are plotted in Figs. \ref{func_back} to \ref{dynamical2}, where
we exhibit the properties of the critical solutions for a sample of values
of $k$ in five dimensions.  Similar plots could have been shown in other
dimensions.

In order to find the Choptuik exponent for the critical collapse of a perfect fluid in $d$ 
dimensions we need to study the corresponding critical solution. In our case this is given by
a spherically symmetric self-similar solution of the Einstein equations sourced by a perfect fluid.
We begin therefore by obtaining the corresponding Einstein equations.

It is important to stress that one of the difficulties in determining the critical solution
in gravitational collapse is that it does not depend on any
parameter; or in modern parlance, it generically has no moduli
apart from the speed of sound and the dimension of space-time.
As a consequence, 
we have to deal with a system of non-linear differential equations
with three singular points that do not seem amenable  to
analytic treatment in spite of (our) many unsuccessful efforts.
In order to obtain numerical solutions that are regular at the singular points
we have to toil through a number of steps that we have
tried to make as palatable as possible.  Once the critical
solutions are determined we proceed to study their
perturbations and to compute their Lyapunov exponents in the next Section.

\subsection{The Einstein equations}

Our starting point is then the spherically symmetric $d$-dimensional line element 
\begin{eqnarray}
ds^{2}=-\alpha(t,r)^{2}dt^{2}+a(t,r)^{2}dr^{2}+R(t,r)^{2}d\Omega_{d-2}^{2},
\label{line_element}
\end{eqnarray}
where $d\Omega_{d-2}^{2}$ is the line element on the $(d-2)$-dimensional sphere 
$S^{d-2}$. We look for solutions to the Einstein equations sourced by a perfect fluid
with energy-momentum tensor
\begin{eqnarray}
T_{\mu\nu}=(\rho+p)u_{\mu}u_{\nu}+p\,g_{\mu\nu}
\label{em_tensor}
\end{eqnarray}
and equation of state
\begin{eqnarray}
p=k\,\rho, \hspace*{1cm} 0\leq k\leq 1.
\label{eos}
\end{eqnarray}
We have written the line element (\ref{line_element}) using comoving coordinates
where the components of the fluid velocity are $u^{\mu}=(u^{t},\vec{0}\,)$. This is
quite standard in the literature.  The  basic advantage is that the integration of
the Bianchi identities is straightforward. 

\paragraph{The equations of motion.}
The details of the calculation of the Einstein equations for the metric (\ref{line_element}) can 
be found in Appendix A. As explained there it is enough to consider the 
$(t,t),(t,r),(r,r)$ components given by
\begin{eqnarray}
R_{,tr}-{R_{,t}\alpha_{,r}\over \alpha}-
{R_{,r}a_{,t}\over a} &=& 0, \nonumber \\[0.2cm]
\left[R^{d-3}\left(1+{R_{,t}^{2}\over \alpha^{2}}-{R_{,r}^{2}\over a^{2}}\right)\right]_{,r} &=& {8\pi G_{N}
\over d-2}\rho R_{,r}R^{d-2},
\label{ee} \\[0.2cm]
\left[R^{D-3}\left(1+{R_{,t}^{2}\over \alpha^{2}}-{R_{,r}^{2}\over a^{2}}\right)\right]_{,t} &=& {8\pi G_{N}
\over d-2}p R_{,t}R^{d-2} ,\nonumber 
\end{eqnarray}
where $G_N$ represents Newton's constant.
The first equations implements the comoving condition with vanishing 
space-like velocity components for the fluid.
These equations have to be supplemented by the Bianchi identities $\nabla_{\mu}T^{\mu\nu}=0$
\begin{eqnarray}
{\alpha_{,r}\over \alpha}&=& -{p_{,r}\over \rho+p}, \nonumber \\
{a_{,t}\over a} &=& -{\rho_{,t}\over \rho+p}-(d-2){R_{,t}\over R}.
\label{bi}
\end{eqnarray}

One of the basic simplifying features of studying spherically symmetric
gravitational collapse is the fact that gravitational waves are not generated,
hence we can write the metric components in terms of the energy-momentum explicitly.
General collapse, without spherical symmetry is a far more difficult problem that
has not been analyzed yet in the search of critical phenomena.
As in the four-dimensional case \cite{harada_maeda}, the equations in (\ref{ee}) can be rewritten in a 
simpler form by introducing the function $m(t,r)$ defined by
\begin{eqnarray}
2G_{N}m &=& R^{d-3}\left(1+{R_{,t}^{2}\over \alpha^{2}}-{R^{2}_{,r}\over a^{2}}\right).
\label{mass_function}
\end{eqnarray}
Together with the Bianchi identities (\ref{bi}) we arrive at the following system of equations
\begin{eqnarray}
2m_{,r} &=& {16\pi \over d-2}\,G_N\,\rho R_{,r}R^{d-2}, \nonumber \\[0.2cm]
2m_{,t} &=& -{16 \pi \over d-2}\, G_N\,p R_{,t} R^{d-2}, \nonumber \\[0.2cm]
2G_{N}m &=& R^{d-3}\left(1+{R_{,t}^{2}\over \alpha^{2}}-{R^{2}_{,r}\over a^{2}}\right),\label{eom}
\\[0.2cm]
{\alpha_{,r}\over \alpha} &=& - {p_{,r}\over \rho+p} , \nonumber 
 \\[0.2cm]
{a_{,t}\over a} &=& -{\rho_{,t}\over \rho+p}-(d-2){R_{,t}\over R}. \nonumber 
\end{eqnarray}
These equations plus the equations of state (\ref{eos}) make a total of six equations to determine the five unknown 
functions, $\alpha(t,r)$, $a(t,r)$, $R(t,r)$, $\rho(t,r)$ and $p(t,r)$. There are therefore only five
independent equations. 
Their physical meaning is transparent. The first one indicates how the mass function $m$ defined in 
Eq. (\ref{mass_function}) grows with
the positive energy density $\rho$.  The second implies the decrease of this same mass function induced by
a positive pressure.  The third gives a measure of the formation of an apparent horizon.  Notice
that equation (\ref{mass_function}) can be written as the condition when the surface $R(t,r)={\rm constant}$ 
become null.  In these coordinates it represents an integral of motion.

The Bianchi identities can actually be written as
\begin{eqnarray}
(\log \alpha)_{,r}&=&-{k\over k+1}(\log\rho)_{,r} \nonumber \\
(\log a)_{,t}&=& -{1\over k+1}(\log \rho)_{,t}-(d-2)(\log{R})_{,t},
\end{eqnarray}
hence
\begin{eqnarray}
\alpha(t,r)=c_{\alpha}(t)\rho^{-{k\over k+1}}, \hspace*{1cm}
a(t,r)=c_{a}(r)\rho^{-{1\over k+1}}R^{2-d},
\label{alpha_a}
\end{eqnarray}
where $c_{\alpha}(t)$, $c_{a}(r)$ are arbitrary functions of $t$ and $r$ respectively. The
function $c_{a}(r)$ can be absorbed by a redefinition of the radial coordinate $r$. Thus we set 
$c_{a}(r)=1$ in the following.

\paragraph{Special velocities.}

There are two families of surfaces in the geometry given by the line element
(\ref{line_element}). The first ones are 
\begin{eqnarray}
R(t,r)=\mbox{constant},
\end{eqnarray}
defining spheres of constant area. The second family of surfaces comes about because we 
are interested eventually in self-similar solutions. They are defined by
\begin{eqnarray}
r=-zt, \hspace*{1cm} z=\mbox{constant}.
\end{eqnarray}
One can define now the velocity of these surfaces as the velocity of a fiducial observer lying on one
of them with respect to the fluid. 
This is given by the ratio between the proper distance $d\ell=a dr$ that this observer 
covers and the proper time $d\lambda=\alpha dt$ it takes. In the case of the constant-$R$ surface we find
\begin{eqnarray}
0={R_{,t}\over \alpha}d\lambda+{R_{,r}\over a}d\ell \hspace*{1cm} \Longrightarrow \hspace*{1cm}
V_{R}=-{a R_{,t}\over \alpha R_{,r}},
\label{VR}
\end{eqnarray}
whereas for surfaces of constant $z$ the velocity is
\begin{eqnarray}
0=-{z\over \alpha}d\lambda+{1\over a}d\ell, \hspace*{1cm} \Longrightarrow \hspace*{1cm}
V_{z} = -{a z\over \alpha}.
\label{Vz}
\end{eqnarray}

\paragraph{Changing coordinates.}
Since eventually we are interested in self-similar solutions it is convenient to define the new coordinates
$(\tau,z)$ by
\begin{eqnarray}
\tau=-\log\left(-{t\over \ell_{s}}\right), \hspace*{1cm} z=-{r\over t},
\label{tau,z}
\end{eqnarray}
where we have introduced a length scale $\ell_{s}$ in order to make the argument of the logarithm dimensionless.
Notice that the collapse takes place for $t<0$, the critical black hole being formed in the limit $t\rightarrow 0^{-}$.
This corresponds to the limit $\tau\rightarrow \infty$. In addition, we introduce the 
dimensionless functions $\eta(t,r)$,  $S(t,r)$ and $M(t,r)$ defined respectively by \cite{harada_maeda}
\begin{eqnarray}
\eta(\tau,z)&=&8\pi G_{N} r^{2}\rho(t,r), \nonumber \\[0.2cm]
S(\tau,z)&=& {R(t,r)\over r}, 
\label{newfunctions}\\[0.2cm]
M(t,r)&=&{2G_{N}m(t,r)\over r^{d-3}} . \nonumber
\end{eqnarray}

Up to this point we have written the Newton constant $G_{N}$ explicitly. From now on we use units 
in which $G_{N}=1$. In this system of units, $\rho$ and $p$ have dimensions of (length)$^{2}$, whereas $m$
is measured in units of (length)$^{d-3}$. The powers of the Newton constant 
can be restored in all expressions by replacing $m\rightarrow G_{N}m$,
$\rho\rightarrow G_{N}\rho$ and $p\rightarrow G_{N}p$. 

Denoting now derivatives with respect to $\tau$ and $\log{z}$ respectively by a dot  and a prime,
\begin{eqnarray}
\dot{ }\,\,\,\equiv {\partial\over\partial \tau}, \hspace*{1cm} {}'\equiv {\partial\over\partial\log z},
\end{eqnarray}
the Einstein equations become
\begin{eqnarray}
{M'\over M}+(d-3) &=& {d-3\over y}\left(1+{S'\over S}\right), \nonumber \\[0.2cm]
{\dot{M}\over M}+{M'\over M} &=& -{(d-3)k\over y}\left( {S'\over S}+{\dot{S}\over S}\right) ,
\label{ee2_1}\\[0.2cm]
a^{2} S^{-2}\left({M\over S^{d-3}}-1\right) &=& V_{z}^{2}\left({\dot{S}\over S}-{S'\over S}\right)^{2}
-\left(1+{S'\over S}\right)^{2} ,\nonumber 
\end{eqnarray}
where the dimensionless function $y(\tau,z)$ is defined by
\begin{eqnarray}
y={1\over 2}(d-2)(d-3){M\over \eta S^{d-1}}
\label{y}
\end{eqnarray}
and $V_{z}$ is given by Eq. (\ref{Vz}). Both $\alpha(t,r)$ and $a(t,r)$ have been integrated in (\ref{alpha_a})
and can now be expressed in terms of the new functions as
\begin{eqnarray}
\alpha=c_{\alpha}(\tau)\left({z^{2}\over \eta}\right)^{k\over k+1}, \hspace*{1cm}
a=\eta^{-{1\over k+1}}S^{2-d} .
\end{eqnarray}
Thus the velocity $V_{z}$ is known and
we are left with three functions $M(\tau,z)$, $S(\tau,z)$ and $y(\tau,z)$ [or $\eta(\tau,z)$] to be determined
from the three differential equations in (\ref{ee2_1}).

\paragraph{Self-similar solutions.}
We now impose self-similarity in the metric (\ref{line_element}) with respect to the conformal 
Killing vector field
\begin{eqnarray}
v=t{\partial_{t}}+r{\partial_{r}}
\end{eqnarray} 
that satisfies Eq. (\ref{conformal}). In the new coordinates (\ref{tau,z}) the
action of the homothety generated by $v$
simply shifts the new time variable $\tau\rightarrow\tau+
\mbox{constant}$. Thus, the condition of self-similarity translates into
the condition that all dimensionless functions
in the problem are independent functions of $\tau$ and depend only of the scaling coordinate $z$.
This is the case for the functions defined in Eqs. (\ref{newfunctions}) and (\ref{y}). Then, the Einstein 
equations are (cf. \cite{harada_maeda})
\begin{eqnarray}
{M'\over M}+(d-3) &=& {(d-3)\over y}\left(1+{S'\over S}\right) ,\nonumber \\[0.2cm]
{M'\over M} &=& -{(d-3)k\over  y}\,{S'\over S} ,
\label{ee2}\\[0.2cm]
a^{2} S^{-2}\left({M\over S^{d-3}}-1\right) &=& V_{z}^{2}\left({S'\over S}\right)^{2}
-\left(1+{S'\over S}\right)^{2} .\nonumber 
\end{eqnarray}
Using the first two equations we can solve for $M'$ and $S'$. The
third one can be interpreted as an equation defining $\eta(z)$ as a function of
$S(z)$ and $M(z)$. Taking its derivative with respect to $\log{z}$ and after a
tedious computation one arrives at the following system of Einstein equations 
\begin{eqnarray}
{d\log{M}\over d\log{z}} &=& {(d-3)k\over k+1}\left({1\over {y}}-1\right), \nonumber \\[0.2cm]
{d\log{S}\over d\log{z}} &=& {1\over k+1}({y}-1) ,
\label{ee_ss} \\[0.2cm]
{d\log{\eta}\over d\log{z}}&=& {1\over V_{z}^{2}-k}\left[{(1+k)^{2}\over d-2}\eta^{k-1\over k+1}
S^{4-2d}-(d-2)(y-1)V_{z}^{2}-2k\right] .\nonumber 
\end{eqnarray}
Since in deriving the last equation we have taken a derivative these equations have to 
supplemented by the third equation in (\ref{ee2})
at least at a particular point, and also by the expressions for $\alpha(z)$, 
$a(z)$ and $V_{z}(z)$
\begin{eqnarray}
a^{2} S^{-2}\left({M\over S^{d-3}}-1\right) &=& V_{z}^{2}\left({S'\over S}\right)^{2}
-\left(1+{S'\over S}\right)^{2} ,\nonumber \\[0.2cm]
\alpha(z) &=& c_{\alpha}\left({ z^{2}\over \eta}\right)^{k\over k+1} ,
\label{suppl}\\[0.2cm]
a(z) &=& \eta^{-{1\over k+1}}S^{2-d}, \nonumber \\[0.2cm]
V_{z}(z) &=& -c_{\alpha}^{-1}\left({z\over \eta}\right)^{1-k\over 1+k}S^{2-d} ,\nonumber 
\end{eqnarray}
where we have fixed the gauge by setting $c_{\alpha}(\tau)$ to a constant $c_{\alpha}$. 
Notice that the system still has a residual gauge freedom corresponding to constant
rescalings of the time coordinate $t\rightarrow \zeta t$, with $\zeta\neq 0$. This implies a rescaling 
of the constant $c_{\alpha}$ by $c_{\alpha}\rightarrow \zeta c_{\alpha}$. Without loss
of generality we require $\alpha(0)=1$.  A very important equation is the
form of the velocity field $V_R$ for self-similar solutions.  From (\ref{VR}) and the
self-similar ansatz  we can write
\begin{equation}
V_R\,=\,-V_z\,{{\dot S}+S'\over S\,+\,S'}\,=\,-V_z\,{1-y\over k+y}.
\label{vRincss}
\end{equation}
This equation implies in particular that, if $V_{z}\neq 0$, the zeroes of $V_R$ are in one-to-one
correspondence with the values of $z$ where $y=1$.

\paragraph{The sonic point.}
The Einstein equations (\ref{ee_ss}) present a number of singularities.  In principle
the possible singularities at $z=0$ and $z=\infty$ are expected from the fact that we
are looking for self-similar solutions. As in the case of regular singularities 
of linear differential equations, these singularities imply that the solutions
have a Frobenius behavior in their neighborhood.  Hence they behave as $z^a f(z)$,
where the power $a$ depends on whether we are at $z=0$ or $z=\infty$, but $f(z)$
is in any case an analytic function.

The singularity at $V_{z}^{2}=k$, the sonic point or sonic surface, is more delicate.
It represents the locus where the velocity of the lines 
of constant $z$ reach the speed of sound.  At this place shock waves can form.
This means that if we start integrating the solution from the origin (or from
infinity), we may not be able to cross the sonic surface analytically.
Geometrically, the line $z=z_{\rm sp}$ where $V_{z}(z_{\rm sp})^{2}=k$  
is a Cauchy horizon.  Choptuik's critical solution is everywhere analytic,
hence we need to impose stringent conditions at the sonic point so that 
the solution passes through smoothly.  These conditions will be explained in detail
later.

A generic solution of the system (\ref{ee_ss}) depends on three
initial conditions.  When requiring analyticity throughout, we
start by imposing it  at the origin.  As we will show below this
reduces the family of solutions to a one-parameter family.
If  we also require analyticity at the sonic point, we find
that only a discrete set of values of this parameter yield
admissible solutions.  Generically, the critical solutions
have no moduli.  They are classified by the zeroes of the
velocity field (\ref{VR}). In particular, the critical
solution we are interested in, the so-called Hunter-A solution \cite{hunter}, 
is characterized by having a single zero of 
$V_R$ \cite{ori_piran,harada_maeda}.  The lack of moduli characterizing
critical behavior is one of the main difficulties in trying an analytic
approach to the solutions of interest.  We can easily compute their
asymptotic behavior, but close to the sonic point, unfortunately, 
they only seem to be amenable to numerical analysis.

It is reasonably easy to compute the asymptotic properties of the
solution at the origin and at infinity.  However if we want to study
the solution near the sonic point, we need to impose some restrictions
there. In particular regularity of 
the third equation in (\ref{ee_ss}) 
at $z=z_{\rm sp}$ imposes the additional
constraint
\begin{eqnarray}
\left.\left[{(1+k)^{2}\over d-2}\eta^{k-1\over k+1}
S^{4-2d}-(d-2)(y-1)V_{z}^{2}-2k\right]\right|_{z=z_{\rm sp}}=0.
\label{regcond}
\end{eqnarray}
This, however, is not enough to determine $(\log\eta)'$ at the sonic point, 
since the right-hand side of the corresponding equation is still indeterminate.
In order to evaluate this derivative we have to define the right-hand side at
the sonic  point using l'H\^opital rule
\begin{eqnarray}
\left.{d\log{\eta}\over d\log{z}}\right|_{z_{\rm sp}}
=\left.{1\over (V_{z}^{2})'}\left[{(1+k)^{2}\over d-2}\eta^{k-1\over k+1}
S^{4-2d}-(d-2)(y-1)V_{z}^{2}-2k\right]'\right|_{z=z_{\rm sp}} .
\label{ee3}
\end{eqnarray} 

As we will show in detail in Subsection \ref{ambiguity}, 
evaluating the derivatives on the right-hand side of Eq. (\ref{ee3}) and 
using the other equations of motion leads to a quadratic equation to be 
satisfied by  $(\log{\eta})'$ at the  sonic
point.  This means that one has to determine which of the two solutions provides
the correct analytic solutions connecting smoothly the $z=0$ to $z=\infty$
through the sonic point. The ``wrong" solution would show catastrophic implosion
to form a black hole, an explosion to shed all collapsing matter to
infinity  or something in-between.  

In the critical solutions, on the other hand, there
is a balance between the  matter that collapses and the matter that
is ejected, so that eventually we only form a black hole of ``zero" mass.
It can be shown that once the first derivative  
of $(\log{\eta})'_{\rm sp}$ is determined, all higher derivatives can be
calculated unambiguously at the  sonic point through linear 
equations obtained by repeated application of the 
l'H\^opital rule.   
Indeed, the function $\log{\eta}$ can be expanded  
near $z=z_{\rm sp}$ in a power series in $\log(z/z_{\rm sp})$
\begin{eqnarray}
(\log\eta)'=\sum_{n=0}^{\infty}\mathcal{C}_{\eta}^{(n)}(y_{\rm sp})\log^{n}\left({z\over z_{\rm sp}}\right),
\label{series_sp}
\end{eqnarray}
where $y_{\rm sp}\equiv y(z_{\rm sp})$. 
The coefficients $\mathcal{C}_{\eta}^{(n)}(y_{\rm sp})$ are calculable from
(\ref{ee3}) and the equations obtained by taking further derivatives  
with respect to $\log{z}$ in the third equation in (\ref{ee_ss}). In Subsection \ref{ambiguity} we evaluate
the first coefficient of the expansion, $\mathcal{C}_{\eta}^{(0)}(y_{\rm sp})=(\log\eta)_{\rm sp}'$. The details of the
calculation of $\mathcal{C}_{\eta}^{(1)}(y_{\rm sp})=(\log\eta)_{\rm sp}''$ are given in Appendix B.

From the previous discussion it follows that one of the main problems we 
have to solve is to determine the correct sign in
the quadratic equation for $(\log{\eta})'_{\rm sp}$.  This input is crucial 
in the numerical evaluation of the critical solution.  Before
we get there, however,  we need to get acquainted with some asymptotic properties
of the solutions to Eqs. (\ref{ee_ss}).

\paragraph{Behavior near $\boldsymbol{z=0}$ and $\boldsymbol{z=\infty}$.}

A good place to start the numerical integration of (\ref{ee_ss})
in the search of the critical solution is the origin (or infinity).
Hence it is worth studying our system of differential equations in these regions.
Alternatively, we could start the numerical integration at the sonic point and integrate towards $z=0$ and $z=\infty$.
Although there are some advantages in doing this, we stick here to $z=0$  
and defer the analysis of this other case to a future work \cite{future_sp}.

We start the analysis by finding the 
conditions imposed by analyticity at $z=0$. By demanding regularity
at the origin, we find that the three-parameter 
family of general solutions to (\ref{ee_ss}) reduce to
a one-parameter family. 

The asymptotic behavior as $z\rightarrow 0^{+}$ can be found from the
first equation in (\ref{eom}). 
Since for a regular solution $R(t_{0},r)$ is expected  to be a single valued 
function of $r$, Eq. (\ref{ee}) can be formally
integrated at fixed time $t=t_{0}<0$
\begin{eqnarray}
m(t_{0},r)= {8\pi \over (d-1)(d-2)}\int_{0}^{R(r)}d(R^{d-1})\rho(t_{0},R^{d-1}).
\end{eqnarray}
For small values of the $r$-coordinate, $r\rightarrow 0^{+}$, one 
expects the coordinate $R(t,r)$ to vanish. Then 
the integral can be approximated by
\begin{eqnarray}
m(t_{0},r\sim 0) \approx {8\pi \rho(t_{0},0)\over (d-1)(d-2)}R(t_{0},r\sim 0)^{d-1}.
\end{eqnarray}
Writing now the function $y(t,r)$ defined in Eq. (2.18) in terms of $m(t,r)$, $\rho(t,r)$ and $R(t,r)$,
we find its limit when $z\rightarrow 0^{+}$ to be
\begin{eqnarray}
y(t,r) = (d-2)(d-3){m \over 8\pi \rho R^{d-1}} \hspace*{1cm} \Longrightarrow \hspace*{1cm}
y(t_{0},0^{+}) = {d-3\over d-1}.
\label{limitingy}
\end{eqnarray}
It is important to notice that this behavior of the function $y(t,r)$ as $r\rightarrow 0$ 
is completely general. 
It does not require the solution to be self-similar, and it is valid as 
well for the perturbations to the self-similar solution.  It only depends on
the regularity of the solution at the origin.

Once the limiting value of $y(z)$ as $z\rightarrow 0^{+}$ has been 
obtained it is possible to solve the first two equations 
in the system (\ref{ee_ss}) for small $z$. The fact that the value 
$y(0^{+})$ is universal, together with the quadratic relation
(\ref{ee2}), leads to the fact that the asymptotic behavior 
of the functions $M(z)$, $S(z)$ and $\eta(z)$ 
depends only on a single free parameter. Following \cite{harada_maeda}, 
we use the constant $D$ defined by
\begin{eqnarray}
2D=\lim_{r\rightarrow 0}\Big[8\pi t^{2}\rho(t,r)\Big].
\end{eqnarray}

From Eqs. (\ref{ee_ss}) in the limit $z\rightarrow 0^{+}$, 
together with the gauge condition on the metric
coefficient  $\alpha(z)\rightarrow 1$ in 
the same limit, one finds the asymptotic form of $M(z)$, $S(z)$ and $\eta(z)$ 
near $z\rightarrow 0^{+}$ as
\begin{eqnarray}
M(z) &\simeq &z^{2k\over k+1}
\left\{ {(2D)^{k\over k+1}\over (d-2)}\left[{k+1\over (d-1)k+d-3}\right]
+\sum_{n=1}^{\infty} M^{(n)}_{0} z^{2n{d-3+k(d-1)\over (d-1)(k+1)}}\right\}, \nonumber \\[0.3cm]
S(z) &\simeq & z^{-{2\over (d-1)(k+1)}} 
\left\{\left[{(2D)^{1\over k+1}\over k+1}\left(k+{d-3\over d-1}\right)\right]^{1\over 1-d}
+\sum_{n=1}^{\infty} S^{(n)}_{0} z^{2n{d-3+k(d-1)\over (d-1)(k+1)}}\right\},
\label{asymptotic_back_0}\\[0.3cm]
\eta(z) &\simeq & z^{2}\left[2D+\sum_{n=1}^{\infty} \eta_{0}^{(n)}z^{2n{d-3+k(d-1)\over (d-1)(k+1)}}\right].
 \nonumber 
\label{ic1}
\end{eqnarray}
Apart from an overall power of $z$, all functions admit an 
expansion in integer powers of the variable 
$z^{2{d-3+k(d-1)\over (d-1)(k+1)}}$. It should be noticed 
that the exponent of the expansion variable never vanishes
for physical values of $k$. 
Incidentally, the constant $c_{\alpha}$ can be written in terms of 
$D$ as $c_{\alpha}=(2D)^{k\over k+1}$. It is clear from 
this expression that $D$ is a gauge dependent quantity with respect to the
residual gauge transformations $z\rightarrow \lambda^{-1}z$, $c_{\alpha}\rightarrow 
\lambda c_{\alpha}$. 
We conclude from this exercise that analyticity at the  
origin gives a family of solutions depending on a 
single parameter $D$.

A further restriction on the possible values of $D$ comes from 
requiring that the solution is analytic at the other 
singular value of $z$, the sonic point $z=z_{\rm sp}$. 
Generically we find that only a discrete set of values of 
$D$ lead to a regular solution for all finite values of $z$. 
These correspond to the higher-dimensional
generalization of flat-Friedmann, general relativistic 
Larson-Penston and Hunter-A/B solutions \cite{harada_maeda}.
How these solutions are found numerically is discussed in 
detail in Sec. (\ref{numerics_back}).  

The behavior of the functions in the limit $z\rightarrow \infty$ can 
also be obtained using similar techniques. 
It can be shown that all functions $M(z)$, $S(z)$ and $\eta(z)$ tend 
to constants in that limit and that from the equations
of motion (\ref{ee_ss}) it follows that $y(z)\rightarrow 1$ when 
$z\rightarrow \infty$. Looking at the subleading 
terms, the expansion is of the form
\begin{eqnarray}
Z(z)=Z_{\infty}+Z_{\infty}^{(1)}z^{-{1-k\over 1+k}}+\ldots,
\label{beh_inf_back}
\end{eqnarray}
for any of the relevant functions collectively represented 
here by $Z(z)$.
All functions near $z=\infty$ can 
be expanded in inverse power series in the variable $z^{1-k\over 1+k}$. 
This expansion parameter is different from the one 
appearing when expanding the functions around $z=0$. It is
important to notice that this expansion parameter is independent
of the dimension.  Furthermore, if we look at the expression for
$V_z$ in (2.23), we find that its absolute value diverges
at $z=\infty$ precisely as $z^{1-k\over 1+k}$.  In fact, $V_z^2$
provides a good uniformizing variable for the system of equations
under study.  Indeed, $V_{z}^{2}$ is a monotonic function in the whole region
between $z=0$ and $z=\infty$.  This variable is essentially the
one used in the analysis of Ref. \cite{bicknell_henriksen}. 
We turn now to study the form of the field equations in
terms of this variable.  This  will be helpful in determining
the correct sign of $ (\log\eta)'$ at the sonic point.

\paragraph{The uniformization variable.}

One unpleasant aspect of the equations obtained so far 
is the presence of generically irrational exponents in both the 
equations and the initial conditions. This problem can 
be solved by choosing an appropriate uniformization variable
in terms of which integer powers are obtained. This can be 
done by noticing that the expansion of $V_{z}^{2}$
around $z=0$ has the form
\begin{eqnarray}
V_{z}^{2}=z^{2{d-3+k(d-1)\over (d-1)(k+1)}}
\left\{\left[{(2D)^{-{1\over (d-2)(k+1)}}\over (k+1)^{2}}\left(k+{d-3\over d-2}\right)
\right]^{2{d-2\over d-1}}+\sum_{n=1}^{\infty} V_{0}^{(n)} z^{2n{d-3+k(d-1)\over (d-1)(k+1)}}\right\}.
\end{eqnarray}
We have seen before how all functions, apart from 
an overall power of $z$, can be written as series expansion in 
$z^{2{d-3+k(d-1)\over (d-1)(k+1)}}$. Therefore, by 
redefining the functions properly in such a way that all the Frobenius
factors are absorbed, it should be possible to write them 
as a series expansion in integer powers of $V_{z}^{2}$. 
Moreover, for critical solutions $V_{z}(z)^{2}$ is a 
monotonous function of $z$ that can be used as the independent 
variable. 

We begin by defining a new function $\widehat{S}(z)$ as
\begin{eqnarray}
\widehat{S}=-(2D)^{-{1\over k+1}}z^{k-1\over k+1}V_{z}S^{2-2d}.
\end{eqnarray}
Using this new function together with $y(z)$ and $V_{z}(z)$, 
the system (\ref{ee_ss}) can be rewritten as
\begin{eqnarray}
{d\log{\widehat{S}}\over d\log{z}} &=& -{2(d-2)\over k+1}(y-1) \nonumber \\[0.2cm]
& & \,\, -\,\,
{(k-1)\over (k+1)(V_{z}^{2}-k)}\left[2k+(d-2)V^{2}_{z}(y-1)-2D {(k+1)^{2}\over d-2}\widehat{S}\right],
\nonumber \\[0.2cm]
{d\log{y}\over d\log{z}} &=& {k(d-3)\over k+1}\left({1\over y}-1\right)-{(d-1)\over k+1}(y-1) \nonumber \\[0.2cm]
& &\,\,+\,\,{1\over V_{z}^{2}-k}\left[2k+(d-2)V^{2}_{z}(y-1)-2D{(k+1)^{2}\over d-2}
\widehat{S}\right] ,
\label{eq}\\[0.2cm]
{d\log{V_{z}^{2}}\over d\log{z}} &=& {2-2k\over k+1}-{2(d-2)\over k+1}(y-1) \nonumber \\[0.2cm]
& & \,\,-\,\, {2(k-1)\over (k+1)(V_{z}^{2}-k)}\left[2k+(d-2)V^{2}_{z}(y-1)-2D {(k+1)^{2}\over d-2}
\widehat{S}\right] .
\nonumber 
\end{eqnarray}
After the introduction of $\widehat{S}(z)$ all 
functions appear with integer exponents on the right-hand
side of the equations. At the same time the parameter $D$ appears 
explicitly in the system to be solved.

The velocity $V_{z}(z)^{2}$ vanishes in the limit $z\rightarrow 0^{+}$. 
Similarly, the function $\widehat{S}$ has
a simple behavior in that limit as a function of $V_{z}^{2}$, namely
\begin{eqnarray}
\widehat{S} \sim V_{z}^{2}.
\end{eqnarray}
This shows that $V_{z}^{2}$ is a good candidate to be used as 
independent variable. With this  choice, the functions
$\widehat{S}(V_{z}^{2})$ and $y(V_{z}^{2})$ satisfy the following system of differential equations
\begin{eqnarray}
{d\log{\widehat{S}}\over d\log{V_{z}^{2}}} &=&
{-{2(d-2)}(y-1)-{k-1\over V_{z}^{2}-k}\left[2k+(d-2)V^{2}_{z}(y-1)-2D {(k+1)^{2}\over d-2}\widehat{S}\right]
\over 
{2-2k}-{2(d-2)}(y-1)- {2(k-1)\over V_{z}^{2}-k}\left[2k+(d-2)V^{2}_{z}(y-1)-2D {(k+1)^{2}\over d-2}
\widehat{S}\right]} ,\nonumber \\[0.2cm]
 & & \label{eqs_background}\\[0.2cm]
 {d\log{y}\over d\log{V_{z}^{2}}}  &=& {
 {k(d-3)}\left({1\over y}-1\right)-{(d-1)}(y-1)
+ {k+1\over V_{z}^{2}-k}\left[2k+(d-2)V^{2}_{z}(y-1)-2D{(k+1)^{2}\over d-2}\widehat{S}\right]  \over 
{2-2k}-{2(d-2)}(y-1)-
{2(k-1)\over V_{z}^{2}-k}\left[2k+(d-2)V^{2}_{z}(y-1)-2D {(k+1)^{2}\over d-2}
\widehat{S}\right]}.
\nonumber 
\end{eqnarray}

This system can be integrated in principle by requiring 
regularity at the sonic point $V_{z}^{2}=k$. Notice that
in these variables the regularity at these points leads again to 
the condition (\ref{regcond}). As a bonus, using these
variables we eliminate the constant $D$ from the initial 
conditions at $V_{z}^{2}=0$ to make it appear 
explicitly in the equations. In integrating the system numerically
we can use either Eqs. (\ref{ee_ss}) or Eqs. (\ref{eqs_background}). Both
systems are equivalent, although  the conceptual advantage of
the $V^2_z$ parametrization is that near the regular singular
points the functions can be expanded in integer powers of this variable.

Once the functions $\widehat{S}(V_{z}^{2})$ and $y(V_{z}^{2})$ are known, 
it is possible to find their dependence on the 
scaling coordinate $z$ by integrating the last equation in (\ref{eq}) 
to find $V_{z}=V_{z}(z)$. 
This procedure gives the same values of $D$ for the regular solutions 
as the ones obtained integrating the equations in terms 
of the independent variable $z$. The fact that the values of $D$ are 
the same is a consequence of the fact that in 
writing the equations (\ref{eqs_background}) we have 
not changed the gauge fixing conditions.

\vspace*{0.7cm}

\noindent
{\bf Shooting from the sonic point\footnote{This paragraph
can be read independently of the main argument line.  The reader can skip
this paragraph without losing the thread of the arguments. In a future
publication \cite{future_sp}
we plan to present the numerical integration from the sonic
point towards the origin and infinity. }.}
The system of ordinary differential equations (\ref{ee_ss}) can also be integrated from the 
sonic point $z=z_{\rm sp}$. In this case it is convenient
to fix the residual gauge freedom, $t\rightarrow \zeta t$, 
by locating the position of the sonic point at a given
value of the scaling coordinate, $z_{\rm sp}=\xi>0$. In this case 
the constant $c_{\alpha}$ is fixed by 
the equation defining the sonic point, $V_{z}(z_{\rm sp})^{2}=k$
\begin{eqnarray}
c_{\alpha}^{2}=k^{-1}\xi^{2{1-k\over 1+k}}\eta_{\rm sp}^{-2{1-k\over 1+k}}S_{\rm sp}^{4-2d},
\label{calpha}
\end{eqnarray}
where the subscript ``sp" indicate that the corresponding 
function is evaluated at the sonic point.

In the same way as the behavior of the solutions near the origin 
is determined by a single parameter $D$, 
the requirement of regularity at the sonic point forces now that the 
values of the functions at this point depend on
just one free parameter. We can take this to be the value of 
the function $y(z)$ at the sonic point, $y_{\rm sp}$. 
To see this we notice that there are two conditions that the 
functions should satisfy at the sonic point, namely 
the regularity condition (\ref{regcond}) and the quadratic equation in (\ref{ee2}) 
\begin{eqnarray}
2k+(d-2)k(y_{\rm sp}-1)-{(k-1)^{2}\over d-2}\eta_{\rm sp}a_{\rm sp}^{2} &=& 0 ,\nonumber \\[0.2cm]
(1+k)^{2}\left(1-{M_{\rm sp}\over S_{\rm sp}^{d-3}}\right)+k{S_{\rm sp}^2\over a_{\rm sp}^{2}}(1-y_{\rm sp})^{2}
-{S_{\rm sp}^{2}\over a_{\rm sp}^{2}}(k+y_{\rm sp})^{2}&=& 0 .
\label{regularity+quadratic}
\end{eqnarray}
The quantity $M_{\rm sp}$ can be now written in 
terms of $\eta_{\rm sp}$ by using the definition of the function 
$y(z)$ in Eq. (\ref{y}), whereas $a_{\rm sp}=\eta_{\rm sp}^{-{1\over k+1}}S_{\rm sp}^{2-d}$. 
In this way it is possible
to solve the previous two equations for 
$\eta_{\rm sp}$ and $S_{\rm sp}$ as a function of $y_{\rm sp}$. 
Moreover, by requiring analyticity of the functions at 
$z_{\rm sp}$ it is possible to show that
all derivatives of the functions at that point are also 
determined by $y_{\rm sp}$. 

Thus, we have found then that there is a one-parameter family of 
solution to the system (\ref{ee_ss}) 
regular at the sonic point. Only when we require in addition that the 
corresponding functions also show the behavior 
described previously as $z\rightarrow 0^{+}$ the set of allowed
functions restrict to a finite set of values of the initial 
condition $y_{\rm sp}$. These allowed values for each 
$k$ correspond to different types of solutions associated 
with the values of $D$ giving regular solutions
everywhere.

It is interesting to relate the initial condition $y_{\rm sp}$ at 
the sonic point with the parameter $D$ governing
the asymptotic behavior of the regular solution at $z=0$. 
The key ingredient to do this is to notice that 
the condition $c_{\alpha}=(2D)^{k\over k+1}$  
together with Eq. (\ref{calpha}) gives the following relation 
between $D$ and $y_{\rm sp}$
\begin{eqnarray}
2D=\xi^{1-k\over k}\left({1\over\sqrt{k}}S_{\rm sp}^{2-d}\right)^{k+1\over k}\eta_{\rm sp}^{k-1\over k},
\end{eqnarray}
where as shown above $S_{\rm sp}$ and $\eta_{\rm sp}$ 
are given in terms of $y_{\rm sp}$. The explicit
dependence of this expression on the  ``gauge fixing parameter" $\xi$ 
shows once more that $D$ is a gauge-dependent
quantity. In addition, from this expression it is immediate to verify 
that the combination $D\, z_{\rm sp}^{k-1\over k}$ is 
gauge invariant.

\subsection{Dealing with the sign ambiguity.}
\label{ambiguity}

Let us go back to the determination of $(\log\eta)'$ at the sonic point.  
The last step that we need to solve before embarking in the numerical 
solution of the self-similar Einstein equations is to fix somehow the ambiguity
associated with the choice of the right solution to the quadratic equation satisfied
by $(\log\eta)'_{\rm sp}$. This leads to a long and involved calculation. 
This is, in fact, one of the prices one has to pay for using comoving
coordinates.  In Schwarzschild coordinates a similar problem
appears \cite{hara_koike_adachi}
but it is technically easier to solve.  If the reader is willing
to give us the credit that we have solved the problem, this subsection
can be skipped.  The result however is very important in the numerical
determination of the critical solution.

To see how the quadratic equation for $(\log\eta)'_{\rm sp}$ comes about we go back to Eq. (\ref{ee3}).
In order to evaluate the right-hand side of this equation we need the expressions for 
$(V^{2}_{z})'$, $(a^{2}\eta)'$ and $y'$ evaluated at the sonic point. For the first one we use
the expression of $V_{z}^{2}$ in Eq. (\ref{suppl}) to write
\begin{eqnarray}
(V_{z}^{2})'_{{\rm sp}}&=&\left.\left\{2\left[{1-k\over 1+k}-{1-k\over 1+k}{\eta'\over \eta}+
(2-d){S'\over S}\right]V_{z}^{2}\right\}\right|_{z=z_{\rm sp}} \nonumber \\[0.2cm]
&=& {2k\over k+1}\left[{1-k}+
{(2-d)}(y_{\rm sp}-1)-{(1-k)}(\log\eta)'_{\rm sp}\right],
\label{vzeq}
\end{eqnarray}
where in the second line we have used the field equations (\ref{ee2}) to write the logarithmic derivatives
in terms of the values of the functions at the sonic point. For $\eta a^{2}$ we find also from Eq. (\ref{suppl})
\begin{eqnarray}
(\eta a^{2})'_{{\rm sp}}={1\over k+1}\eta_{\rm sp}a_{\rm sp}^{2}\Big[{(k-1)}(\log \eta)'_{\rm sp}+{2(2-d)}(y_{\rm sp}-1)\Big].
\end{eqnarray}
Finally, the derivative of the function $y$ can be written as
\begin{eqnarray}
(\log y)'_{{\rm sp}}={k(d-3)\over k+1}\left({1\over y_{\rm sp}}-1\right)-{d-1\over k+1}(y_{\rm sp}-1)-(\log\eta)'_{\rm sp}
\end{eqnarray}
Substituting these expressions in Eq. (\ref{ee3}) leads to the following quadratic equation for $(\log\eta)'_{\rm sp}$
\begin{eqnarray}
 {2k\over k+1}\left[{1-k}+
{(2-d)}(y_{\rm sp}-1)-{(1-k)}(\log\eta)'_{\rm sp}\right]\Big[(\log\eta)_{\rm sp}' +(d-2)(y_{\rm sp}-1)\Big]\nonumber \\[0.2cm]
= \,\,\,
{k+1\over (d-2)}\eta_{\rm sp}a_{\rm sp}^{2}\Big[{(k-1)}(\log \eta)'_{\rm sp}+{2(2-d)}(y_{\rm sp}-1)\Big] \hspace*{3cm}
\label{quadratic_eq_logeta1}\\[0.2cm]
+\,\,\,{(d-2)k\over  k+1}\left(y_{\rm sp}-1\right) \Big[(d-1)\left(y_{\rm sp}+k\right)-2k\Big]
+(d-2)k\,y_{\rm sp}(\log\eta)'_{\rm sp}.\hspace*{0.6cm}
\nonumber 
\end{eqnarray}

As discussed above, the values of all functions at the sonic point can be expressed in terms of a single parameter
that we can choose to be $y_{\rm sp}$. This means that the coefficients of the quadratic equation (\ref{quadratic_eq_logeta1})
can be written in terms of this single parameter. To do this we solve Eqs. (\ref{regularity+quadratic}) 
for the combinations $a_{\rm sp}^{2}\eta_{\rm sp}$ and $a_{\rm sp}^{2}S_{\rm sp}^{-2}$ with the result
\begin{eqnarray}
\eta_{\rm sp} a_{\rm sp}^{2}&=& {k(d-2)\over (k+1)^{2}}\Big[(d-2)y_{\rm sp}-(d-4)\Big] , \nonumber \\[0.2cm]
(1+k)^{2}a^{2}_{\rm sp} S^{-2}_{\rm sp} &=& (k+y_{\rm sp})^{2}-k(1-y_{\rm sp})^{2}+{2 k\over d-3}y_{\rm sp}
\Big[(d-2)y_{\rm sp}-(d-4)\Big] .
\label{ysp}
\end{eqnarray}
Substituting now these expressions in Eq. (\ref{quadratic_eq_logeta1}) we arrive at
\begin{eqnarray}
 {2}\Big[{1-k}+
{(2-d)}(y_{\rm sp}-1)-{(1-k)}(\log\eta)'_{\rm sp}\Big]\Big[(\log\eta)_{\rm sp}' +(d-2)(y_{\rm sp}-1)\Big]\nonumber \\[0.2cm]
= \,\,\,\Big[(d-2)y_{\rm sp}-(d-4)\Big]\Big[{(k-1)}(\log \eta)'_{\rm sp}+{2(2-d)}(y_{\rm sp}-1)\Big]\hspace*{1.5cm}\\[0.2cm]
+\,\,\,{(d-2)}\left(y_{\rm sp}-1\right) \Big[(d-1)\left(y_{\rm sp}+k\right)-2k\Big]
+(d-2)(k+1)y_{\rm sp}(\log\eta)'_{\rm sp}.
\nonumber 
\end{eqnarray}
As advertised, all coefficients depend only on $y_{\rm sp}$.

This quadratic equation can be written in the canonical form as
\begin{eqnarray}
[(\log\eta)'_{\rm sp}]^{2}+\mathcal{A}(y_{\rm sp})(\log\eta)'_{\rm sp}+\mathcal{B}(y_{\rm sp})=0,
\end{eqnarray}
where the coefficients are given by
\begin{eqnarray}
2(k-1)\mathcal{A}(y_{\rm sp})&=& 3d-2-(d+2)k-4(d-2)y_{\rm sp}. \nonumber \\
2(k-1)\mathcal{B}(y_{\rm sp}) &=& (d-2)(y_{\rm sp}-1)\Big[6-(d-1)(y_{\rm sp}+k)\Big].
\end{eqnarray}
The first thing to be noticed is that the equation for $(\log{\eta})'_{\rm sp}$ is independent of $\xi$ and 
therefore it is gauge invariant, as it is to be expected. Solving the equation 
\begin{eqnarray}
(\log\eta)'_{\rm sp} = -{\mathcal{A}(y_{\rm sp})\over 2}\pm \sqrt{{\mathcal{A}(y_{\rm sp})^{2}\over 4}-\mathcal{B}(y_{\rm sp})}.
\label{solutionquadratic}
\end{eqnarray}
There is therefore an ambiguity in the value of $(\log\eta)'_{\rm sp}$ given by the two possible signs in this equation.

The numerical integration of the equations (\ref{ee3}) from the origin $z=0$ shows that, for both 
the Hunter-A/B solutions, $(\log\eta)_{\rm sp}'$ is determined by the minus branch in Eq. (\ref{solutionquadratic}).
For these types of solutions, this branch is automatically selected by the system when regularity
at both the origin and the sonic point is imposed, {\it i.e.} when the constant $D$ takes the 
corresponding critical values.
The situation changes if we start the integration from the sonic point. In this case we find that the 
``correct" branch in Eq. (\ref{solutionquadratic}) has to be chosen in order to determine the right
initial conditions for the numerical integration. 
This situation will be addressed in detail in a separate publication
\cite{future_sp}.

\subsection{Numerical analysis}
\label{numerics_back}

After this long study of the properties of the equations of motion (\ref{ee_ss}) for the
self-similar background, we can proceed to integrate the equations 
numerically. To do so we use a fourth-order Runge-Kutta routine that 
integrates the system starting at $z=0$. As explained before, 
the main difficulty lies in finding the right
value of $D$ providing an analytic solution at the sonic point. 
For a generic value of $D$ the routine
gives a solution that diverges at this point.

In order to find the right value we perform a scan in $D$ to find
the smooth numerical solution 
at the sonic point $z_{\rm sp}$. Since we look for solutions that can be
interpreted as describing critical gravitational collapse of a perfect fluid, we are 
interested in the solution with a single unstable mode. In dimension larger than four,  
this corresponds
to the higher-dimensional generalization of the general relativistic 
Hunter-A solution, which are characterized by 
single zero of the function $V_{R}$. 

The numerical analysis
is a bit delicate due to the singular character of the sonic point, 
where minor numerical noise can easily lead
one astray. Here it is important to use the analytical information
we have at hand: the behavior of the solution at the origin
and the value of  the derivatives of the relevant functions
at the sonic point.  After long numerical work,
we have determined the critical solutions for different
values of $k$ in various dimensions, in particular $d=4,5,6$ and $7$.
We have included 
the $d=4$ computation in our analysis in order to compare with previous work, 
in particular Ref. \cite{hara_koike_adachi}. 

\begin{figure}[H]
\begin{center}
 \includegraphics[width=3.2in,angle=-90]{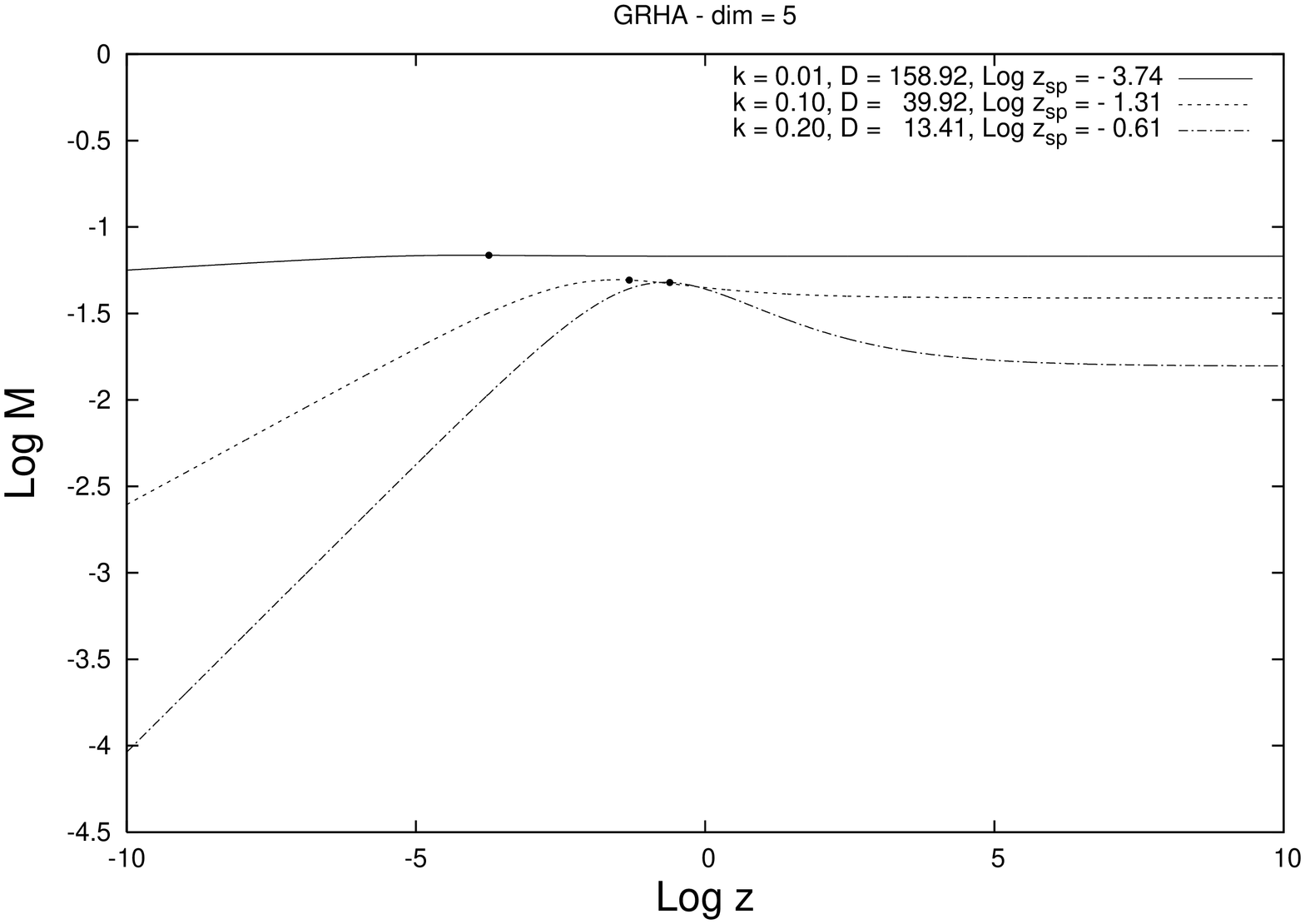} 
\\[0.4cm]
\includegraphics[width=3.2in,angle=-90]{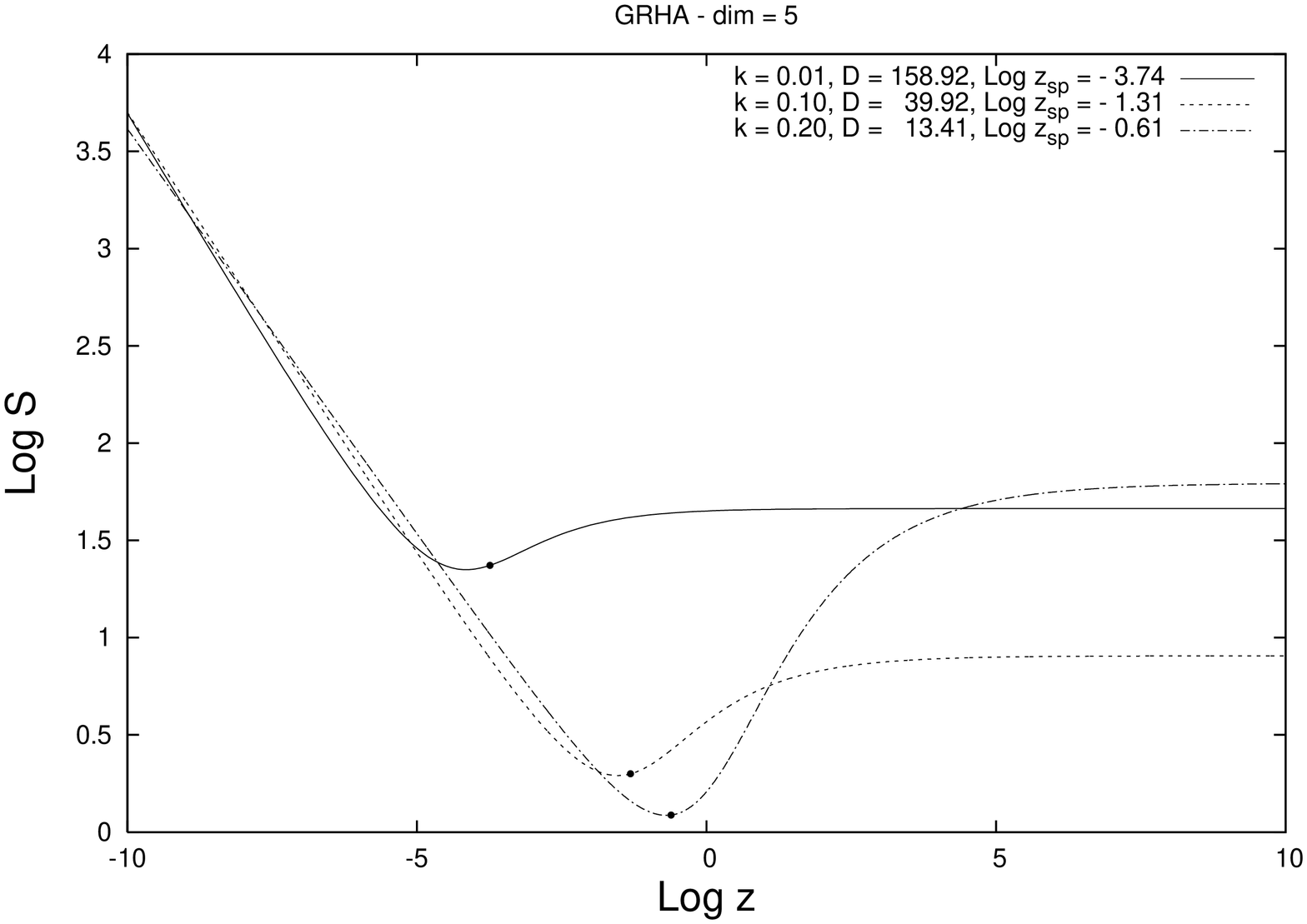}
\end{center}

 \caption{Plot of $M(z)$ and $S(z)$ for $k=0.01$, $k=0.1$ and $k=0.2$ 
 in five dimensions. The dot on 
 the curves indicates the position of the sonic point.}
\label{func_back}
\end{figure}

\begin{figure}[H]
\begin{center}
 \includegraphics[width=3.2in,angle=-90]{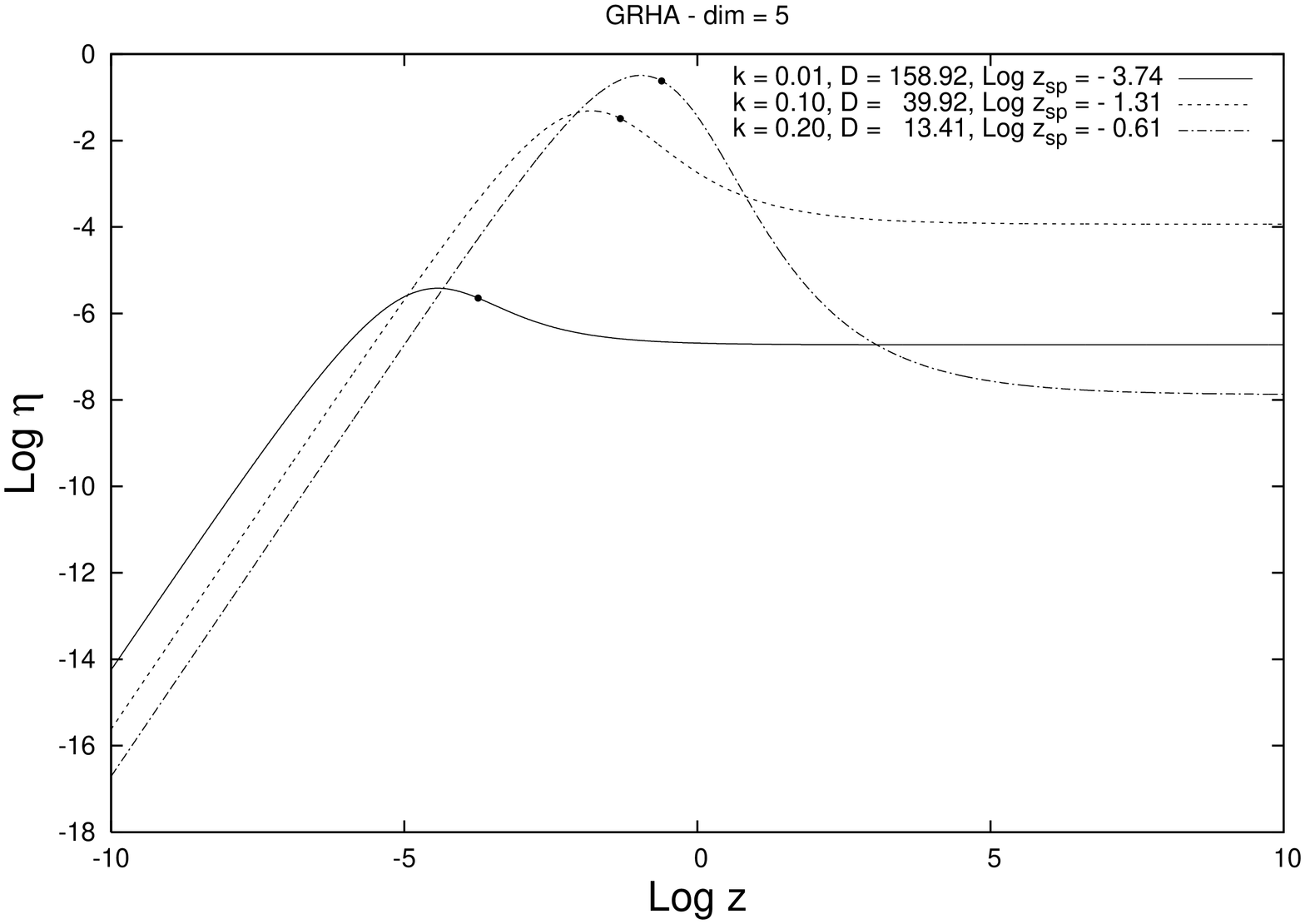}
 \\[0.4cm]
  \includegraphics[width=3.2in,angle=-90]{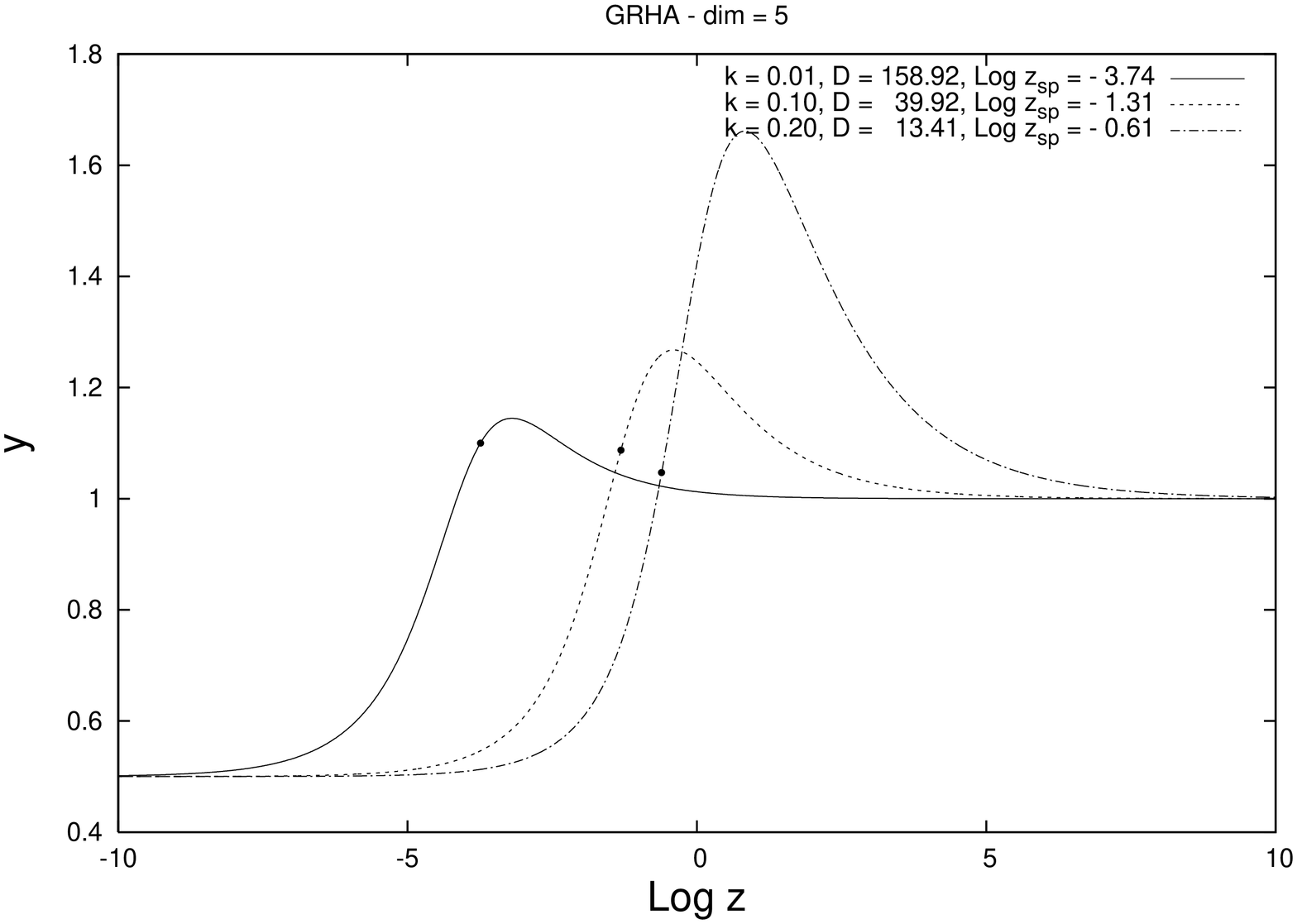}
  \end{center}
 \caption{Plot of  $\eta(z)$ and $y(z)$ for $k=0.01$, $k=0.1$ and $k=0.2$ 
 in five dimensions. The dot on 
 the curves indicates the position of the sonic point.}
  \label{func_back2}
\end{figure}

In Figs. \ref{func_back} and \ref{func_back2} we have plotted the profile of the
main functions for three values of $k$ in five dimensions. The position of the 
sonic point is denoted by a dot on the corresponding curve.  
We see that the curves have a structure completely similar to their four-dimensional counterparts,
as found for instance in Ref. \cite{harada_maeda}.

\begin{figure}[H]
 \begin{center}
 \includegraphics[width=3.2in,angle=-90]{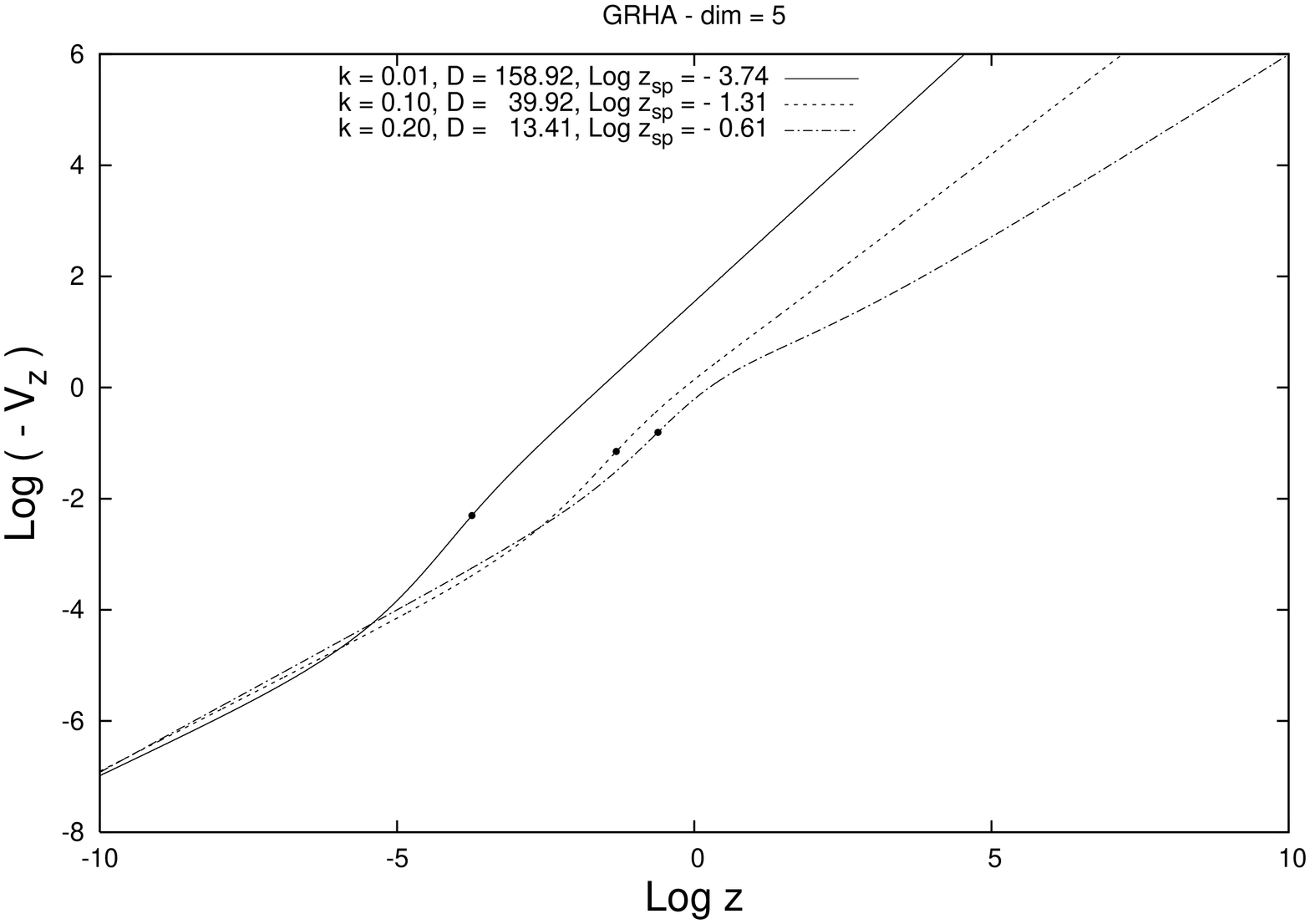} 
 \\[0.4cm]
  \includegraphics[width=3.2in,angle=-90]{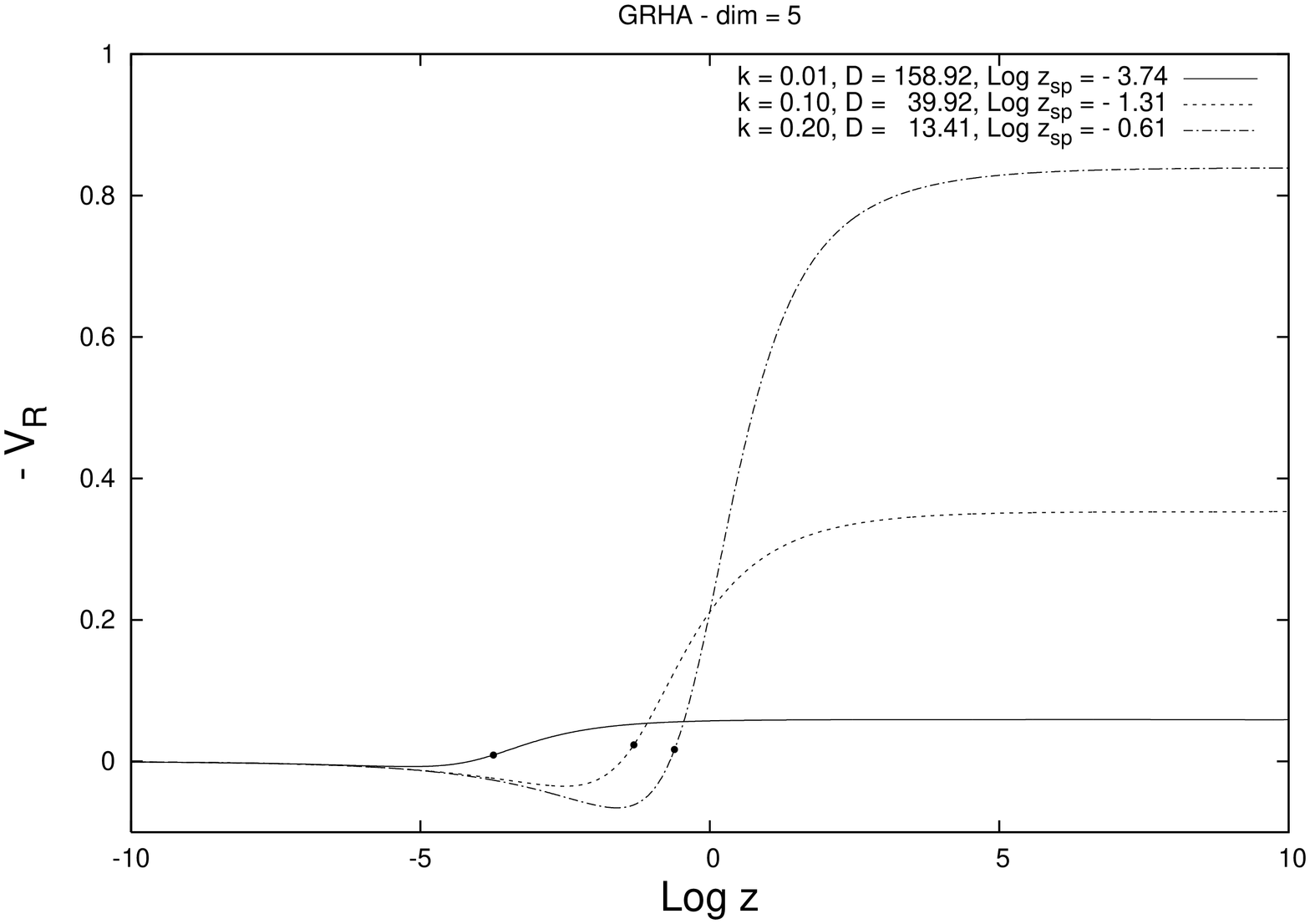}
  \end{center}
  \caption{Plot of the special velocities $V_{z}$ and $V_{R}$ 
  versus $z$ for $k=0.01$, $k=0.1$ and $k=0.2$ in $d=5$.}
    \label{thirdfigure}
\end{figure}

Fig. \ref{thirdfigure} shows how the function $V_{z}^{2}$ is monotonic in the whole range of 
values of $z$. This demonstrates that this function can be used as a good independent variable to describe
the evolution of the system.

\begin{figure}[H]
 \begin{center}
 \includegraphics[width=3.2in,angle=-90]{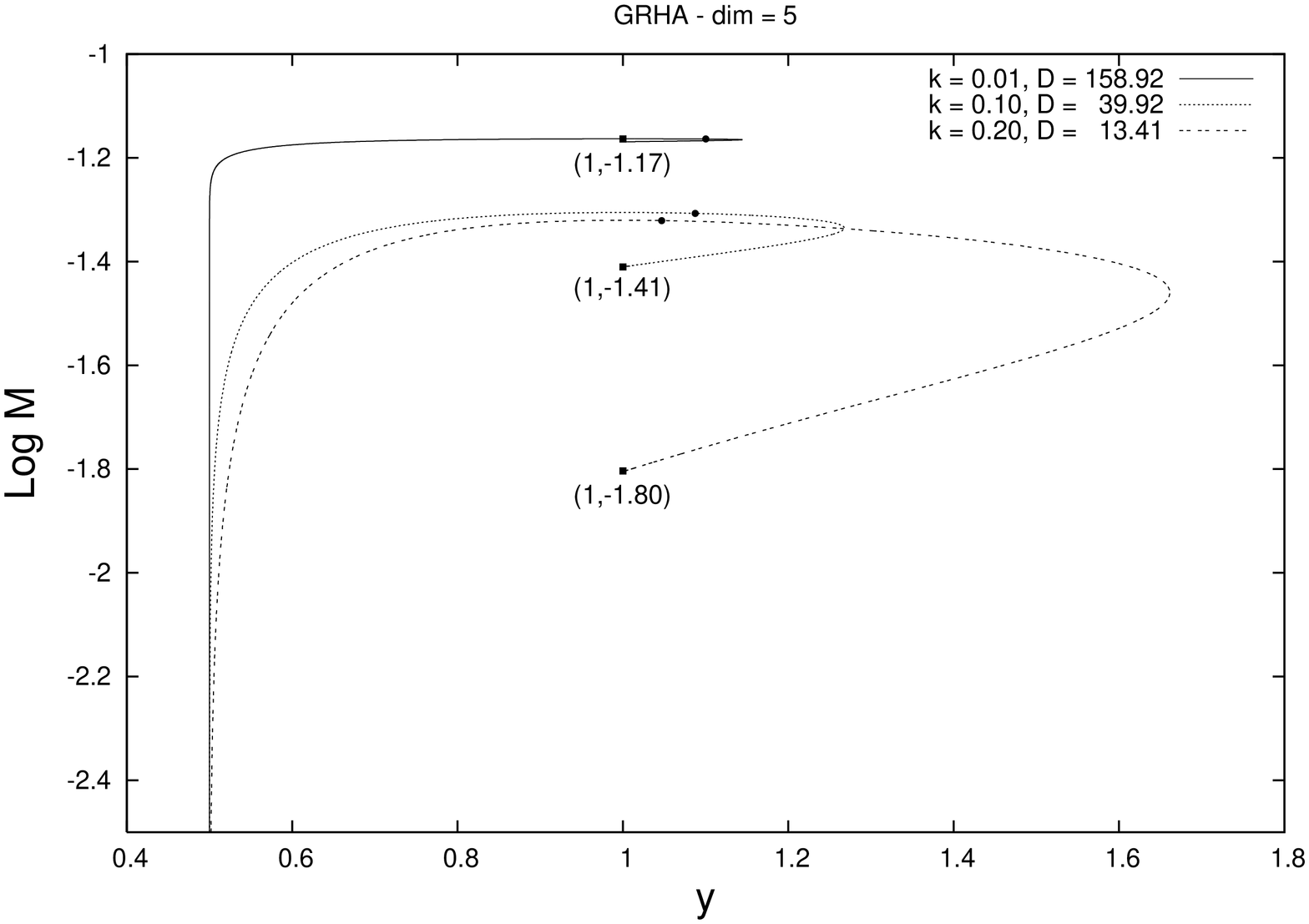}
 \\[0.4cm] 
 \includegraphics[width=3.2in,angle=-90]{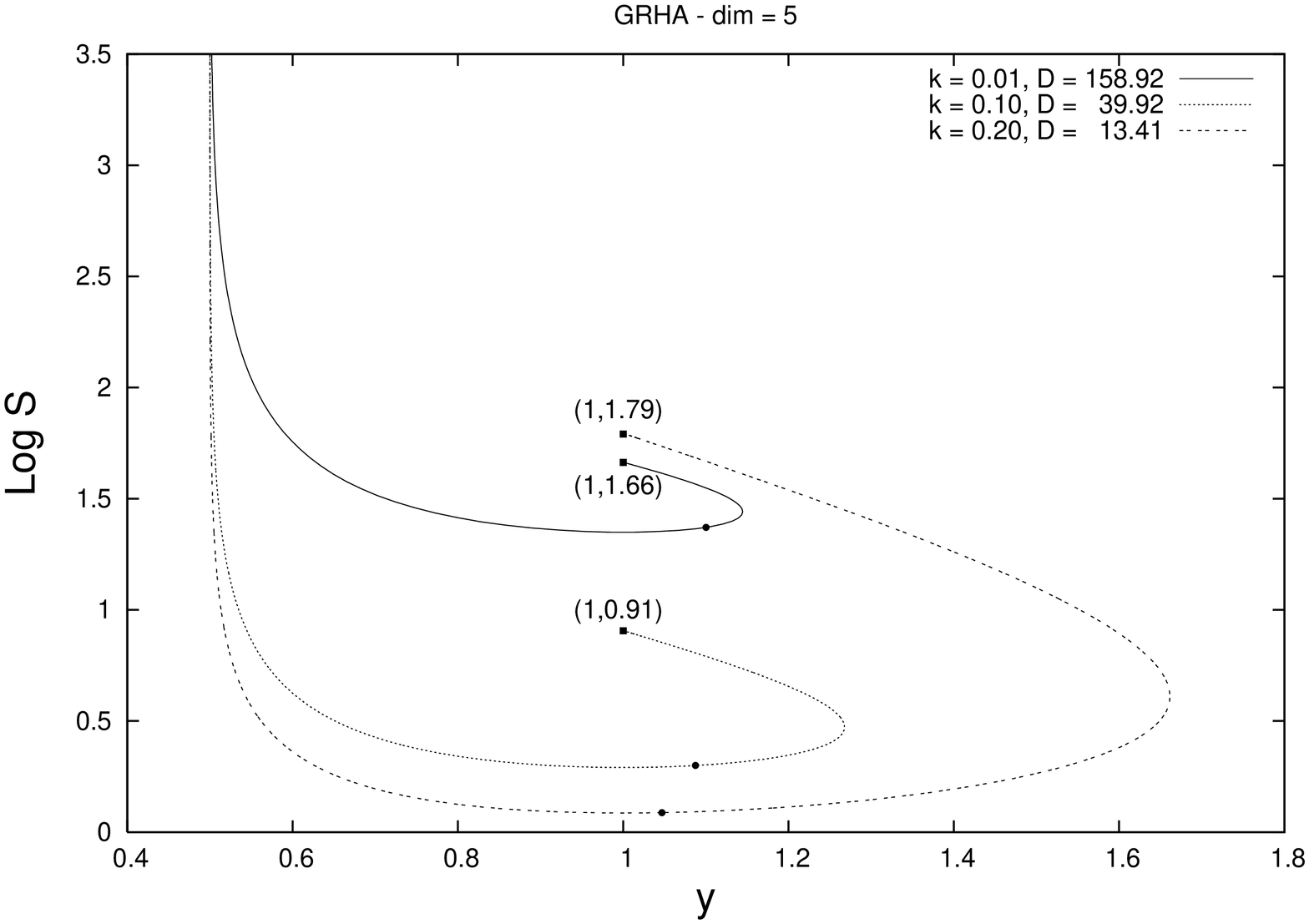}\hspace*{1cm}
 \end{center}
  \caption{Plot of $M$ and $S$ versus $y$ for $k=0.01$, $k=0.1$ and $k=0.2$ in five dimensions.
 The circular dot indicates on each curve the position of the sonic point. The 
 corresponding coordinates indicate the position at which the system is attracted as $z\rightarrow\infty$.}
 \label{dynamical}
 \end{figure}

\begin{figure}[H]
 \begin{center}
 \includegraphics[width=3.2in,angle=-90]{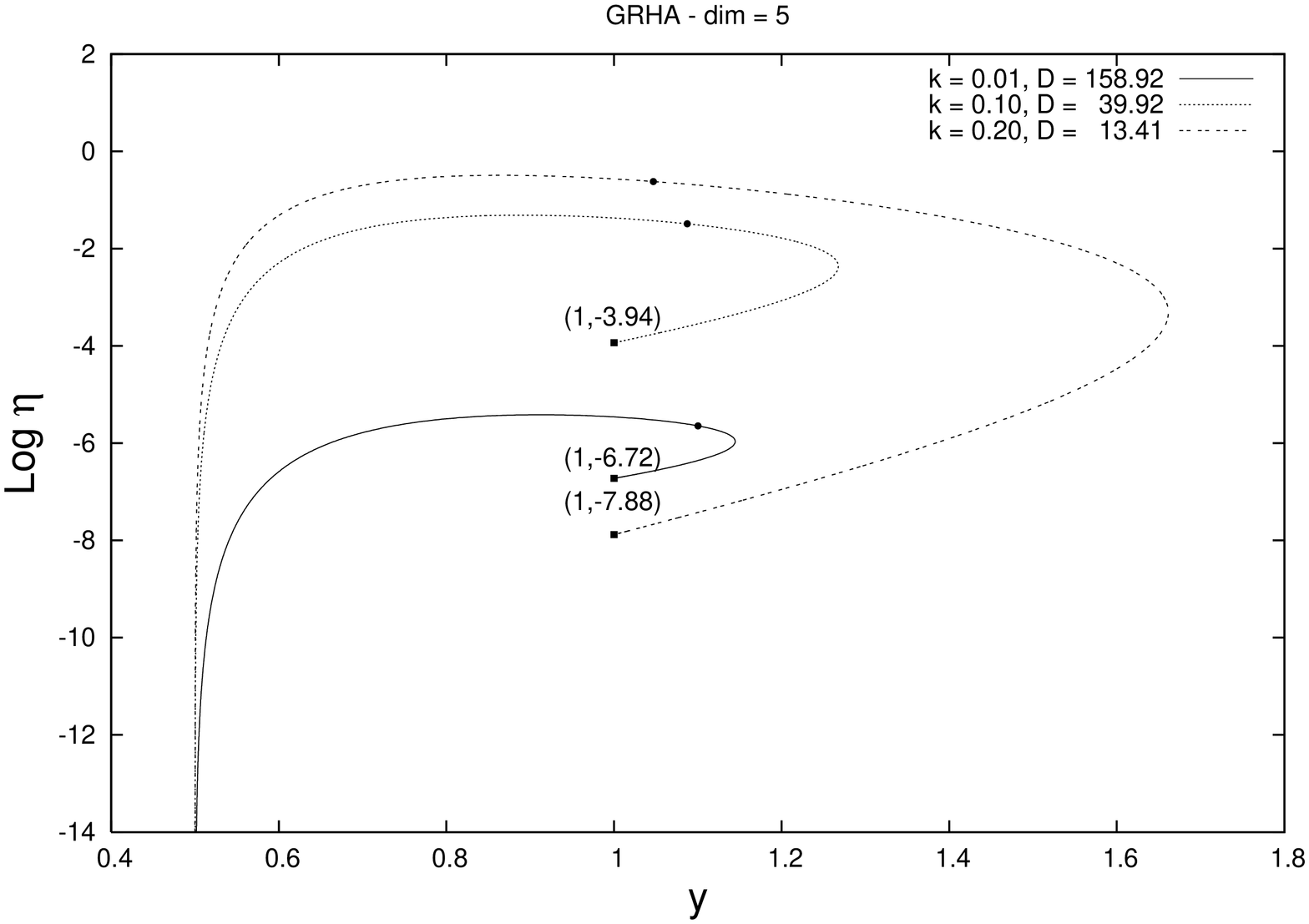}
 \\[0.4cm]
  \includegraphics[width=3.2in,angle=-90]{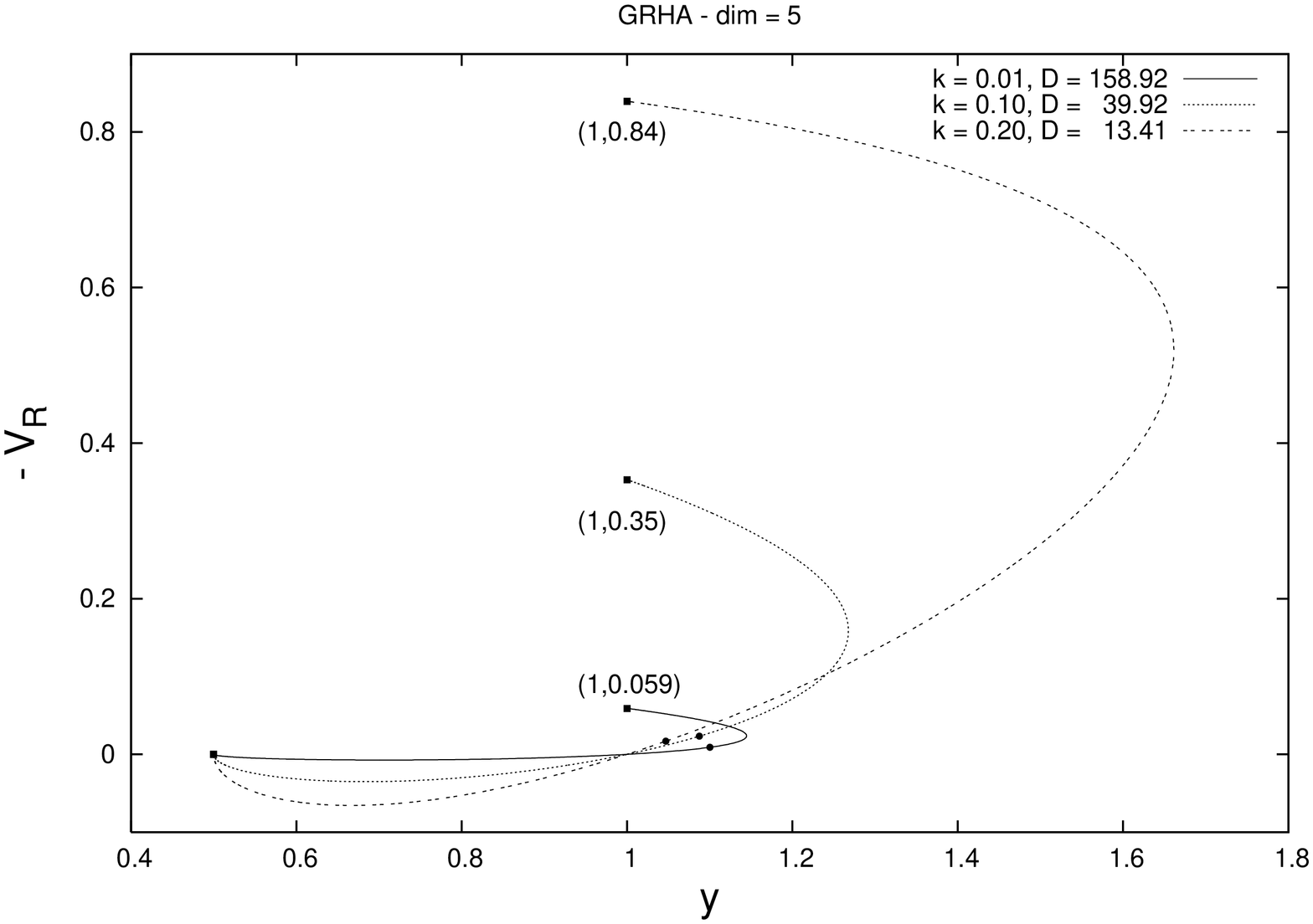}
  \end{center}
 \caption{Plot of $\eta$ and $V_{R}$ versus $y$ for $k=0.01$, $k=0.1$ and $k=0.2$ in five dimensions.
 The circular dot indicates on each curve the position of the sonic point. The 
 corresponding coordinates indicate the position at which the system is attracted as $z\rightarrow\infty$.}
  \label{dynamical2}
\end{figure}

Finally, in Figs. \ref{dynamical} and \ref{dynamical2} we plot the various functions of interest against $y$ instead of $\log z$.
This  is useful to show that we have indeed found the critical  solution.  As explained before the
critical solutions are characterized by the zeroes of the velocity field $V_R$ (see Fig. 3).
What Figs. \ref{dynamical} and \ref{dynamical2} show is a form of  critical  damping.  The solution tends to
its asymptotic value at $z=\infty$ after turning  around it  only once.  Critical solutions
with more than one zero in $V_R$ would spiral around their value at infinity a number of
times given by the number of zeroes of this velocity field.  Notice that $y$ 
cannot be used to uniformize the problem,  because it is not monotonous as a function of $z$. 
However, the fact that $y(z)$
only crosses the value $y=1$ (at which $V_{R}=0$) at a single finite value of $z$ before monotonously decreasing  to
$y=1$ as $z\rightarrow\infty$ is also a sign that the solution is critical.  Indeed, the  relation between
$V_R$ and $y$ given in Eq. (\ref{vRincss}) shows that for the critical solution can be also characterized as the one
for which the function $y(z)$ crosses the value $y=1$ at a single finite value of $z$.

\clearpage

\clearpage

\clearpage

\clearpage

 \clearpage
 
\clearpage

\section{Perturbations of the self-similar solutions}
\label{sec_pert}

Once we have obtained the critical self-similar solutions we 
proceed to study their stability by considering
perturbations around them. 
The perturbations break 
self-similarity and therefore have an explicit $\tau$ dependence. 
In the context of critical gravitational collapse
we are interested in solutions with a single growing mode.
The analysis is long, and in particular  we need to use
a very relevant equation that does not seem to appear
in the literature (see the paragraph on the ${\dot\eta}$-equation).  
The sonic point,
together with a new singularity that appears in the perturbation
equations at $y(z)=1$, play an important r\^ole in the analysis.
Similarly to the calculation of the self-similar brackground, 
analyticity at all singular points 
gives the key to the computation of critical exponents.

\subsection{Analysis of the perturbations.}

\paragraph{The equations for the perturbations.}
Let us denote by $Z(\tau,z)$ any of the functions involved in the problem. 
We consider perturbations around the self-similar 
solution $Z(z)$ of the form
\begin{eqnarray}
Z(\tau,z)=Z(z)\left[1+\epsilon\,e^{\lambda\tau}\,Z_{1}(z)+\ldots\right],
\label{perturbations}
\end{eqnarray}
where $\lambda$ is the corresponding Lyapunov exponent. 
Since we are interested in the mode that grows as
$\tau\rightarrow \infty$, the unstable mode is associated 
with a positive value of $\lambda$. The dots in the equation 
include the contribution of stable modes with negative Lyapunov exponents. 

In order to write a set of differential equations for the 
perturbations $M_{1}(z)$, $S_{1}(z)$ and $\eta_{1}(z)$ we have
to go back to the full equations of motion (\ref{ee_ss}) and 
plug in the ansatz (\ref{perturbations}). Then we collect all 
the terms linear in  the expansion parameter $\epsilon$. From the first two equations in (\ref{ee2_1}) 
we find the equations of the perturbations $M_{1}(z)$ and $S_{1}(z)$ to be
\begin{eqnarray}
(1+k)yM_{1}' &=& (3-d)ky_{1}-\lambda\left[M_{1}y+(d-3)kS_{1}\right], \nonumber \\[0.2cm]
(1+k)S_{1}' &=& y y_{1}-{\lambda\over d-3}\left[M_{1}y+(d-3)k S_{1}\right].
\label{pert1}
\end{eqnarray}
We still lack one more equation for the perturbation of a third function that we chose to be $y_{1}(z)$. 
To find it we have to use the third equation in 
(\ref{ee2_1}) and expand to first order
in $\epsilon$. After a long calculation one arrives at
\begin{eqnarray}
\mathcal{Q}y_{1} = \mathcal{D}_{M}M_{1}+\mathcal{D}_{S}S_{1},
\label{pert2}
\end{eqnarray}
where all the ``coefficients'' are functions of $z$ given by 
\begin{eqnarray}
\mathcal{Q} &=& (V_{z}^{2}-k)(1-y)(k+y), \nonumber \\[0.2cm]
\mathcal{D}_{M} &=& k V_{z}^{2}(1-y)^{2}-(k+y)^{2}-{(1+k)^{3}\over (d-2)(d-3)}y\,\eta^{k-1\over k+1}
S^{4-2d} \nonumber \\[0.2cm]
& & \,\,+\,\, {\lambda(k+1)\over (d-3)}y\left[V^{2}_{z}(1-y)+k+y\right] ,
\\[0.2cm]
\mathcal{D}_{S} &=& [1-k(d-2)]V_{z}^{2}(1-y)^{2}-k(d-1)(k+y)^{2}+{(1+k)^{3}\over (d-2)}y\,\eta^{k-1\over k+1}
S^{4-2d}\nonumber \\[0.2cm]
& & \,\,+\,\,\lambda(k+1)\left[V_{z}^{2}(y-1)+k(k+y)\right] .\nonumber 
\end{eqnarray}

Expressions (\ref{pert1}) and (\ref{pert2}) give the relevant equations governing the perturbations.
We can actually reduce them to a system of two differential equations by 
plugging the value of the perturbation 
$y_{1}$, obtained from Eq. (\ref{pert2}), into Eqs. (\ref{pert1}). As a result one gets the following 
system of two first order differential equations for the perturbations $M_{1}(z)$ and $S_{1}(z)$
\begin{eqnarray}
M_{1}' &=& \mathcal{A}_{M}M_{1}+\mathcal{B}_{M}S_{1} ,\nonumber \\[0.2cm]
S_{1}'&=& \mathcal{A}_{S}M_{1}+\mathcal{B}_{S} S_{1},
\label{pert22}
\end{eqnarray}
where $\mathcal{A}_{M}$, $\mathcal{B}_{M}$, $\mathcal{A}_{S}$ and $\mathcal{B}_{S}$ are given 
respectively by
\begin{eqnarray}
\mathcal{A}_{M}&=& -{(d-3)k\mathcal{D}_{M}+\lambda \mathcal{Q} y\over (k+1)\mathcal{Q}y},
\nonumber \\[0.2cm]
\mathcal{B}_{M}&=& -(d-3)k{\mathcal{D}_{S}+\lambda\mathcal{Q}\over (k+1)\mathcal{Q} y}, \nonumber \\[0.2cm]
\mathcal{A}_{S}&=& y{(d-3)\mathcal{D}_{M}-\lambda\mathcal{Q}\over (d-3)(k+1)\mathcal{Q}} , \\[0.2cm]
\mathcal{B}_{S}&=&{y\mathcal{D}_{S}-\lambda k\mathcal{Q}\over (k+1)\mathcal{Q}}. \nonumber
\end{eqnarray}

\paragraph{Singularities and the $\boldsymbol{\dot{\eta}}$-equation.}

As in the case of the background equations one has to be careful about potential singularities in the 
equations for the perturbations. In particular, by looking at 
the expression of $y_{1}$ in Eq. (\ref{pert2}) and the form
of the coefficients $\mathcal{Q}$, $\mathcal{D}_{M}$ and $\mathcal{D}_{S}$, we find that 
there are singular points corresponding to the zeroes
of $\mathcal{Q}$. This can 
happen at three points. The first one is again
the sonic point $z_{\rm sp}$ where $V_{z}^{2}=k$.
The second point is the value of $z_{a}$ at which
$y(z_{a})=1$. The third possibility, $y(z)=-k$, 
is never realized for physical solutions with $k>0$ because the 
function $y(z)$ is always positive definite. 

In the case of the self-similar background,
regularity at the sonic point implied the
extra condition (\ref{regcond}) to be satisfied 
by the physically admissible solutions.
In the case of the perturbations an extra condition 
can be also extracted by considering the perturbations of
the quadratic equation in (\ref{ee2}) in a consistent way. 
In particular we take the derivative of this
equation with respect to $\tau$ and express the 
result in terms of the functions $S(\tau,z)$, $\eta(\tau,z)$
and $y(\tau,z)$. After these manipulations the 
resulting expression can be written as a linear equation for 
$\dot{\eta}$ of the form 
\begin{eqnarray}
\mathcal{C}_{1}\dot{\eta}+\mathcal{C}_{2}=0,
\label{etadot}
\end{eqnarray}
where the coefficients $\mathcal{C}_{1}$ and 
$\mathcal{C}_{2}$ are given in terms of the functions 
$M(\tau,z)$, $S(\tau,z)$ and their derivatives. 

For the self-similar ansatz, Eq. (\ref{etadot}) reduces 
to the second equation in (\ref{ee_ss}). If, on the other hand,
we consider perturbations of the self-similar solution this expression contains 
important information about
the regularity of the perturbed solution at the points at which $\mathcal{Q}$ vanishes. 
To see this, we insert the perturbations (\ref{perturbations})
into (\ref{etadot}) and use Eqs. (\ref{pert1}) and (\ref{pert2}) 
to obtain a relation between $M_{1}$ and $S_{1}$.
A long computation yields
\begin{eqnarray}
\mathcal{G}_{M}M_{1}+\mathcal{G}_{S}S_{1}=0,
\label{identity}
\end{eqnarray}
where $\mathcal{G}_{M}$ and $\mathcal{G}_{S}$ 
are respectively given by
\begin{eqnarray}
\mathcal{G}_{M}&=& {2(k+1)\over d-2}\eta^{k-1\over k+1}S^{4-2d}\mathcal{A}_{S}
-{2\lambda\over k+1}\Big[V_{z}^{2}+k-(d-3)(d-2)(V_{z}^{2}-1)y\Big]\mathcal{A}_{S} \\[0.2cm]
& & \hspace*{-2cm}
+\,\,{2\lambda\over (k+1)^{3}}\left({\mathcal{D}_{M}-\mathcal{Q}\over \mathcal{Q}}\right)\Big\{
(1-k)V_{z}^{2}\Big[(d-3)(d-2)y-1\Big]^{2}-(k+1)^{2}S^{2-2d}\eta^{-{2\over k+1}}
\Big(2S^{2}\eta y-1\Big)  \Big\}
 \nonumber
\end{eqnarray}
and
\begin{eqnarray}
\mathcal{G}_{S} &=& 2\eta^{k-1\over k+1}\Big[(k+1)\mathcal{B}_{S}+(d-3)(d-2)y-1\Big]S^{4-2d}+
{2\lambda\over k+1}\left[{\mathcal{D}_{S}-(d-1)k\mathcal{Q}\over \mathcal{Q}}\right]
\eta^{-{2\over k+1}}S^{2-2d} \nonumber \\[0.2cm]
&+& {2\lambda\over k+1}\eta^{k-1\over k+1}\left[k(k+1)+(d-2)(3kd+d-5k-3)y-2(d-2){\mathcal{D}_{S}\over
\mathcal{Q}}y\right]S^{4-2d} \\[0.2cm]
&-&{2\lambda\over k+1}\mathcal{B}_{S}\Big[V_{z}^{2}+k-(d-3)(d-2)(V_{z}^{2}-1)y\Big] 
-{2\lambda\over k+1}V_{z}^{2}\Big[(d-3)(d-2)y-1\Big]
\nonumber \\[0.2cm]
&+&{2\lambda\over (k+1)^{3}} V_{z}^{2}\Big[(d-2)(d-3)y-1\Big]^{2}\left[2dk-3k-1+
(k-1){\mathcal{D}_{S}\over \mathcal{Q}}\right].
\nonumber 
\end{eqnarray}

The coefficients $\mathcal{Q}$, $\mathcal{D}_{M}$, $\mathcal{D}_{S}$, $\mathcal{B}_{S}$ and 
$\mathcal{A}_{S}$ are the ones defined above. We see that the only poles in 
$\mathcal{G}_{S}$ and $\mathcal{G}_{M}$ come from the zeros of
$\mathcal{Q}$, {\it i.e.}  the points at which $y=y_{\rm sp}$, $y=1$ and $y=-k$. 
Thus, Eq. (\ref{identity}) can be written as a sum of terms 
potentially divergent 
at $\mathcal{Q}=0$ plus those which are regular at 
those points. After some manipulations we find that this equation has a structure of the form
\begin{eqnarray}
\mathcal{F}\,{\mathcal{D}_{M}M_{1}+\mathcal{D}_{S}S_{1}\over \mathcal{Q}}
+\mbox{regular terms}=0,
\label{identity2}
\end{eqnarray}
where we have introduced the coefficient $\mathcal{F}$ defined by
\begin{eqnarray}
\mathcal{F}&=&{2\over d-2}y\eta^{k-1\over k+1}S^{4-2d}-{2\lambda\over (k+1)^{2}}
y\Big[V_{z}^{2}+k-(d-3)(d-2)(V_{z}^{2}-1)y\Big] \\[0.2cm]
&-& {2\lambda\over (k+1)^{3}}\left\{(k-1)V_{z}^{2}\Big[(d-2)(d-3)y-1\Big]^{2}+(k+1)^{2}S^{2-2d}\eta^{-{2\over k+1}}
\Big(2S^{2}y\eta-1\Big)\right\}.
\nonumber 
\end{eqnarray}

In order to preserve the identity (\ref{identity2}) 
one must require that the whole expression is regular
at the zeroes of $\mathcal{Q}$. Since numerically it 
can be shown that $\mathcal{F}$ never vanishes
at those points we arrive at the conclusion that 
\begin{eqnarray}
\left.\Big(\mathcal{D}_{M}M_{1}+\mathcal{D}_{S}S_{1}\Big)\right|_{\mbox{{\tiny zeroes of }}\mathcal{Q}}=0.
\label{condition_for_perturbation}
\end{eqnarray}
In particular, this equation has to be satisfied 
at the sonic point and provides a nontrivial condition to 
be satisfied by the perturbations there.

Notice that this condition has been derived here as a straightforward consequence of the perturbations 
of Eq. (\ref{etadot}), which becomes trivial after 
imposing the self-similar ansatz.  It should be said that
Eq. (\ref{condition_for_perturbation}) plays  a basic r\^ole in the determination of
the Lyapunov exponents of the critical solutions, and
that to our  knowledge it has never appeared in previous
literature.  Without including it, the path to understanding
the properties of critical perturbations is rather treacherous.
A systematic computation of the  Choptuik exponents for
different values of $k$ and $d$ relies heavily on it.

\paragraph{Behavior of the perturbations at the sonic point.}

In order to negociate the singularities of the differential equation (\ref{pert2}) 
we impose analyticity at the singular points.
For this we have to use again the l'H\^opital rule to define 
the perturbation $y_{1}(z)$ in terms of the corresponding ones for the other functions. 
This means that in the regions near the points 
where $\mathcal{Q}(z)= 0$ we can replace Eq. (\ref{pert2}) by
\begin{eqnarray}
\mathcal{Q}'y_{1}=(\mathcal{D}_{M}M_{1}+\mathcal{D}_{S}S_{1})'.
\end{eqnarray}
At the same time, instead of using Eq. (\ref{pert22}) we consider the modified system
\begin{eqnarray}
M_{1}'&=& \widetilde{\mathcal{A}}_{M}M_{1}+\widetilde{\mathcal{B}}_{M}S_{1}, \nonumber \\[0.2cm]
S_{1}'&=& \widetilde{\mathcal{A}}_{S}M_{1}+\widetilde{\mathcal{B}}_{S} S_{1}.
\label{reduced_systempert}
\end{eqnarray}
The new coefficients $\widetilde{\mathcal{A}}_{M}$, $\widetilde{\mathcal{B}}_{M}$, 
$\widetilde{\mathcal{A}}_{S}$ and $\widetilde{\mathcal{B}}_{S}$ are given by
\begin{eqnarray}
\mathcal{R}\,\widetilde{\mathcal{A}}_{M} &=& 
-{(d-3)k\over y}\mathcal{D}_{M}'+\lambda\left({k+y\over k+1}\right)\mathcal{D}_{S},
\nonumber \\[0.2cm]
\mathcal{R}\,\widetilde{\mathcal{B}}_{M} &=& 
-{(d-3)k\over y}\left\{\mathcal{D}_{S}'+\lambda\left[\mathcal{Q}'-\left({k+y\over k+1}
\right)\mathcal{D}_{S}\right]\right\} ,
\label{coefficients_pert} \\[0.2cm]
\mathcal{R}\,\widetilde{\mathcal{A}}_{S} &=& 
y\left\{\mathcal{D}_{M}'-\lambda\left[{\mathcal{Q}'\over d-3}+{k+y\over (k+1)y}
\mathcal{D}_{M}\right]\right\} ,\nonumber \\[0.2cm]
\mathcal{R}\,\widetilde{\mathcal{B}}_{S} &=& 
y\mathcal{D}_{S}'-\lambda k\left[\mathcal{Q}'+(d-3){k+y\over (k+1)y}\mathcal{D}_{M}
\right],
\end{eqnarray}
where the function $\mathcal{R}(z)$ is defined by
\begin{eqnarray}
\mathcal{R}=(k+1)\mathcal{Q}'-y\mathcal{D}_{S}+(d-3){k\over y}\mathcal{D}_{M}.
\end{eqnarray}
In order to solve the previous system, the derivatives of the different coefficients are required.
To keep here the exposition simple, these expressions are given in Appendix C.

\paragraph{Decoupling of the perturbations.}

In this paragraph we show how the perturbations $M_{1}$ and $S_{1}$ can be decoupled from
one another. Although outside the mail line of the paper, the analysis presented here may 
provide a different way to understand the perturbations to the critical solutions. In any case
the paragraph can be skipped in a first reading.

Going back to the equations for the perturbation (\ref{pert22}), we can formally integrate each of the
equations to find the following system of integral equations for $M_{1}$ and $S_{1}$ (keep in mind
that the prime indicates derivation with respect to $\log{z}$)
\begin{eqnarray}
M_{1}(z)&=& e^{\int_{-\infty}^{\log{z}}dx\,\mathcal{A}_{M}(x)}\left[M_{1}(-\infty)+
\int_{-\infty}^{\log{z}}dx\,
\mathcal{B}_{M}(x)S_{1}(x)\, e^{-\int_{-\infty}^{x}dx'\,\mathcal{A}_{M}(x')}\right] ,\nonumber \\[0.2cm]
S_{1}(z)&=& e^{\int_{-\infty}^{\log{z}}dx\,\mathcal{B}_{S}(x)}\left[S_{1}(-\infty)+
\int_{-\infty}^{\log{z}}dx\,
\mathcal{A}_{S}(x)M_{1}(x)\, e^{-\int_{-\infty}^{x}dx'\,\mathcal{B}_{S}(x')}\right].
\end{eqnarray}
Using again (\ref{pert22}) they  become
\begin{eqnarray}
M_{1}(-\infty)+
\int_{-\infty}^{\log{z}}dx\,
\mathcal{B}_{M}(x)S_{1}(x)\, e^{-\int_{-\infty}^{x}dx'\,\mathcal{A}_{M}(x')}&=&  
e^{-\int_{-\infty}^{\log{z}}dx\,\mathcal{A}_{M}(x)}\left[{S_{1}'-\mathcal{B}_{S}S_{1}\over
\mathcal{A}_{S}}\right],
\nonumber \\[0.2cm]
S_{1}(-\infty)+
\int_{-\infty}^{\log{z}}dx\,
\mathcal{A}_{S}(x)M_{1}(x)\, e^{-\int_{-\infty}^{x}dx'\,\mathcal{B}_{S}(x')} &=& 
e^{\int_{-\infty}^{\log{z}}dx\,\mathcal{B}_{S}(x)}\left[{M_{1}'-\mathcal{A}_{M}M_{1}\over
\mathcal{B}_{M}}\right].\,\,\,\,\,
\end{eqnarray}
Taking now a derivative with respect to $\log{z}$ on both equations, and after some manipulations, we
arrive at a set of two decoupled second-order linear differential 
equations for the perturbations
$M_{1}$ and $S_{1}$
\begin{eqnarray}
M_{1}''+\mathcal{P}_{M}M_{1}'+\mathcal{Q}_{M}\,M_1 &=& 0 ,\nonumber \\
S_{1}''+\mathcal{P}_{S}S_{1}'+\mathcal{Q}_{S}\, S_1 &=& 0,
\label{decoupledsys}
\end{eqnarray}
where the coefficients are given by
\begin{eqnarray}
\mathcal{P}_{M} &=& -\left(\mathcal{A}_{M}+\mathcal{B}_{S}+{\mathcal{B}_{M}'\over 
\mathcal{B}_{M}}\right), \nonumber \\[0.2cm]
\mathcal{P}_{M} &=& -\left(\mathcal{A}_{M}+\mathcal{B}_{S}+{\mathcal{A}_{S}'\over 
\mathcal{S}_{S}}\right), \\[0.2cm]
\mathcal{Q}_{M} &=& \mathcal{A}_{M}\mathcal{B}_{S}-\mathcal{A}_{S}\mathcal{B}_{M}-\mathcal{A}_{M}'
+\mathcal{A}_{M}{\mathcal{B}_{M}'\over \mathcal{B}_{M}}, \nonumber \\[0.2cm]
\mathcal{Q}_{S} &=& \mathcal{A}_{M}\mathcal{B}_{S}-\mathcal{A}_{S}\mathcal{B}_{M}-\mathcal{B}_{S}'
+\mathcal{B}_{S}{\mathcal{A}_{S}'\over \mathcal{A}_{S}}.\nonumber 
\end{eqnarray}

The system of equations (\ref{decoupledsys}) can be written in a more suggestive form
by a further function redefinition
\begin{eqnarray}
M_{1}(z)&=&\psi_{M}(z) e^{-{1\over 2}\int_{-\infty}^{\log{z}}dx\,\mathcal{P}_{M}(x)}, \nonumber \\[0.2cm]
S_{1}(z)&=&\psi_{S}(z) e^{-{1\over 2}\int_{-\infty}^{\log{z}}dx\,\mathcal{P}_{S}(x)},
\end{eqnarray}
where the new functions $\psi_{M}(z)$ and $\psi_{S}(z)$ satisfy one-dimensional
Schr\"odinger equations
\begin{eqnarray}
\psi_{M}''+\mathcal{V}_{M}\,\psi_{M}&=& 0, \nonumber \\[0.2cm]
\psi_{M}''+\mathcal{V}_{S}\,\psi_{S}&=& 0.
\label{schroedinger}
\end{eqnarray}
The potentials for these equations are given in terms of the coefficients of Eq. (\ref{decoupledsys})
by
\begin{eqnarray}
\mathcal{V}_{M} &=& \mathcal{Q}_{M}-{1\over 2}\mathcal{P}_{M}'-{1\over 4}\mathcal{P}_{M}^{2}, \nonumber\\[0.2cm]
\mathcal{V}_{M} &=& \mathcal{Q}_{S}-{1\over 2}\mathcal{P}_{S}'-{1\over 4}\mathcal{P}_{S}^{2} .
\end{eqnarray}
Hence, we can obtain the perturbations to the self-similar background by solving a system of
two Schr\"odinger equations with the appropriate boundary conditions. The potential associated
with these equations is completely determined by the self-similar solution. 
Once (\ref{schroedinger}) is solved, the perturbations for the other functions, 
$y_{1}(z)$ and $\eta_{1}(z)$, can be written in terms of $\psi_{M}(z)$ and $\psi_{S}(z)$.  This formulation
of the  pertubations can provide a good arena for qualitative analysis.  We will however
proceed with a direct study of the original equations.

\paragraph{Asymptotic behavior of the perturbations near $\boldsymbol{z=0}$ and $\boldsymbol{z=\infty}$.}

In order to solve the perturbations $M_{1}(z)$, $S_{1}(z)$ and $y_{1}(z)$ by integration of the 
system of differential equations (\ref{pert22}) starting at $z=0$ we need to determine
the behavior of these functions at the origin. To do so, we start by noticing that the function
$y(t,r)$ (including the perturbation) should have the limiting value determined in Eq. (\ref{limitingy}).
Because this condition is satisfied already by the self-similar background solution we conclude that
the perturbation $y_{1}(z)$ has to vanish in this limit
\begin{eqnarray}
y_{1}(0^{+})=0.
\label{conditionpert1}
\end{eqnarray}
Hence, studying the system (\ref{pert1}) around $z=0$ one arrives 
at the identity
\begin{eqnarray}
M_{1}'(z)+k(d-1)S_{1}'(z)=-\lambda\Big[M_{1}(z)+k(d-1)S_{1}(z)\Big] \hspace*{1cm} (z\rightarrow 0^{+}).
\end{eqnarray}
For positive $\lambda$ (as it is the case of the growing mode we are 
interested in), the integration of
this equation results in the combination $M_{1}(z)+k(d-1)S_{1}(z)$ 
blowing up at $z\rightarrow 0^{+}$ 
as $z^{-\lambda}$. Therefore, in order to impose regularity of 
the perturbations at the origin we require the numerical coefficient of this divergent terms to be 
zero. This leads to the condition
\begin{eqnarray}
\left.\Big[M_{1}(z)+k(d-1)S_{1}(z)\Big]\right|_{z=0^{+}}=0.
\label{conditionpert2}
\end{eqnarray}
Equations (\ref{conditionpert1}) and (\ref{conditionpert2}) 
give the initial conditions for the numerical
integration of the system of differential equations (\ref{pert1}) at $z=0$. 
Notice that since we have a linear system, only the ration $M_1/S_1$ at
the origin is relevant.  Furthermore, the same conclusions can 
be reached from the equations determining $y_1$ from $M_1$ and $S_1$
together with standard asymptotic analysis at $z=0$.

The functions $M_{1}(z)$ and $S_{1}(z)$ can be written, 
around $z=0$, as a power series.
From the reduced system of equations (\ref{reduced_systempert}) 
it can be seen that they can be
expressed as
\begin{eqnarray}
M_{1}(z) &=& M_{1}(0)+\sum_{n=1}^{\infty} M_{1}^{(n)} z^{2n{d-3+k(d-1)\over (d-1)(k+1)}}, \nonumber \\[0.2cm]
S_{1}(z) &=& S_{1}(0)+\sum_{n=1}^{\infty} S_{1}^{(n)} z^{2n{d-3+k(d-1)\over (d-1)(k+1)}}.
\end{eqnarray}
Comparing with Eq.  (\ref{asymptotic_back_0}) we learn that 
the expansion variable here is the same as 
the one for the self-similar background solution.

We now consider the behavior of the perturbations at infinity. 
Since in that limit the function $y(z)$ tends to one,
we find that the system of equations (\ref{pert1}) simplifies for large $z$ to
\begin{eqnarray}
(1+k)M_{1}'&=& (3-d)ky_{1}-\lambda[M_{1}+(d-3)kS_{1}] ,\nonumber \\[0.2cm]
(1+k)S_{1}'&=& y_{1}-{\lambda\over d-3}[M_{1}+(d-3)kS_{1}],
\end{eqnarray}
leading to 
\begin{eqnarray}
\Big[M_{1}(z)+k(d-3)S_{1}(z)\Big]'=-\lambda\Big[M_{1}(z)+k(d-3)S_{1}(z)\Big] .
\end{eqnarray}
For the growing mode, $\lambda>0$, this means that the 
combination $M_{1}(z)+k(d-3)S_{1}(z)$
vanishes at infinity as
\begin{eqnarray}
M_{1}(z)+k(d-3)S_{1}(z) \sim z^{-\lambda} \hspace*{1cm} (z\rightarrow \infty).
\end{eqnarray}

To get the individual behavior of $M_{1}(z)$ and $S_{1}(z)$ 
we look at the asymptotic form of $y_{1}(z)$
for large $z$. From Eq. (\ref{pert2}), using the asymptotic 
expansion of the background functions
when $z\rightarrow\infty$, we arrive at 
\begin{eqnarray}
y_{1}(z) \sim {\lambda\over d-3}\Big[M_{1}(z)-(d-3)S_{1}(z)\Big]\hspace*{1cm} (\mbox{when $z\rightarrow \infty$)}.
\end{eqnarray}
Plugging this in Eq. (\ref{pert1}) we get the following equations, valid for large $z$,
\begin{eqnarray}
M_{1}'(z)&=& -\lambda M_{1}(z), \nonumber \\[0.2cm]
S_{1}'(z)&=& -\lambda S_{1}(z),
\end{eqnarray}
implying
\begin{eqnarray}
M_{1}(z) &=&M_{1}^{(\infty)}z^{-\lambda}+\ldots \nonumber \\[0.2cm]
S_{1}(z)&=&S_{1}^{(\infty)}z^{-\lambda}+\ldots
\end{eqnarray}

To summarize, we have found that for $\lambda>0$ (the growing mode) 
all perturbations $M_{1}(z)$, $S_{1}(z)$ and $y_{1}(z)$ 
go to zero at infinity as powers of $z$. On the other hand, for small $z$ the 
only perturbation that vanishes is $y_{1}(z)$, while $M_{1}(z)$
and $S_{1}(z)$ tend to constants in that limit satisfying the condition (\ref{conditionpert2}).

\paragraph{The gauge mode.}

In studying the perturbations to the self-similar solutions 
of the Einstein equations it is very important to 
eliminate those perturbations which can be written as 
mere gauge transformations of the background solutions.
In particular, by performing changes of coordinates on 
the self-similar solutions it is possible to introduce 
an explicit dependence on time that is just a gauge artifact. 
One can consider reparametrizations
in time \cite{hara_koike_adachi}, leaving invariant the 
form of the line element (\ref{line_element}). Infinitesimally
\begin{eqnarray}
r\longrightarrow r, \hspace*{1cm} t\longrightarrow t+\varepsilon(t).
\label{coordinate_change}
\end{eqnarray}
This changes the scaling coordinate $z$ by 
\begin{eqnarray}
\tau\longrightarrow \tau+\delta(\tau), \hspace*{1cm}
z\longrightarrow z\Big[1+\delta(\tau)\Big],
\end{eqnarray}
where the function $\delta(\tau)$ is defined in terms of $\varepsilon(t)$ as
\begin{eqnarray}
\delta(\tau)=-{\varepsilon\left(-\ell_{s}e^{-\tau} \right)\over \ell_{s}e^{-\tau}}.
\end{eqnarray}

The functions $M(t,r)$, $S(t,r)$ and $y(t,r)$ only change under (\ref{coordinate_change}) 
through their dependence on $t$. In the case of the self-similar solutions we find
\begin{eqnarray}
Z(z)\longrightarrow Z(z)+\delta(\tau)Z'(z)=Z(z)\Big\{1+\delta(\tau)[\log Z(z)]'\Big\},
\end{eqnarray}
where $Z$ stands for $M$, $S$ and $y$. This has precisely the form of the perturbation (\ref{perturbations})
if we choose the arbitrary function $\delta(\tau)$ of the form
\begin{eqnarray}
\delta(\tau)=\epsilon\,{\Delta\over 2} e^{\lambda \tau},
\end{eqnarray}
with $\Delta$ any real constant. The perturbations 
$M_{1}(z)$, $S_{1}(z)$ and $y_{1}(z)$ are then given by
\begin{eqnarray}
M_{1}(z)_{\rm gauge}&=&{\Delta\over 2}\Big[\log{M(z)}\Big]', \nonumber \\[0.2cm]
S_{1}(z)_{\rm gauge}&=& {\Delta\over 2}\Big[\log{S(z)}\Big]' ,
\label{gauge_modep}\\[0.2cm]
y_{1}(z)_{\rm gauge}&=& {\Delta\over 2}\Big[\log{y(z)}\Big]' .\nonumber
\end{eqnarray}

Using the behavior of the self-similar solutions near $z=0$ it 
is straightforward to see that the perturbations
(\ref{gauge_modep}) automatically satisfy the boundary condition 
(\ref{conditionpert2}). However, in order to 
satisfy the boundary condition at $z=\infty$ we find from 
(\ref{beh_inf_back}) that the Lyapunov exponent
$\lambda$ has to be
\begin{eqnarray}
\lambda_{\rm gauge}={1-k\over 1+k}.
\end{eqnarray}
In particular, we see that the gauge mode is independent of the space-time dimension.

\subsection{Numerical study of the perturbations and calculation of the Choptuik exponent}

As in the case of the self-similar solution, the equations for 
the perturbations have to be numerically 
integrated using the background solutions obtained numerically in 
Sec. \ref{numerics_back}.
We use a strategy similar to the one applied there. 
We numerically integrate the system (\ref{pert22}) 
starting at $z=0$ using a fourth-order Runge-Kutta routine. 
We find that for generic values of the 
Lyapunov exponent $\lambda$ the perturbation is singular both 
at the sonic point and at the value
$z=z_{a}$ where $y(z_{a})=1$. We then scan in $\lambda$ to find 
a critical value for which the solution smoothly crosses both points. 
At the same time we ignore the value of $\lambda$ associated with the gauge mode.
The singularities of  the equations lead to some numerical instabilities,
but the use of l'H\^opital rule explained above gives a guidance to the numerical
integration.  A good deal of work leads to the results for the  Choptuik exponents
in Table \ref{tab1} for dimensions $d=4,5,6$ and $7$.  As an important check of our results
we agree with previous results in $d=4$ \cite{hara_koike_adachi,harada_maeda}.

\begin{table}
\begin{center}
\begin{tabular}{||l|r|r|r|r||}
\hline
$k$ & $\lambda_{d=4}$ & $\lambda_{d=5}$ & $\lambda_{d=6}$ & $\lambda_{d=7}$ \\ \hline 
0.01  &  8.747 & 4.435 & 3.453 & 3.026  \\
0.02 &  8.140 &  4.288 & 3.376 & 2.974 \\ 
0.03 &  7.617 & 4.152 & 3.302  & 2.924 \\
0.04 &  7.163 &  4.027  & 3.233  & 2.876 \\
0.05 &  6.764 & 3.911 & 3.169  & 2.831 \\
0.06 &  6.412 & 3.804 & 3.107  & 2.788 \\
0.07 &  6.099 & 3.703 & 3.049  & 2.746 \\
0.08 &  5.818 &  3.609 & 2.993  & 2.706 \\
0.09 & 5.565 & 3.521  & 2.940  & 2.668 \\
0.10 & 5.334 &  3.438  & 2.890  & 2.631 \\
0.11 & 5.124 & 3.360  & 2.841  & 2.595 \\
0.12 & 4.932 & 3.286  & 2.795  & 2.561 \\
0.13 & 4.756 &  3.216  & 2.751  & 2.527 \\
0.14 & 4.593 &  3.149  & 2.708  & 2.494 \\
0.15 & 4.442 &  3.086   & 2.667  & 2.464 \\
0.16 &  4.301 & 3.026  & 2.627  & 2.433 \\
0.17 & 4.170 & 2.968  & 2.589 & 2.414 \\
0.18 & 4.048 &  2.913  & 2.552 & 2.377 \\
0.19 &  3.933 & 2.860  & 2.517 & 2.348 \\
0.20 &  3.825 & 2.809  & 2.482 & 2.321 \\
0.21 &  3.723 & 2.760  & 2.449 & 2.297 \\ 
0.22 &  3.627 &  2.713  & 2.417 & 2.272 \\
0.23 &  3.536 & 2.668  & 2.386 & 2.246 \\
0.24 &  3.449 &  2.625  & 2.355 & 2.224 \\
0.25 & 3.367 & 2.583   & 2.325 & 2.202 \\ 
\hline
\end{tabular}\hspace*{2cm}
\begin{tabular}{||l|r|r|r|r||}
\hline
$k$ & $\gamma_{d=4}$ & $\gamma_{d=5}$ & $\gamma_{d=6}$ & $\gamma_{d=7}$ \\ \hline 
0.01& 0.114  &0.225 & 0.290 & 0.330\\
0.02 & 0.123 & 0.233 & 0.296 & 0.336 \\
0.03 & 0.131 & 0.241  & 0.303 & 0.342 \\
0.04 & 0.140 & 0.248 & 0.309 & 0.348 \\
0.05 & 0.148 & 0.256  & 0.316 & 0.353 \\
0.06 & 0.156 & 0.263  & 0.322 & 0.359 \\
0.07 & 0.164 & 0.270 & 0.328 & 0.364 \\
0.08 & 0.172 & 0.277 & 0.334  & 0.369 \\
0.09 & 0.180 & 0.284 & 0.340 & 0.375 \\
0.10 & 0.187 & 0.291  & 0.346 & 0.380 \\
0.11 & 0.195 & 0.298 & 0.352 & 0.385 \\
0.12 & 0.203 & 0.304 & 0.358  & 0.390 \\
0.13 & 0.210 & 0.311 & 0.364 & 0.396 \\
0.14 & 0.218 & 0.318 & 0.369  & 0.401 \\
0.15 & 0.225 & 0.324 & 0.375 & 0.406 \\
0.16 & 0.232 & 0.330 & 0.381 & 0.411 \\
0.17&  0.240 & 0.337 & 0.386 & 0.416 \\
0.18 & 0.247 & 0.343  & 0.392 & 0.421 \\
0.19 & 0.254 & 0.347 & 0.397 & 0.426\\
0.20 & 0.261 & 0.356 & 0.403 & 0.431\\
0.21 & 0.259 & 0.362 & 0.408 & 0.435 \\
0.22 & 0.276 & 0.368 & 0.414 & 0.440 \\
0.23 & 0.283 & 0.375 & 0.419 & 0.445 \\
0.24 & 0.290 & 0.381 & 0.425 & 0.450 \\
0.25 & 0.297 & 0.387 & 0.430 & 0.454 \\
\hline
\end{tabular}
\end{center}
\caption{Values of the Lyapunov exponent (left) and the Choptuik exponent (right) as a function 
of $k$ for $d=$4, 5, 6 and 7. All numbers are calculated with a precision $\pm 0.001$.}
\label{tab1}
\end{table}

\begin{figure}[H]
 \begin{center}
 \includegraphics[width=3.5in,angle=-90]{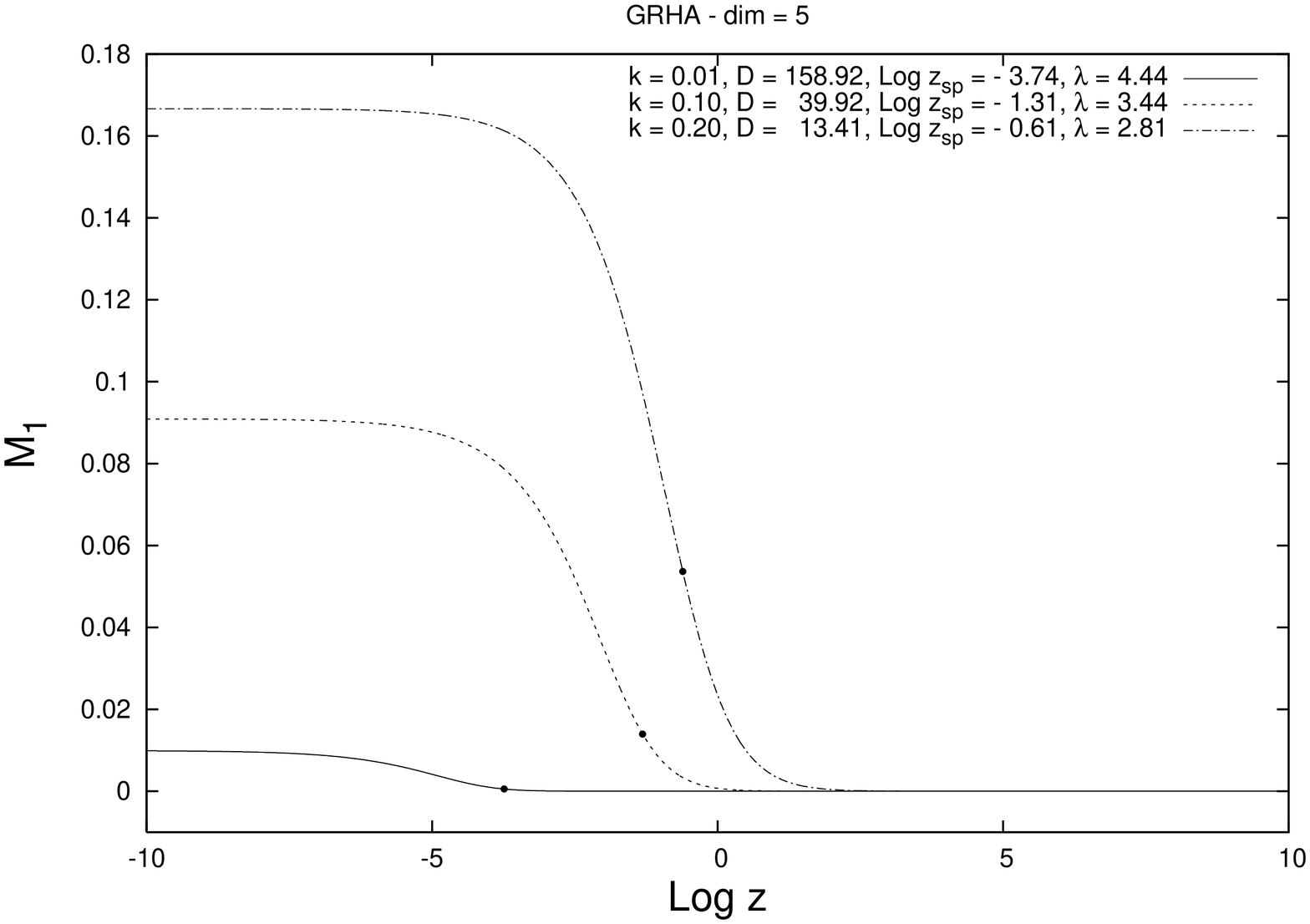} 
 \\[0.4cm]
 \includegraphics[width=3.5in,angle=-90]{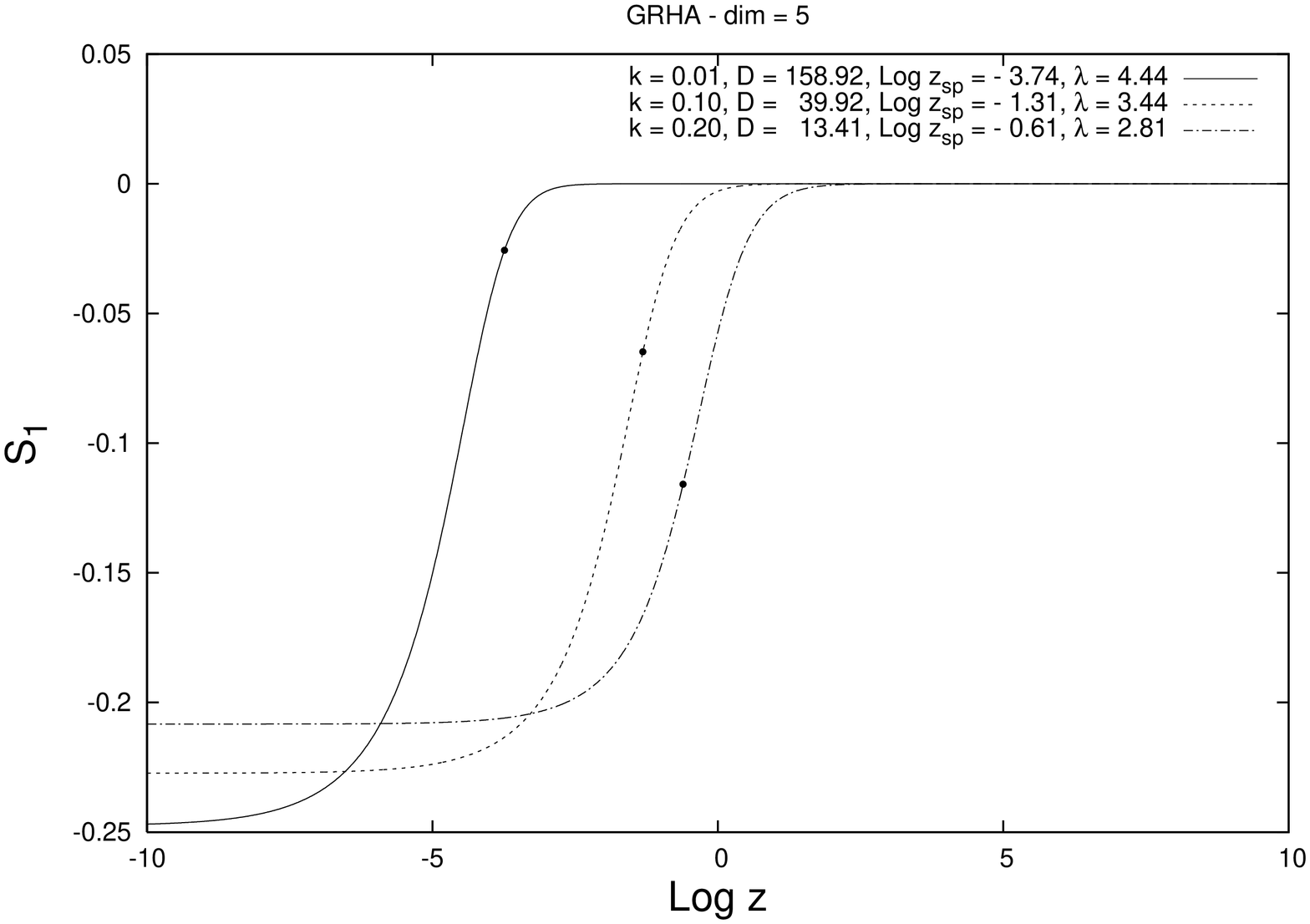}
 \end{center}
 \caption{Plot of the perturbations $M_{1}(z)$ and $S_{1}(z)$ versus $\log{z}$ for 
$k=0.01$, $k=0.1$ and $k=0.2$ in five dimensions.
 The dot gives the position of the sonic point.}
  \label{pert_plots1}
\end{figure}

\begin{figure}[H]
\centerline{ \includegraphics[width=3.2in,angle=-90]{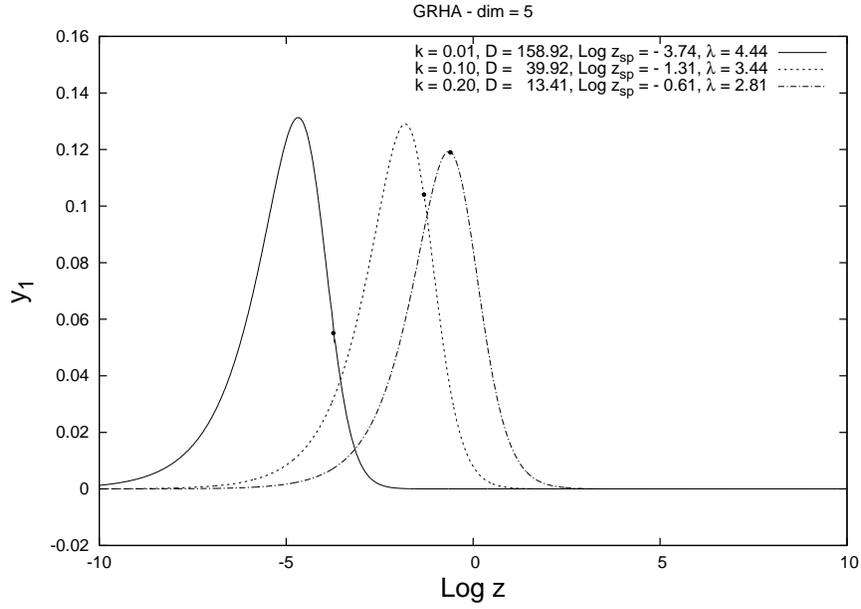} }
 \caption{Plot of the perturbation $y_{1}(z)$ as a function of $\log{z}$, both for 
$k=0.01$, $k=0.1$ and $k=0.2$. 
 As usual the dot indicates the position of the sonic point on each curve.}
  \label{pert_plots2}
\end{figure}

In Figs. \ref{pert_plots1} and \ref{pert_plots2} we 
show the plots of $M_{1}$, $S_{1}$
and $y_{1}$ versus $\log{z}$ for three different values of $k$. 
As expected, 
all perturbations go to zero at large $z$, whereas only $y_{1}$ 
vanishes in the limit $z\rightarrow 0^{+}$.
This means that, for constant $\tau$, self-similarity is mostly 
broken near $z=0$ whereas the symmetry 
is restored at large values of $z$. In particular, we see that 
the position of the sonic point $z_{\rm sp}$
(marked by a dot in the plots in Figs. \ref{pert_plots1} and 
\ref{pert_plots2}) approximately 
indicates the value of $z$ where $M_{1}$, $S_{1}$
and $y_{1}$ begin to decay towards zero. 

Notice, however, that the perturbations to the self-similar solution come multiplied by 
$e^{\lambda\tau}$ where $\lambda>0$ is the positive Lyapunov exponent. This means that
the region where CSS is broken grows as $\tau$ increases until at some point the linear
perturbation regime becomes invalid. The system goes then into a full nonlinear regime
that, for supercritical solutions, results in the formation
of a black hole in the asymptotic future.

\begin{figure}[H]
\centerline{\includegraphics[width=3.2in,angle=-90]{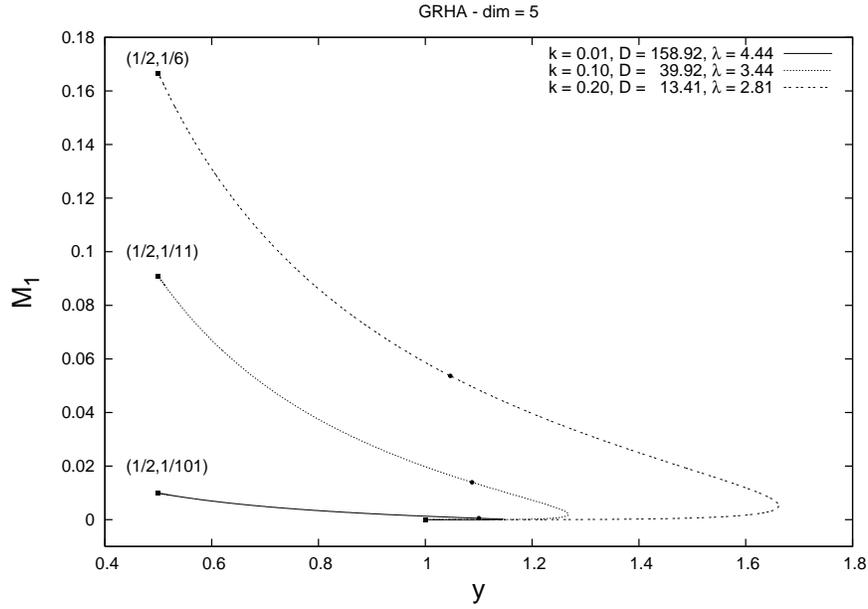} }
\caption{Plot of $M_1$ versus 
 $y$ for $k=0.01$, $k=0.1$ and $k=0.2$ in five dimensions.}
  \label{dynamicalpert1}
 \end{figure}

\begin{figure}[H]
\begin{center}
\includegraphics[width=3.2in,angle=-90]{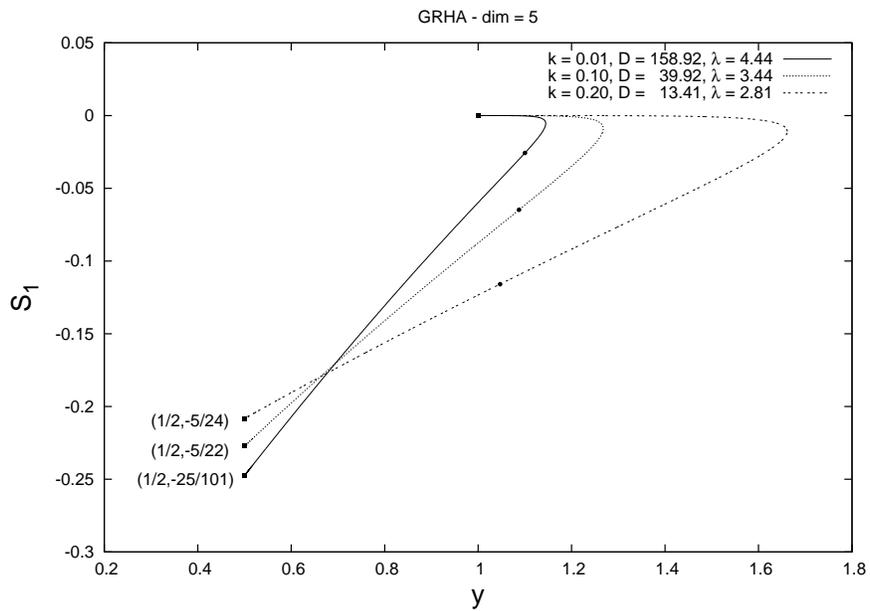} \\[0.4cm]
\includegraphics[width=3.2in,angle=-90]{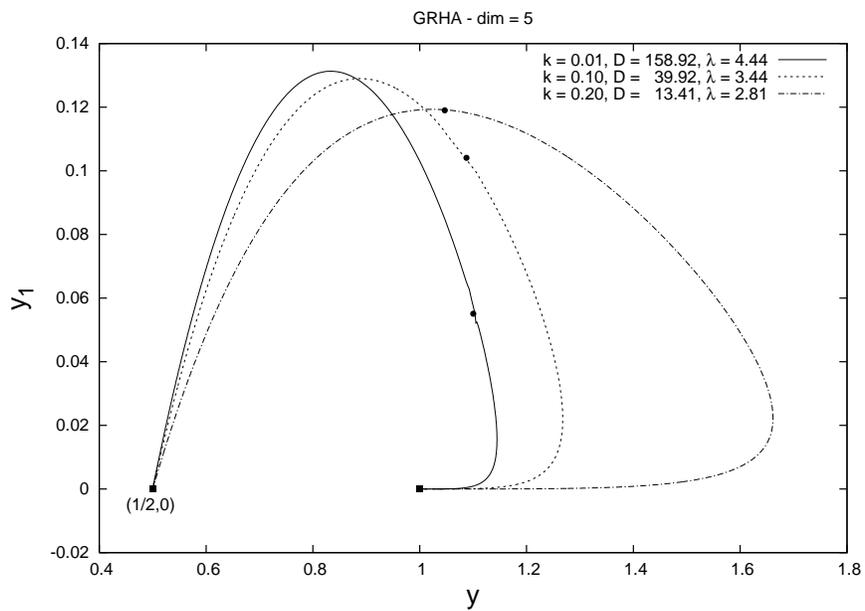}
\end{center}
 \caption{Plots of $S_1$, and $y_1$ versus 
 $y$ for $k=0.01$, $k=0.1$ and $k=0.2$ in five dimensions.}
  \label{dynamicalpert2}
\end{figure}

In Figs. \ref{dynamicalpert1} and \ref{dynamicalpert2} we plot the perturbations $M_1,S_1,y_1$ in terms of
$y$.  As in the case of the critical solution, this shows that the perturbations
reach their asymptotic values at $z=\infty$ after a single turn.  Once again
this  shows that  we have correctly  identified the most relevant perturbations
leading to the computation of the critical exponents.

The regularity of the perturbations at the sonic point fully determine the value of 
the Lyapunov exponent $\lambda$. In Table \ref{tab1} we show the values of both $\lambda$
and the Choptuik exponent $\gamma=\lambda^{-1}$ for various values of $k$ in four, five, six
and seven space-time dimensions. All values shown in this table are given with a precision of
$\pm 0.001$. These values of the Lyapunov and Choptuik exponents are the main concrete result of  this  paper.
Looking at the value obtained for the  conformal fluid in five dimensions $k=1/4$,
$\lambda=2.58$, we find it is reasonably close to the value for  a massless scalar field.
This situation is very similar to the four-dimensional case, where the  Choptuik exponent for
a conformal fluid is very close to the value of a massless scalar field. This is also the case
for the conformal fluids in $d=6$ ($k=1/5$) and $d=7$ ($k=1/6$), as compared to the values
of the Choptuik exponents for the collapse of a massless scalar field in these dimensions \cite{scalar_hD}.
Furthermore the value of the Choptuik exponent for the conformal perfect fluid in $d=5$ is
also close to the BFKL exponent as calculated in \cite{us}.
To what extent this numerical agreement is serendipity or a sign of something
deep will be analyzed in the next Section.

\clearpage

\clearpage

\clearpage

\clearpage

\section{A holographic glimpse through the Regge region}
\label{sec_holo}
   
\subsection{Nonperturbative physics at weak 't Hooft coupling}
\label{subsec_holo1}

Probably the most important lesson we have learned from Maldacena's 
conjecture~\cite{maldacena} 
is that the QCD string, {\it i.e.} the effective string defining the 
representation of Yang-Mills Wilson loops as sums over random surfaces, is 
the fundamental string in some higher dimensional 
geometry~\cite{higher_dimensional}. In spite of 
many efforts during the last ten years, we are still lacking the concrete 
string background geometry for QCD. However, many of the results so far 
obtained hint at the existence of a close relation between nonperturbative 
effects in QCD and black hole dynamics in the gravity dual picture. 

An interesting perturbative regime where stringy aspects of QCD naturally 
appear is the Regge limit of scattering amplitudes. In this region the 
center of mass energy is much larger than all the other Mandelstam invariants 
in the process under consideration, $s \gg -t$. There also exists some hard 
scale, $Q^2 \gg \Lambda^2_{\rm QCD}$, which allows for a perturbative 
treatment of the problem. Although the coupling is small, the amplitudes are 
dominated by logarithms of $s$, making a resummation of terms 
of the form $\left[{\bar \alpha}_s \log(s/Q^2)\right]^n$ mandatory. Note that 
${\bar \alpha}_s = \alpha_s N_c/\pi$ is the usual 't Hooft coupling. This 
resummation can be performed using the 
BFKL~\cite{pomeron} 
formalism which predicts a Regge 
behavior of the type $s^{\omega (t)}$ for total cross sections, with 
$\omega(t)$ being some effective Regge trajectory. 
When the quantum numbers exchanged in the $t$-channel are those of the vacuum, 
the dominant trajectory is that of the BFKL Pomeron (also known as hard 
Pomeron since it is of perturbative origin). 

In the leading logarithmic 
approximation (LLA), where coupling and large logarithms carry the same power 
in the resummed terms, at asymptotically large energies and in the physical 
region of negative $t$ the BFKL Pomeron trajectory goes like 
$\omega (t) \simeq  {\bar \alpha}_s 4 \log{2}$. This power-like behavior 
of the cross sections calls for modification since it violates unitarity 
bounds at very high energies. In the context of deep inelastic scattering 
(DIS) and evolution equations for parton distribution functions, the onset of 
unitarity can be interpreted in terms of ``saturation" corrections which tame 
the power-like growth with $s$. The physics of saturation is driven by 
perturbative degrees of freedom, because we are far away from the 
confinement region. However, the treatment of the problem is nonperturbative 
since it 
corresponds to the description of a very dense system. The main target of this 
Section is to provide a holographic or gravity dual description of these non 
perturbative saturation effects.

The gravity dual interpretation of the BFKL Pomeron has been previously 
addressed in~\cite{brower_et_al}. Using a string target space in the ultraviolet of type 
AdS$_5$ it is possible to study the hard Pomeron at strong 't Hooft coupling 
${\bar \alpha}_s \gg 1$. This is done by computing perturbative string 
amplitudes in the Regge limit with ${\ell_s^2 \over R^2}\log{s}$ being the small 
perturbation parameter, where $\ell_s$ is the string scale, $R$ the curvature 
radius and $\ell_s^2 / R^2 = 1/ \sqrt{{\bar \alpha}_s} \ll 1$. In this weak 
gravity limit the BFKL diffusion in transverse momentum is mapped into 
propagation along the curved holographic direction. The physics of diffusion 
in $\log{s}$ time at weak and strong 't Hooft coupling is qualitatively the 
same but with different coefficients. 

Let us remark that in this holographic correspondence the BFKL Green's 
function in the weak gravity side, computed in a small curvature 
gravitational background, is mapped to the large ${\bar \alpha}_s$ BFKL 
Green's function in the gauge theory side computed around the saddle point 
``background" $\gamma = 1/2 $ of the BFKL kernel eigenvalue $\chi(\gamma) 
= 2 \Psi(1) - \Psi(\gamma) -\Psi(1-\gamma)$. As we will explain in the next 
Subsections, the nonperturbative phenomena of saturation requires to move 
to a different saddle point background $\gamma_s > 1/2$ where amplitudes are 
not power-like anymore at large $s$ and enjoy ``geometrical scaling". 
In the gravity dual picture this change of saddle point ``BFKL" background 
should correspond to a nonperturbative change of gravitational background. 
This new background would require to take into account off shell gravity 
effects and therefore is beyond the semiclassical weak gravity limit. 

In Ref.~\cite{Danilov:2006fv} multi--Regge kinematics was extended to include 
nonperturbative effects in the region of small momentum transfers of the 
BFKL equation in a $d=4$ string theory. In Ref.~\cite{Kotikov:2004er}, 
assuming that the hard Pomeron intercept in $\mathcal{N}=4$ SYM equals the graviton one 
in AdS$_5\,\,\times\,\, $S$^5$, it was possible to obtain a very good extrapolation 
of DIS anomalous dimensions of twist--2 operators from weak to strong 
coupling. 

After this brief introduction we now move to a more detailed description of 
our holographic picture. We start with an introduction to the calculation  
of scattering amplitudes in the Regge limit of perturbative QCD.

\subsection{Deep inelastic scattering and BFKL}
\label{subsec_holo2}

Let us start with a short introduction to DIS at small values of Bjorken $x$. 
In very simple terms, in DIS we are resolving the structure of a large target, 
e.g. a hadron, with an electromagnetic probe. This probe is typically a 
photon with virtuality $Q^2$ and transverse size $1/Q^2$. In the target's rest 
frame we can think of the photon's wave function as forming perturbatively 
well before the interaction with the target. When the transverse size of the 
photon is small it goes through the target like a ``needle", however, for 
smaller virtualities the leading component of the photon's wave function, a 
quark-antiquark pair, is dressed with a color cloud of higher order gluonic 
components. In this case the $\gamma^*$-hadron cross section behaves very 
similarly to that of hadron-hadron scattering 
(see Ref.~\cite{Bartels:2000ze} for a review).

In a leading twist picture, the virtual photon resolves partons in 
the hadron carrying a fraction $x$ of its longitudinal momentum. This takes 
place with a probability $g(x,Q^2)$, the so-called parton distribution 
function. The evolution with $Q^2$ of $g(x,Q^2)$ is well described in 
perturbation theory using the DGLAP 
formalism~\cite{DGLAP}. As the hadron is a complicated 
strong interacting object it is not possible to calculate the parton 
distribution functions from first principles. Nevertheless, very good fits to 
experimental data have been achieved by assuming a certain nonperturbative 
functional form for the $x$ dependence of the parton distribution at a $Q^2$ 
close to the confinement scale and evolving it to larger scales using 
perturbative evolution equations. Collinear factorization theorems ensure 
that, if the final $Q^2$ is large enough, these parton distribution functions 
are universal: they can be measured in DIS experiments in electron-hadron 
machines and then used for predictions at hadron-hadron colliders. The region 
of $x \rightarrow 1$ is well understood from the experimental and theoretical 
sides.

The situation in the limit $x \rightarrow 0$ is far more complicated. In 
principle, even if $Q^2$ is large enough as to ensure a small value for the 
strong coupling, it turns out that the cross sections or structure functions 
are dominated by logarithms of $x$ which are very large as we enter this 
region. This 
implies that terms of the form $\alpha_s \log{(1/x)}$ should be resummed to all 
orders to get an accurate description of the data. The resummation of these 
terms can be effectively carried out by means of the 
BFKL~\cite{pomeron} 
evolution equation, 
which can be considered as orthogonal to DGLAP in the $(x,Q^2)$ kinematic 
plane. While DGLAP is based on perturbative splittings where one emission is 
ordered in transverse momentum with respect to the previous one, BFKL stems 
from strong ordering in longitudinal components. After integration over the 
phase space of these emissions a power like growth of the form $x^{-\Delta}$ 
is obtained for $g(x,Q^2)$. When the resummation is of leading logarithms in 
$x$, {\it i.e.} if the power in the coupling is the same as that of the logarithm then, 
for $x \rightarrow 0$
\begin{eqnarray}
\Delta &=& \frac{\alpha_s N_c}{\pi} 4 \log{2}.
\label{intercept}
\end{eqnarray}
When one power of the logarithm is lost we are in the next-to-leading 
order approximation and $\Delta$ is smaller. 

The BFKL equation in LLA has been rederived in many different ways in the 
literature. 
The original calculation was based upon the study of scattering amplitudes in 
the so-called multi-Regge kinematics. There are two main ingredients: the 
elastic scattering amplitude for the process $g+g  \rightarrow g+g$ and the 
inelastic amplitude for $g+g \rightarrow g+g+g$. We focus on gluons since 
their splitting function dominates at small $x$. 

\begin{figure}[ht]
\centerline{
\includegraphics[width=5cm,bbllx=115,bblly=413,bburx=440,bbury=720,clip=]{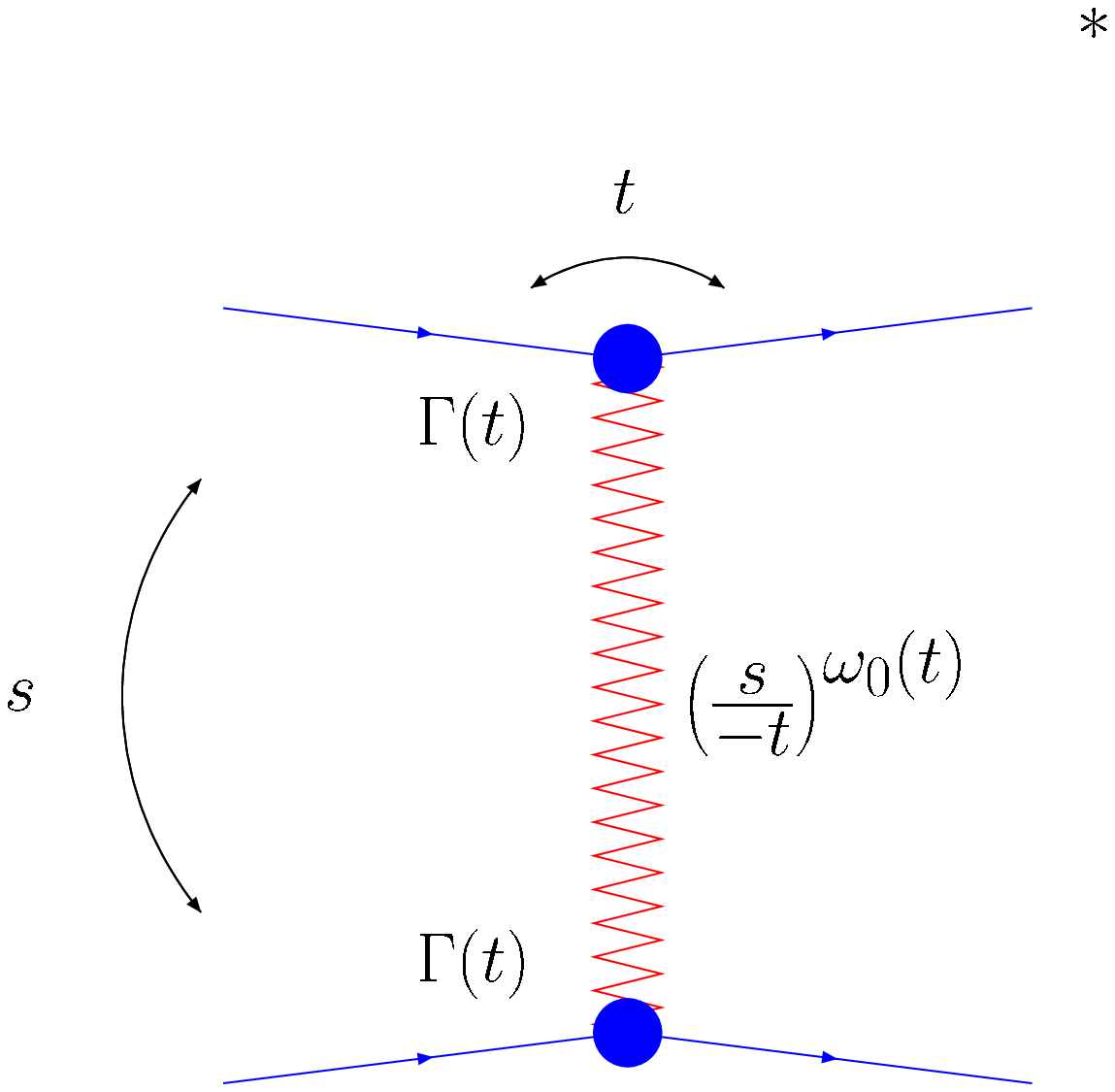}}
\caption{Four point amplitude represented as the exchange of a Reggeized 
gluon in the Regge limit $s \gg -t$.}
\label{Reggeizedgluon}
\end{figure}
When we consider the $M_{2 \rightarrow 2}$ scattering amplitude in the 
Regge limit of $s \gg -t$ 
it has a simple factorized form in terms of two external $t$-dependent 
functions and an energy dependence of the form 
$\left(\frac{s}{-t}\right)^{\omega_0(t)}$. In DIS, $x \sim Q^2/s$ and the limit 
of large $s$ corresponds to small $x$ physics. This exponentiation of the 
logarithms in energy is known as Reggeization of the gluon 
(see Fig.~\ref{Reggeizedgluon}). If we project on 
the gluon's quantum numbers in the $t$-channel this means that, while at 
lowest order in 
$\alpha_s$ we have the usual gluon propagator, when higher orders are taken 
into account this propagator is modified by the Regge factor above--mentioned. 
The gluon then lies on a Regge trajectory, $j(t) = 1 + \omega_0 (t)$, which, 
as usual, reproduces the gluon's spin when $t$ is equal to the gluon mass, 
{\it i.e.} $j(t=0)=1$. In more detail, we have 
\begin{eqnarray}
\omega_0 (t) &=& -\frac{{\bar \alpha}_s \, \vec{q}^{\,\,{2}}}{4 \pi} 
\int \frac{d^2 \vec{k}}{\vec{k}^{2}(\vec{q}-\vec{k})^2} ~\simeq~  
-\frac{{\bar \alpha_s}}{2} \log\left({\frac{\vec{q}^{\,\,2}}{\mu^2}}\right),
\label{trajectory}
\end{eqnarray} 
where $t = - \vec{q}^{\,\,2}$ and 
$\mu$ is an infrared regulator.

\begin{figure}[ht]
\centerline{\includegraphics[width=6cm,bbllx=115,bblly=413,bburx=486,bbury=688,clip=]{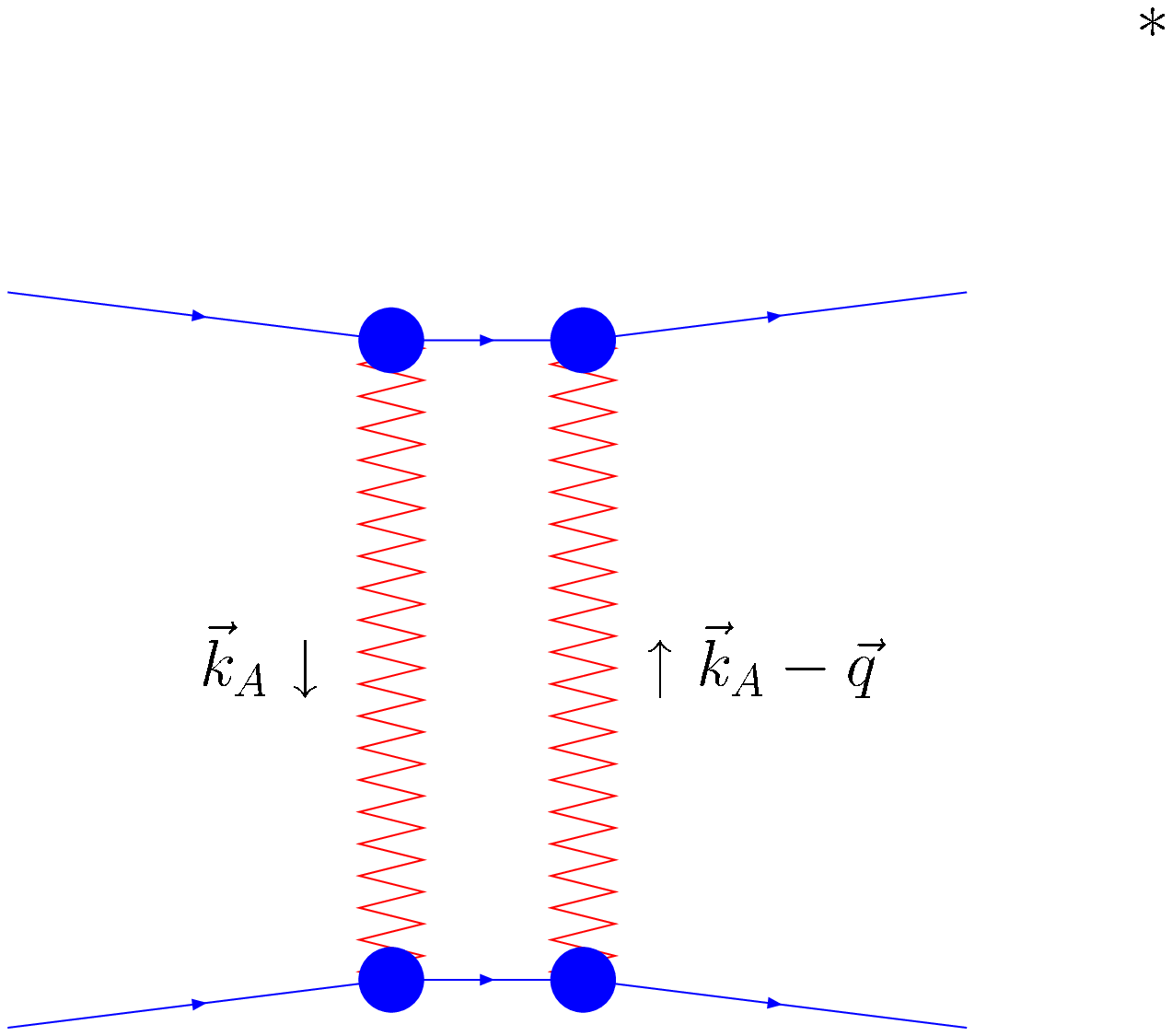}}
\caption{The BFKL Pomeron constructed from two Reggeized gluons exchanged 
in the $t$-channel. Without interactions between them this amplitude is 
infrared divergent and decreases very rapidly with $s$. $\vec{q}$ is the 
momentum transfer.}
\label{2gluons}
\end{figure}
It is possible to construct a perturbative Pomeron with the quantum numbers 
of the vacuum if we exchange two Reggeized gluons and project the amplitude 
on a color singlet. In this case the four-point amplitude can be written as 
\begin{eqnarray}
\varphi \left(\vec{k}_A,\vec{k}_B,\vec{q},{\rm Y}\right) &=& 
e^{\omega_0\left(- \vec{k}_A^2\right){\rm Y}} 
e^{\omega_0\left(-\left(\vec{k}_B-\vec{q}\right)^2\right){\rm Y}} 
\delta^{(2)}\left(\vec{k}_A-\vec{k}_B\right),
\label{fourpoint}
\end{eqnarray}
where $\vec{k}_{i=A,B}$ are the two-dimensional transverse momenta of the 
incoming gluons. 
Note that the transverse components of the four momenta decouple from the 
longitudinal ones. These are integrated out in multi-Regge kinematics with the 
Bjorken variable $x$ playing the r{\^ o}le of a cutoff. Thus we can 
consider ${\rm Y} \sim \log{s} \sim \log{(1/x)}$ as the cutoff of the effective 
theory for the transversal degrees of freedom and the BFKL equation as the 
renormalization group evolution in ``$\log{s}$" time. 

Let us indicate that Eq.~(\ref{fourpoint}) corresponds to the amputated gluon 
Green's function since for zero Y we get the normalization 
\begin{eqnarray}
\varphi \left(\vec{k}_A,\vec{k}_B,\vec{q},{\rm Y}=0\right) &=& 
\delta^{(2)}\left(\vec{k}_A-\vec{k}_B\right)
\end{eqnarray}
which corresponds to the $t$-channel exchange of two normal gluons. 

It is interesting to remark some simple facts derived from 
Eq.~(\ref{fourpoint}). 
The trajectory in Eq.~(\ref{trajectory}) is a negative function, which means 
that the elastic scattering amplitude in Eq.~(\ref{fourpoint}) decreases 
as the energy grows. The delta function in Eq.~(\ref{fourpoint}) imposes 
conservation of transverse momenta but it also means that we can consider 
the evolution of a Reggeized gluon as the action of a diagonal operator 
acting on some basis in the transverse momentum space. 

Being a color singlet, the BFKL Pomeron does not radiate and it is responsible 
for ``diffractive" events, where there are regions in the detector without 
hadronic activity, the so-called ``rapidity gaps". To make connection with 
total cross sections or structure functions we can use the optical theorem. 
In this way we can relate the imaginary part of the forward elastic amplitude 
with the total cross section. This means that we can use Eq.~(\ref{fourpoint}) 
with $\vec{q} = 0$ to obtain a first contribution to $g(x,Q^2)$. 

\begin{figure}[ht]
\centerline{\includegraphics[width=6cm,bbllx=115,bblly=413,bburx=486,bbury=688,clip=]{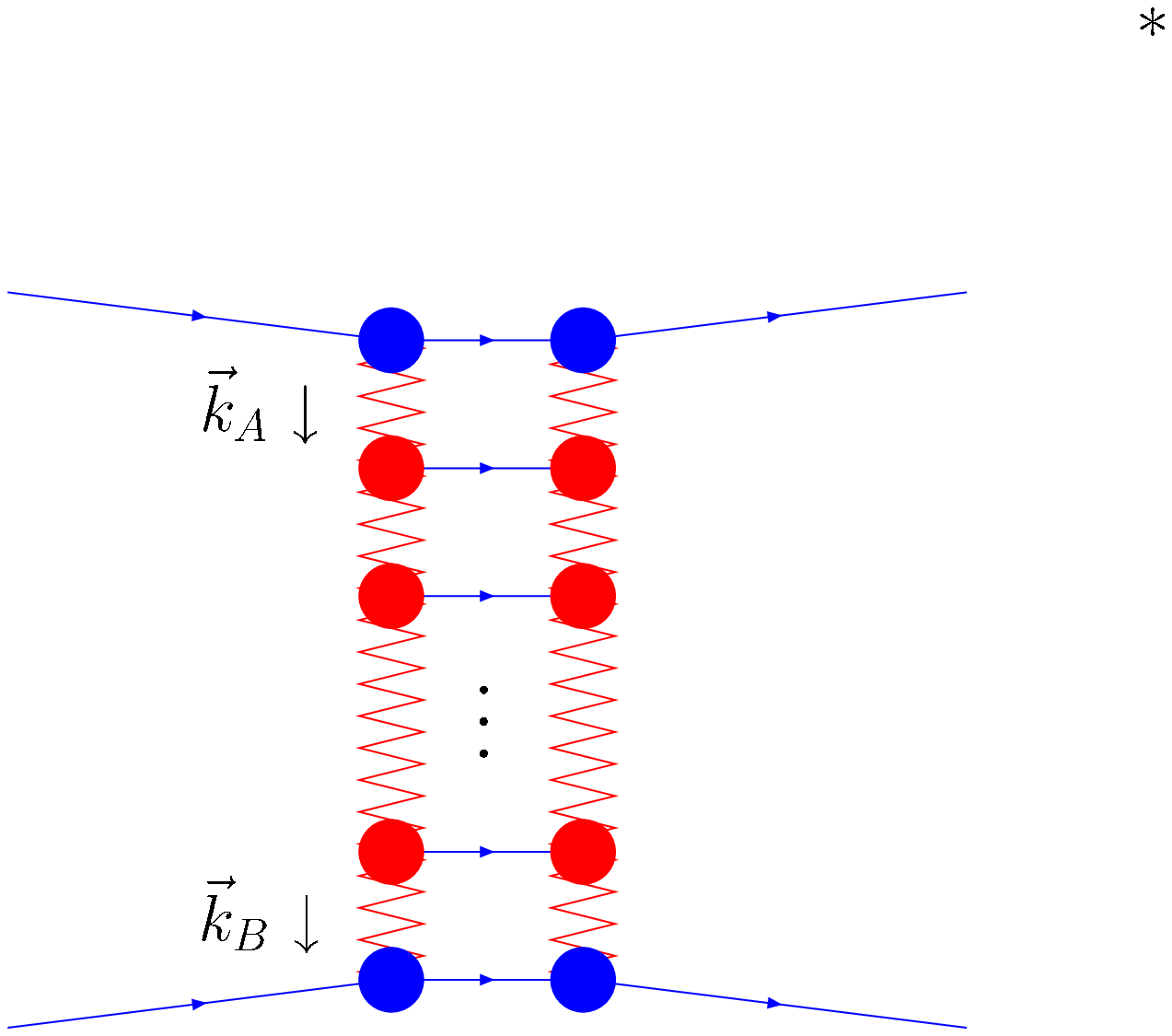}}
\caption{Ladder-like structure for the BFKL Pomeron. The interactions between 
the two Reggeized gluons are built with gauge invariant effective vertices. 
The amplitude is now infrared safe.}
\label{Pomeronladder}
\end{figure}
As it stands, Eq.~(\ref{fourpoint}) is not an infrared safe quantity since it 
has an explicit dependence on $\mu$. This dependence is eliminated when we 
include the second ingredient in the BFKL equation: the inelastic 
$\mathcal{M}_{2 \rightarrow 3}$ amplitude. This amplitude is calculated in multi-Regge 
kinematics where the three outgoing gluons are well separated in rapidity from 
each other. This 
gives rise to a gauge invariant effective emission vertex. When this vertex 
is included we have that the gluon Green's function in the forward case 
now reads
\begin{eqnarray}
\varphi \left(\vec{k}_A,\vec{k}_B,{\rm Y}\right) &=& 
e^{2 \,\omega_0 \left(- \vec{k}_A^2\right){\rm Y}} 
\Bigg[\delta^{(2)}\left(\vec{k}_A-\vec{k}_B\right)\nonumber\\
&&\hspace{-3.5cm}+ \sum_{n=1}^\infty \prod_{i=1}^n {\bar \alpha}_s 
\int \frac{d^2{\vec{k}_i}}{\pi \vec{k}_i^2} 
\theta\left(\vec{k}_i^2-\mu^2\right)
\int_0^{y_{i-1}} \hspace{-0.4cm}dy_i \, e^{2 \, \omega^{(i,i-1)}_0 y_i} 
\delta^{(2)}\left(\vec{k}_A-\vec{k}_B + \sum_{l=1}^n \vec{k}_l\right)
\hspace{-0.2cm}\Bigg],
\label{fourpointforward}
\end{eqnarray}
where
\begin{eqnarray}
\omega^{(i,i-1)}_0 &\equiv&  \omega_0 
\left(- \left(\vec{k}_A+\sum_{l=1}^i \vec{k}_l\right)^2\right)
- \omega_0 
\left(- \left(\vec{k}_A+\sum_{l=1}^{i-1} \vec{k}_l\right)^2\right).
\end{eqnarray}
and $y_0 \equiv {\rm Y}$ (see Fig.~\ref{Pomeronladder}). Below we will 
show a simpler representation where the 
dependence on the infrared regulator explicitly disappears. Before this we 
would like to point out that the action of the real emission kernel is not 
diagonal in transverse space anymore, as can be clearly seen in the second 
line of Eq.~(\ref{fourpointforward}). This will be an important point when 
connecting with gravitational theories. It is also worth noting that the 
Green's function without virtual terms would be 
\begin{eqnarray}
\varphi \left(\vec{k}_A,\vec{k}_B,{\rm Y}\right) &=&   
\delta^{(2)}\left(\vec{k}_A-\vec{k}_B\right) \nonumber\\
&&\hspace{-2cm}+ \sum_{n=1}^\infty \frac{({\bar \alpha}_s {\rm Y})^n}{n!}
\prod_{i=1}^n \int \frac{d^2{\vec{k}_i}}{\pi \vec{k}_i^2} 
\theta\left(\vec{k}_i^2-\mu^2\right)
\delta^{(2)}\left(\vec{k}_A-\vec{k}_B+\sum_{l=1}^n \vec{k}_l\right),
\label{onlyreal}
\end{eqnarray}
this means that the real emission makes $g(x,Q^2)$ grow very rapidly as we 
move towards smaller values of $x$. Let us note that both competing 
effects in Eq.~(\ref{fourpoint}) and Eq.~(\ref{onlyreal}) depend on the 
infrared regulator $\mu$.

\begin{figure}[ht]
\centerline{\includegraphics[width=8cm,angle=-90]{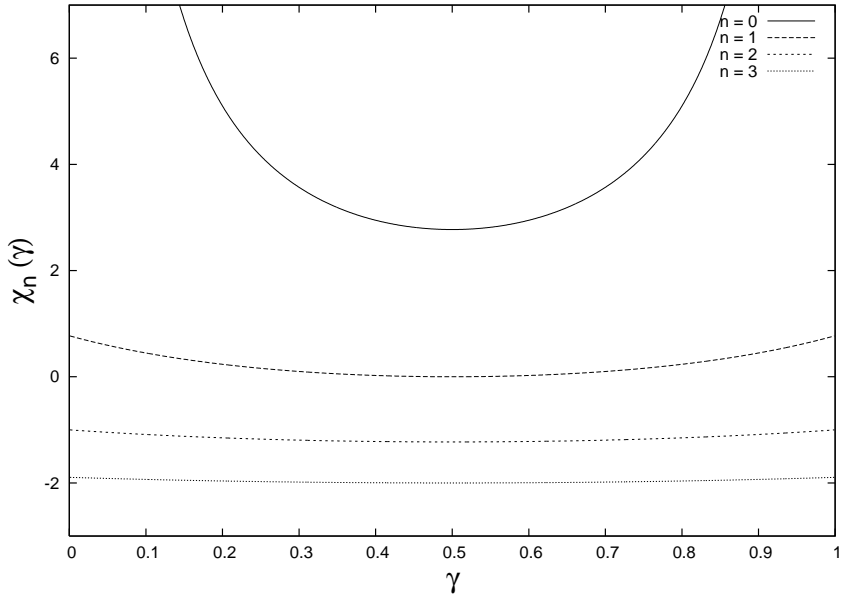}}
\caption{Dependence on $\gamma$ of the eigenvalue of the BFKL kernel for 
different values of $n$.}
\label{kernels}
\end{figure}
Making use of bootstrap, the virtual and real contributions to the 
BFKL amplitude can be combined in such a 
way that the infrared parameter $\mu$ is eliminated. This brings the following 
simpler representation for the Green's function
\begin{eqnarray}
\varphi \left(\vec{k}_A,\vec{k}_B,{\rm Y}\right) =  
\frac{1}{\pi \sqrt{\vec{k}_A^2 \vec{k}_B^2}}
\sum_{n=-\infty}^\infty \int \frac{d\omega}{2 \pi i} 
\int \frac{d\gamma}{2 \pi i} 
\left(\frac{\vec{k}_A^2}{\vec{k}_B^2}\right)^{\gamma-\frac{1}{2}} 
\frac{e^{\omega {\rm Y}+i \, n \,\theta}}{\omega 
- {\bar \alpha}_s \chi_n (\gamma)},
\end{eqnarray}
where
\begin{eqnarray}
\chi_n (\gamma) &=& 2 \Psi(1) 
- \Psi\left(\gamma + \frac{|n|}{2}\right)
- \Psi\left(1-\gamma + \frac{|n|}{2}\right)
\end{eqnarray}
is the eigenvalue of the BFKL kernel and $\Psi$ is the logarithmic 
derivative of 
Euler's Gamma function. $\theta$ is the azimuthal angle formed by the 
two--dimensional vectors $\vec{k}_A$ and $\vec{k}_B$. Its Fourier conjugate 
variable $n$ can be interpreted as a conformal spin in elastic scattering. 
The dominant region of integration is $\gamma \simeq 1/2$. Around this 
point we can see in Fig.~\ref{kernels} that the eigenvalue of the kernel is 
positive only for 
$n=0$. This implies that the dependence on $n$ is only important at low 
energies and we can keep $n=0$ for our arguments. Hence, we average over 
the azimuthal angle
\begin{eqnarray}
{\bar \varphi} \left(k_A,k_B,{\rm Y}\right) &\equiv&  
\frac{1}{2 \pi} \int_0^{2 \pi} \varphi \left(\vec{k}_A,\vec{k}_B,{\rm Y}\right)
\nonumber\\
&=&\frac{1}{\pi k_A k_B}
\int \frac{d\omega}{2 \pi i} \int \frac{d\gamma}{2 \pi i} 
\left(\frac{k_A^2}{k_B^2}\right)^{\gamma-\frac{1}{2}} 
\frac{e^{\omega {\rm Y}}}{\omega - {\bar \alpha}_s \chi (\gamma)},
\label{forwardGGF}
\end{eqnarray}
where $\chi \equiv \chi_0$ and, from now on, $k_i \equiv |\vec{k}_i|$.

Let us now study the structure of this equation for 
very large values of ${\bar \alpha}_s {\rm Y}$. We can write 
\begin{eqnarray}
{\bar \varphi} \left(k_A,k_B,{\rm Y}\right) 
&=&\frac{1}{\pi k_A k_B}\int \frac{d\gamma}{2 \pi i} 
\left(\frac{k_A^2}{k_B^2}\right)^{\gamma-\frac{1}{2}} 
e^{\chi (\gamma){\bar \alpha}_s {\rm Y}},
\label{angavephi}
\end{eqnarray}
and evaluate the integral using the saddle point approximation around the 
minimum of $\chi$ at $\gamma = 1/2$
\begin{eqnarray}
\chi (\gamma) &\simeq& 4 \log{2} + 14 \, \zeta_3 \left(\gamma-\frac{1}{2}\right)^2 
+ \cdots
\label{symmgamma}
\end{eqnarray}
to obtain the expression
\begin{eqnarray}
{\bar \varphi} \left(k_A,k_B,{\rm Y}\right) 
&\simeq&\frac{1}{2 \pi k_A k_B} e^{\Delta {\rm Y}} 
\frac{1}{\sqrt{ 14 \pi \zeta_3{\bar \alpha}_s {\rm Y}}} 
e^{\frac{-t^2}{56 \zeta_3{\bar \alpha}_s {\rm Y}}},
\label{asymptdiff}
\end{eqnarray}
with $t \equiv \log{(k_A^2 / k_B^2)}$. This asymptotic expansion is driven 
by the first exponential, which contains 
the leading behavior in energy, with $\Delta$ being the Pomeron intercept 
given in Eq.~(\ref{intercept}). The function $\chi(\gamma)$ has poles at 
integer values of $\gamma$ with the physical region corresponding to 
$0<\gamma<1$. The poles at $\gamma \simeq 0, 1$ are related to virtualities 
in the internal propagators smaller and larger, respectively, than the 
external scales. The terms containing $\zeta_3$ are 
sensitive to the shape of $\chi$ at its minimum and 
indicate the speed of propagation into the infrared and ultraviolet regions. 
It is easy to verify that the function
\begin{eqnarray}
\Phi \left( k_A, k_B, {\rm Y} \right) &\equiv& k_A \, k_B \,
{\bar \varphi} \left(k_A,k_B,{\rm Y}\right) 
\end{eqnarray}
asymptotically fulfills the diffusion equation
\begin{eqnarray}
\frac{\partial \Phi}{\partial ({\bar \alpha}_s {\rm Y})} &=& 
4 \, \log{2} \, \Phi + 14 \, \zeta_3 \,\frac{\partial^2 \Phi}{\partial t^2}.
\end{eqnarray}
This diffusion in transverse momentum is IR/UV symmetric as a consequence 
of the invariance of $\chi$ under the transformation 
$\gamma \rightarrow 1 - \gamma$.

The gauge/string duality has offered the possibility to explore the 
${\bar \alpha}_s \gg 1$ sector of DIS with the hope of building the correct 
string dual of QCD. In the dipole approximation to DIS a holographic 
description at strong 't Hooft coupling was obtained in terms of a 
semiclassical approximation to the light-like Wilson loop in the target 
metric. At finite temperature and in the quark--gluon phase this was done 
using the AdS black hole metric in \cite{LRW}. 
Another interesting study was carried out in 
Ref.~\cite{polchinski_strassler} where structure functions were calculated at 
strong 't Hooft coupling. The model is based on $\mathcal{N}=4$ supersymmetric 
Yang-Mills (SYM) theory broken down to $\mathcal{N}=1$ at a given mass scale. When going 
to the small $x$ limit the analysis was performed within the single string 
picture. This set up is limited since, as we will see in the coming 
Subsections, new effects in the gauge theory side related to the onset of 
unitarity indicate that one should better operate with multi-string 
configurations. A more complete analysis of this region was presented in 
Ref.~\cite{hatta_iancu_mueller}. In that work, rather than resumming multi--loop 
string amplitudes, inspiration is 
taken from perturbative QCD to understand the flow towards unitarity at 
large ${\bar \alpha}_s$ and the gravity effect on a ``bulk" photon of large 
virtuality is worked out semiclassically. This photon is defined using any 
of the $U(1)$ gauge fields in the bulk associated with the global 
$R$-symmetries of the theory. With increasing virtualities of this probe the 
transition to unitarity changes from a single--Pomeron picture to one with 
multiple graviton exchanges. 

In the present work we investigate unitarization effects in DIS in the 
limit of weak 't Hooft coupling ${\bar \alpha}_s \ll 1$ and very large 
center of mass energy. In this limit the LLA, where terms of the form 
$({\bar \alpha}_s Y)^n$ are resummed, is accurate since higher order 
corrections are always suppressed by larger powers in the coupling, 
e.g. the next--to--leading order approximation (NLLA) resums 
${\bar \alpha}_s ({\bar \alpha}_s Y)^n$ terms. The NLLA corrections in 
$\mathcal{N}=4$ SYM have been studied in 
\cite{NLO}. The program to 
construct even higher order corrections in this theory was initiated 
in \cite{hocorr}.

It is important to note that we are not working with a supersymmetric 
theory from the beginning. It is the dynamics of our process, treated in 
multi--Regge kinematics, what provides a highly symmetric effective theory at 
high energies. As a matter of fact, in the LLA, 
the BFKL kernel is the same in QCD or $\mathcal{N}=1,2,4$ super Yang--Mills theories. 
This is because fermions are suppressed at high energies and their effects are 
only important at higher orders in the approximation. Moreover, we are not 
sensitive to the running of the coupling since the Feynman 
diagrams contributing to it do not belong to the LLA because they bring an 
extra power in ${\bar \alpha}_s$ without generating a logarithm in energy. 
Last, but not least, this theory naturally shows planarity which comes as 
a consequence of the LLA.

We follow a bottom--up approach and find that the 
perturbative description of the saturation line, to be defined in detail 
below, within the BFKL formalism shares many interesting features with the 
critical formation of a tiny black hole in higher dimensions. Our approach to 
the identification of the Regge holography will be mostly constructive. The 
first issue is to identify the holographic extra dimension. Since in the BFKL 
approach to the Regge limit we are working with an effective theory 
parametrized by $\log{s}$, we define the holographic coordinate as  
${\bar \alpha}_s {\rm Y} \sim {\bar \alpha}_s \log{s} \sim {\bar \alpha}_s 
\log{(1/x)}$, the natural energy scale in the problem. In the mapping of BFKL 
dynamics to critical black hole formation we consider the evolution in the 
holographic variable ${\bar \alpha}_s {\rm Y}$ as dual to time evolution in 
the gravitational collapse. At the threshold of formation of the black hole 
the variations in the metric components are large and should be well 
described within a perturbative analysis in the gauge theory side.

The next step in our description should be to identify the geometry on the 
holographic direction. This geometrical 
information for Regge holography should be hidden in the BFKL kernel 
$\chi(\gamma)$, with $\gamma$ parameterizing a particular gravitational 
background. In the AdS/CFT correspondence, in order to describe 
a phase transition, we need to deal with the full quantum gravity canonical 
ensemble of metrics with a given boundary topology. In our case, to describe 
the transit through the saturation region, which corresponds to the flow 
from a dilute partonic regime to a dense system, we need to consider the 
full ensemble of BFKL backgrounds weighted by the kernel $\chi(\gamma)$. 
In the next Subsection we will try to put some substance to these general 
guiding lines of work.

\subsection{Unitarity in DIS and perturbative saturation}
\label{subsec_holo3}

It is well known that the LLA in Eq.~(\ref{asymptdiff}), predicting a 
power-like growth 
in energy for total cross sections, violates the unitarity bound $(\log{s})^{2}$ 
set by Froissart. The problem of unitarization of cross sections at high 
energies is a very complicated one. From a perturbative point of view it 
can be understood by introducing a new set of Feynman diagrams. In the context 
of QCD Reggeon field theory a new vertex arises: that of the transition from 
$2 \rightarrow 4$ Reggeized gluons. This ``triple Pomeron vertex'' is 
invariant under $SL(2,\mathbb{C})$ transformations~\cite{Bartels:1995kf}, 
a M{\" o}bius invariance also present in the BFKL Hamiltonian in the 
non-forward case~\cite{Lipatov:1985uk}. In the large $N_c$ 
limit the triple Pomeron vertex in QCD 
reduces to the vertex used in the Balitsky-Kovchegov (BK) 
equation~\cite{BK}. This 
equation describes a DIS process where the target is an extended and dense 
object. The effect of taking into account unitarity in this equation is 
to introduce a non-linear piece added to the usual BFKL kernel. Neglecting 
impact parameter effects and azimuthal angle dependence, the BK equation can 
be written as
\begin{eqnarray}
\frac{\partial {\Phi} \left(k_A,k_B,{\rm Y}\right)}{\partial ({\bar \alpha}_s {\rm Y})} &=& - \,
{\Phi} \left(k_A,k_B,{\rm Y}\right)^2\nonumber\\
&&\hspace{-3.2cm}+\int_0^1 \frac{dx}{1-x}
\Bigg[{\Phi} \left(\sqrt{x} k_A,k_B,{\rm Y}\right) 
+\frac{1}{x} {\Phi} \left(\frac{k_A}{\sqrt{x}},k_B,{\rm Y}\right)
-2 {\Phi} \left(k_A,k_B,{\rm Y}\right)\Bigg],
\label{BK}
\end{eqnarray}
with the initial condition fixed by 
\begin{eqnarray}
{\Phi} \left(k_A,k_B,{\rm Y} = 0\right) 
&=&\frac{1}{\pi}\int \frac{d\gamma}{2 \pi i} 
\left(\frac{k_A^2}{k_B^2}\right)^{\gamma-\frac{1}{2}}.
\end{eqnarray}
The non--linear term in Eq.~(\ref{BK}) arises from the introduction 
of fan diagrams and it has a 
mild effect at the beginning of the evolution, when the gluon amplitude is 
small. The growth is therefore power-like at early times in 
${\bar \alpha}_s {\rm Y}$. On the contrary, for large values of 
${\bar \alpha}_s {\rm Y}$, when the amplitude is larger, unitarity 
appears in the form of nonlinearities and the quadratic term in Eq.~(\ref{BK}) 
becomes very important since it completely suppresses the growth of the 
amplitude with energy. In terms of parton densities this transition to a 
saturated state can be understood as the crossover from a dilute system of 
partons within the hadron to a very dense one which can be interpreted as a 
``color glass condensate" (CGC) \cite{CGC}. 

As we already remarked, the precise dynamics of  unitarization requires to 
take into account multi-Pomeron vertex effects. However, it is still possible 
to describe within a single Pomeron picture some general features such as 
scaling laws present in the crossover to the saturation region. For this 
we closely follow \cite{saturation}.

\begin{figure}[ht]
\centerline{\includegraphics[width=8cm,angle=-90]{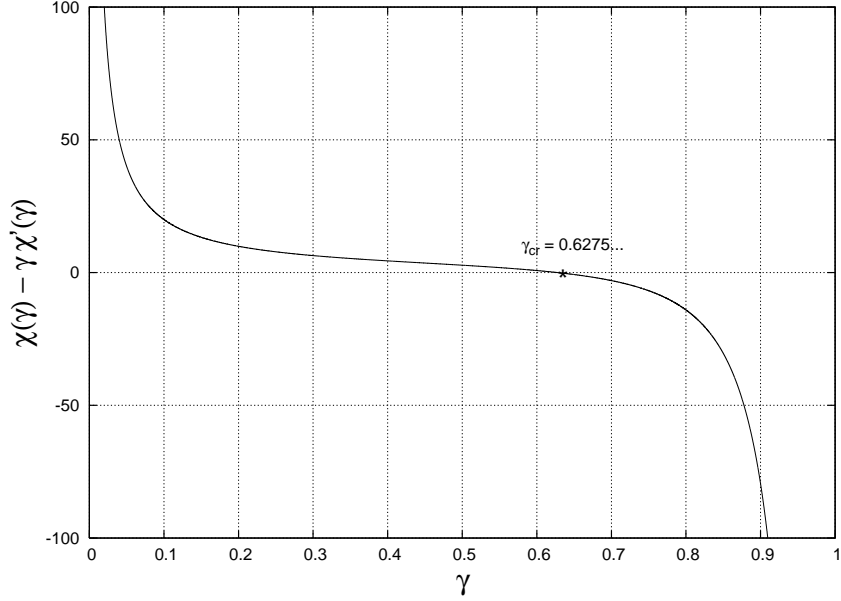}}
\caption{The function of $\gamma$ dominating the energy dependence of the 
gluon Green's function.}
\label{Leg}
\end{figure}

We are considering the limit of very large energies and we can start by 
evaluating the integral in Eq.~(\ref{angavephi}) asymptotically. It is 
important to correctly identify the virtuality corresponding to the target in 
the DIS process, $k_A^2 \equiv Q_{\rm targ}^2$ in Eq.~(\ref{angavephi}), and 
that of the projectile, $k_B^2 \equiv Q_{\rm proj}^2$. We can then 
write
\begin{eqnarray}
{\bar \varphi} \left(Q_{\rm targ},Q_{\rm proj},{\rm Y}\right) 
&=&\frac{1}{\pi Q_{\rm targ}^2}\int \frac{d\gamma}{2 \pi i} 
\left(\frac{Q_{\rm targ}^2}{Q_{\rm proj}^2}\right)^{\gamma} 
e^{\chi (\gamma){\bar \alpha}_s {\rm Y}}.
\end{eqnarray}
At ${\bar \alpha}_s {\rm Y} \gg 1$ it is possible, for fixed values of 
$Q_{\rm proj}^2 = Q_0^2$ and Y, to find the $\gamma$--background 
$\gamma = \gamma_0 (Q_0^2,{\rm Y})$ that dominates asymptotically. It is 
implicitly given by the saddle point condition
\begin{eqnarray}
\chi'(\gamma_0) {\bar \alpha}_s {\rm Y} + \log\left({\frac{Q_{\rm targ}^2}{Q^2_0}}\right)
&=& 0.
\label{saddle}
\end{eqnarray}
To evaluate the Green's function we now expand around this saddle point
\begin{eqnarray}
\chi(\gamma) \simeq \chi(\gamma_0)+ \chi'(\gamma_0) (\gamma - \gamma_0)
+ \frac{1}{2} \chi''(\gamma_0)(\gamma-\gamma_0)^2 + \cdots
\end{eqnarray}
and obtain
\begin{eqnarray}
{\bar \varphi} \left(Q_{\rm targ},Q_{\rm proj},{\rm Y}\right) \simeq 
e^{\gamma_0 t_0+{\bar \alpha}_s {\rm Y} \left(\chi(\gamma_0)
-\gamma_0 \chi'(\gamma_0)\right)}
\frac{e^{\frac{-t_0^2}{2 \chi''(\gamma_0){\bar \alpha}_s {\rm Y}}}}{\pi 
Q_{\rm targ}^2\sqrt{\chi''(\gamma_0) 2 \pi {\bar \alpha}_s {\rm Y}}},
\label{backgrounds}
\end{eqnarray}
where 
$t_0 \equiv \log{(Q_0^2 / Q^2_{\rm proj})}$. The leading behavior of this 
expression is driven by the first exponential. The remaining pieces are of 
diffusive origin and do not need to be considered in a first analysis. The 
growth with energy of Eq.~(\ref{backgrounds}) is dictated by the function 
$\chi(\gamma)-\gamma \chi'(\gamma)$ which we plot in Fig.~\ref{Leg}.
This function is monotonically decreasing as $\gamma$ grows. It has a zero 
at the point
\begin{eqnarray}
\gamma_{\rm cr} &=& {\chi(\gamma_{\rm cr}) \over \chi'(\gamma_{\rm cr})} 
~\simeq~ 0.6275...
\label{gammacritica}
\end{eqnarray}
For any $\gamma_0$ in the physical interval (0,1) the kernel $\chi(\gamma_0)$ 
is positive. For $\gamma_0 < 1/2 < \gamma_{\rm cr}$ we have that 
$\chi'(\gamma_0)$ is negative and Eq.~(\ref{backgrounds}) 
rises very rapidly as Y gets larger. In the range $1/2 < \gamma_0 < 
\gamma_{\rm cr}$, $\chi'(\gamma_0)$ changes sign but its value is not large 
enough as to stop the growth of the Green's function. However, for backgrounds 
with $\gamma_0 > \gamma_{\rm cr}$ there is no growth because 
$\chi'(\gamma_0)$ is so large that it forces the amplitude to decrease 
with energy. The value at which this change in behavior takes place is 
$\chi'(\gamma_{\rm cr}) \simeq 4.8833...$

The value $\gamma_0 = 1/2$ is precisely where the derivative of the kernel cancels, 
$\chi'(1/2) = 0$, and where the usual BFKL expansion happens. It generates a 
rise of the cross sections which must be tamed at some point in energy in 
order to respect unitarity bounds. Unitarity can be enforced in our single 
Pomeron picture if we impose the extra constraint that the dominant region 
of integration at asymptotic Y corresponds to $\gamma_0 = \gamma_{\rm cr}$. 
What we obtain is 
\begin{eqnarray}
{\bar \varphi} \left(Q_{\rm targ},Q_{\rm proj},{\rm Y}\right) 
\simeq 
\left(\frac{Q_{\rm cr} ({\rm Y})}{Q_{\rm proj}}\right)^{2 \gamma_{\rm cr}} 
\frac{e^{\frac{-t_{\rm cr}^2}{2 \chi''(\gamma_{\rm cr}){\bar \alpha}_s {\rm Y}}}}{\pi Q_{\rm targ}^2 \sqrt{ \chi''(\gamma_{\rm cr})2 \pi {\bar \alpha}_s {\rm Y} }},
\label{saturphi}
\end{eqnarray}
with $t_{\rm cr} \equiv \log{[Q_{\rm cr}^2({\rm Y})/Q_{\rm proj}^2]}$, 
$\chi(\gamma_{\rm cr}) \simeq 3.0645...$ and 
$\chi''(\gamma_{\rm cr}) \simeq 48.5176...$ In the kinematic plane of DIS, 
the critical line where the transition to saturation occurs can be read off 
the saddle point condition in Eq.~(\ref{saddle})
\begin{eqnarray}
Q_{\rm cr} ({\rm Y}) &=& Q_{\rm targ} \exp{\left[{\chi'(\gamma_{\rm cr}) 
\over 2} {\bar \alpha}_s {\rm Y}\right]}.
\label{crline}
\end{eqnarray}
\begin{figure}[ht]
\centerline{\includegraphics[width=12cm,angle=0,bbllx=0,bblly=500,bburx=470,bbury=770,clip=]{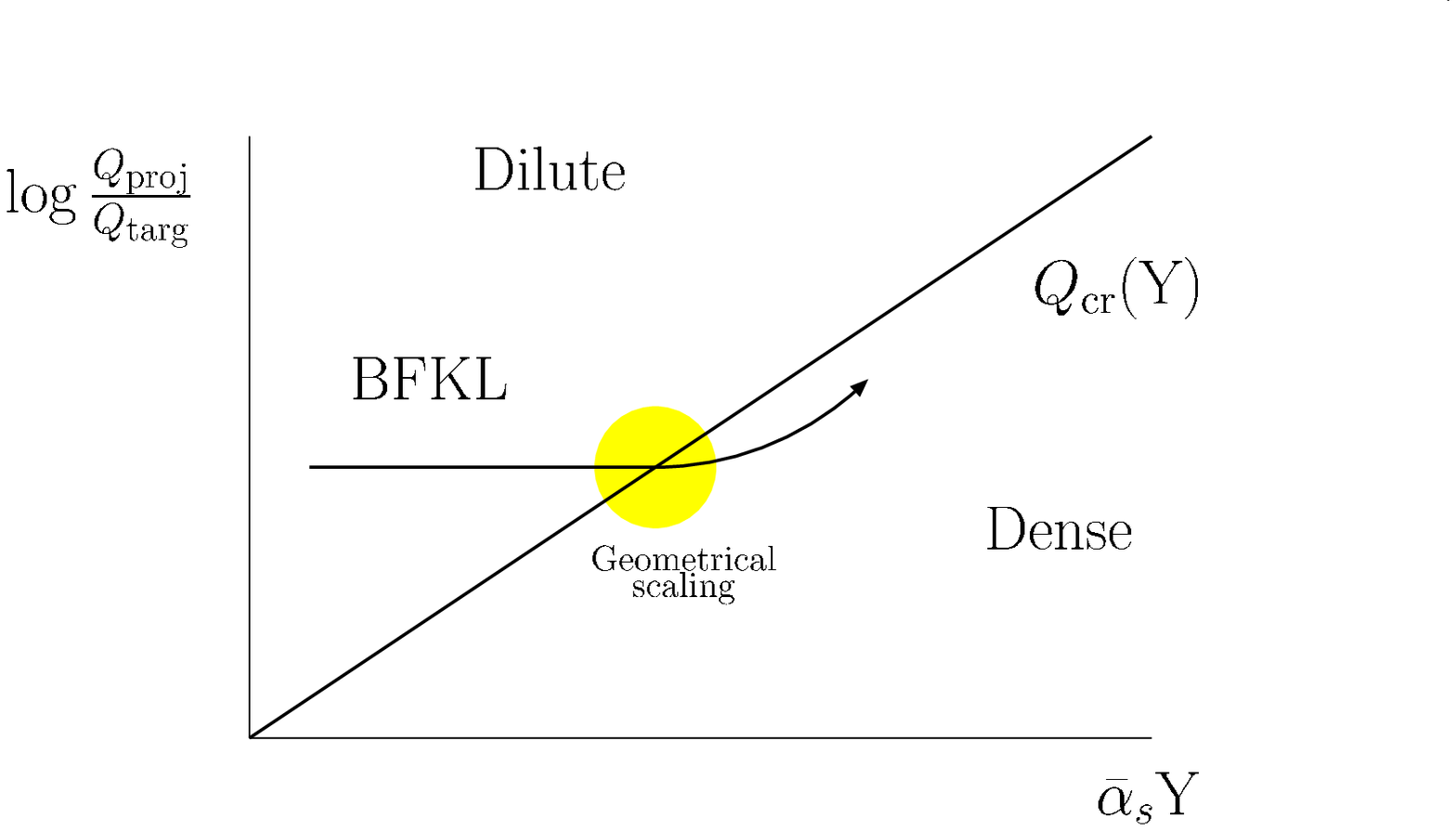}}
\caption{Diagram in the kinematic plane of DIS for the transition from a dilute partonic system to a dense one through BFKL evolution. The transition to a saturated phase takes place crossing a region with scale invariance.}
\label{Saturation}
\end{figure}
This ``saturation line'' is represented in Fig.~\ref{Saturation}. Taking into 
account Eq.~(\ref{crline}) it is easy to show that the saturated amplitude in 
Eq.~(\ref{saturphi}) manifests an extra approximate symmetry at $\gamma_0 = 
\gamma_{\rm cr}$. The appearance of scaling symmetries is common in critical 
phenomena. In our case the BFKL Green's function with saturation is, up to 
very mild subleading corrections, invariant under the scaling transformation
\begin{eqnarray}
{\bar \alpha}_s {\rm Y} &\rightarrow& 
{\bar \alpha}_s {\rm Y} + \log{\lambda},
\nonumber \\[0.2cm]
{Q_{\rm proj} \over Q_{\rm targ}} &\rightarrow&
{Q_{\rm proj} \over Q_{\rm targ}} \lambda^{\frac{\chi'(\gamma_{\rm cr})}{2}}.
\label{YQlambda}
\end{eqnarray}
We can then identify $\chi'(\gamma_{\rm cr})/2 \simeq 2.4417...$ with 
a critical exponent. 

\begin{figure}[ht]
\centerline{\includegraphics[width=10cm,angle=0,bbllx=0,bblly=0,bburx=457,bbury=566,clip=]{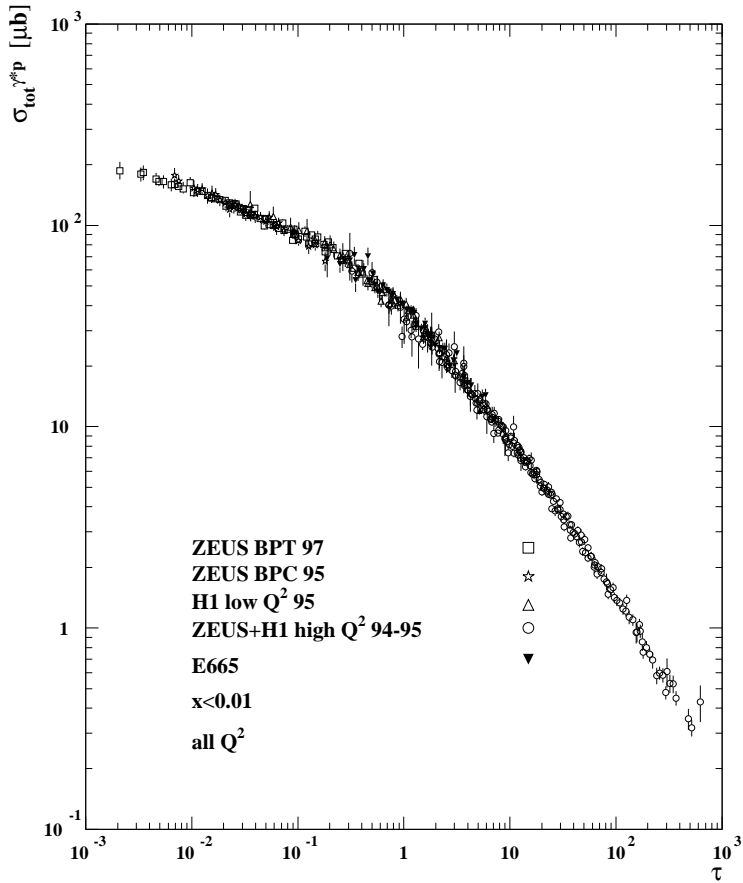}}
\caption{HERA data for $\sigma_{\gamma^* p}$ with $x<0.01$ versus the variable 
$\tau \simeq Q^2 x^{2\lambda}$.}
\label{HERAtau}
\end{figure}

The invariance of scattering amplitudes under this scaling transformation is 
known as ``geometric scaling'' in DIS  at small $x$ 
and has been experimentally observed in HERA data \cite{Stasto:2000er} for the 
$\gamma^* p$ cross sections in the region $x < 0.01$ over a large range in 
$Q{}^{2}$, see Fig. \ref{HERAtau}.

\subsection{Holographic map to critical gravitational collapse}
\label{subsec_hol4}

\begin{figure}[ht]
\centerline{\includegraphics[width=12cm,angle=0,bbllx=78,bblly=538,bburx=558,bbury=770,clip=]{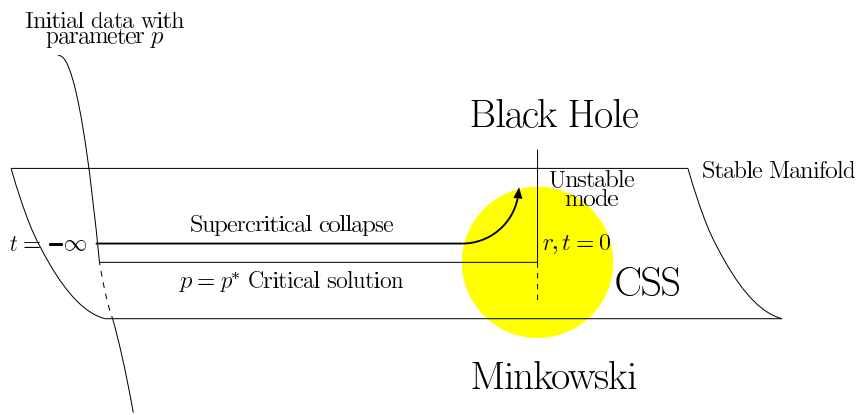}}
\caption{Trajectory in phase space of solutions for supercritical initial 
conditions. The transition to form the black hole is dominated by continuous 
self--similarity.}
\label{CriticalBH}
\end{figure}

Now that we have described an approach to unitarization in high energy QCD we 
have all the ingredients needed to propose a holographic map with the 
formation of a tiny black hole. We identify the region of linear growth with 
${\bar \alpha}_s {\rm Y}$ in the QCD side with that of time evolution on a 
supercritical line just above the critical manifold of codimension one present 
in our gravitational studies of a perfect fluid (see Fig.~\ref{CriticalBH}). 
We set the initial critical density distribution of perfect fluid at time  
$t = - \infty$ and evolve the system following Einstein's and matter equations 
of motion. 

It is also natural to map the radial component $r$ in the gravity side 
to $1/Q$ in the gauge theory sector. Both variables measure the spread of the 
system on a space ``orthogonal'' to the holographic direction. Note that 
the typical virtuality $Q^2$ should be large enough to ensure a perturbative 
description of the scattering. This brings us to the $r \ll 1$ region of the 
critical collapse, precisely the zone where the black hole is formed. 

The competition in QCD between Reggeon exchange, which, as we have 
already explained, tends to reduce the size of the amplitude, and the 
emission of on--shell gluons, with the opposite effect, has an 
interpretation during the collapse: it corresponds to the 
balance between self--gravitational attraction and kinetic energy, or 
pressure, in the perfect fluid. To understand this correspondence in more 
detail we can go back to Eq.~(\ref{gammacritica}) and note that in the 
unitarized solution the dominant background is $\gamma_{\rm cr} > 1/2$. 
This is a non--trivial statement since now we are not at the minimum of 
$\chi(\gamma)$ as in the original BFKL case. The dominant region of 
integration is shifted towards the ultraviolet pole at $\gamma =1$. This 
implies that the net effect of non--linear unitarity corrections 
is to suppress the diffusion towards the infrared region. 

This can be mapped into the dual gravitational collapse framework. 
The infrared modes in QCD have small virtualities and therefore occupy large 
regions in the transverse plane. The physics of saturation is that of a system 
with perturbative degrees of freedom forced to grow in a finite region. The 
critical line $Q_{\rm cr} ({\rm Y})$ in Eq.~(\ref{crline}) shows that, in 
order to accommodate further multiplicity, the preferred configurations are 
those with small transverse size, {\it {\it i.e.}} large virtualities. This is 
exactly what happens when the black hole is formed: the largest part of the 
initial collapsing matter does not confine in a compact size and expands 
infinitely, while only a minute fraction remains within the tiny black hole 
radius in the region $r < r_{\rm BH}$. It is natural to think that the black 
hole contains the bulk of collapsing modes with the smallest spread in the 
radial component. 

At asymptotic energies we can heuristically picture the mechanism of 
suppression of infrared modes as that of a two--dimensional disk packing. It 
is useful to go back to the linear BFKL equation and notice that the evolution 
it describes is not sensitive to the transverse size of the gluons in the 
ladder. This is a consequence of the lack of ordering in transverse scales, 
which generates an IR/UV symmetric random walk in these variables. In some 
sense, the system does not have memory of the transverse size of the previous 
steps in the evolution and is insensitive to the size of the target. Note that 
$\gamma = 1/2$, corresponding to the minimum of $\chi(\gamma)$, is the only 
$\gamma$--background where the saddle point condition in Eq.~(\ref{saddle}) 
does not mix longitudinal and transverse components. As it was shown in 
Eq.~(\ref{fourpointforward}) the gluon Green's function is the sum of many 
configurations with different multiplicities. 
\begin{figure}[ht]
\centerline{\includegraphics[width=12cm,angle=0,bbllx=50,bblly=400,bburx=582,bbury=610,clip=]{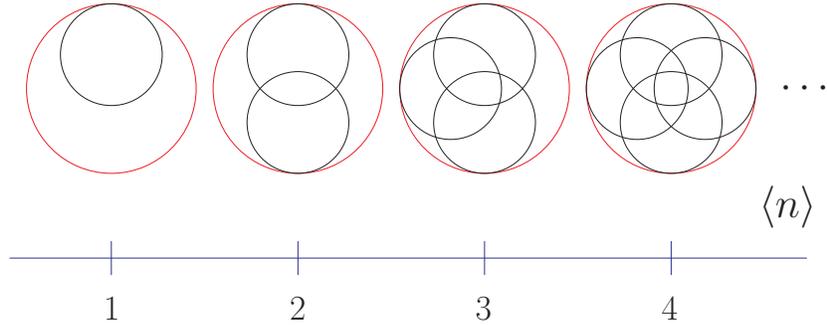}}
\caption{The evolution with energy of the linear BFKL equation is insensitive 
to the target's size and circles are allowed to overlap.}
\label{Linear}
\end{figure}
We can then represent the asymptotic linear evolution as a function of the 
multiplicity $\left<n\right>$ in Fig.~\ref{Linear} 
where the average transverse size of each gluon is represented by the circles, 
which are allowed to overlap. The largest circle corresponds to the target's 
transverse size. 
\begin{figure}[ht]
\centerline{\includegraphics[width=12cm,angle=0,bbllx=50,bblly=400,bburx=582,bbury=610,clip=]{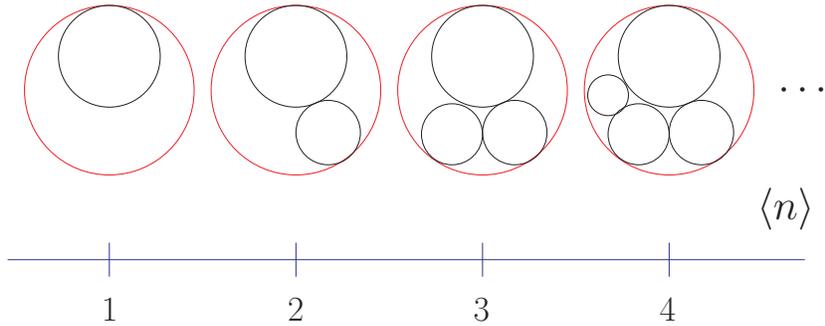}}
\caption{When saturation is introduced the evolution with energy of the BFKL 
equation is sensitive to the target's size and circles are not 
allowed to overlap.}
\label{Nonlinear}
\end{figure}

Nearby the saturation line the parton evolution is constrained by the relation 
in Eq.~(\ref{crline}) which we can write in terms of transverse sizes 
${\cal T} \equiv 1/Q$ as
\begin{eqnarray}
{\cal T}_{\rm cr} &=& {\cal T}_{\rm targ} 
\exp{\left[-\frac{\chi'(\gamma_{\rm cr})}{2} {\bar \alpha}_s {\rm Y}\right]}.
\label{satline}
\end{eqnarray}
This implies that each gluon decreases its transverse size with respect to 
the previous one as the energy 
grows. This introduces ordering in $k_t$ and suppresses the random walks 
towards lower virtualities enhancing the diffusion into ultraviolet modes. 
The system can now resolve the transverse size of the target and it can be 
pictured as a problem of packing many hard disks within a circular boundary 
without allowing overlapping. This is pictorially shown in 
Fig.~\ref{Nonlinear}. The fact that we have CSS indicates that there is a 
fractal structure in the evolution since the system reproduces itself at 
shorter distances. 

In terms of the function in Fig.~\ref{Leg} we can associate gravitational 
potential energy with the term $- \gamma \chi'(\gamma)$ and kinetic energy 
with $\chi(\gamma)$. Thus, for $\gamma_0 < \gamma_{\rm cr}$ kinetic energy 
dominates, with no formation of a black hole. This corresponds to violation 
of unitarity in the Yang--Mills side. The threshold for formation of a 
black hole corresponds to $\gamma_0 = \gamma_{\rm cr}$, when the gravitational 
potential energy is large enough to neutralize its kinetic counterpart. 
This is interpreted as the existence of a unitarization mechanism in QCD.

Let us indicate again that the QCD transition from linear evolution, which 
corresponds to a ``dilute" region in the transverse space, towards the 
``dense" regime is characterized by the critical anomalous dimension 
$\gamma_{\rm cr}$ and the ``saturation exponent", 
$\lambda_{\rm cr} = \chi'(\gamma_{\rm cr}) / 2 \simeq 2.4416...$ 
It is 
remarkable that these are pure numbers independent of any kinematic variable 
or the 't Hooft coupling. 
The interesting phenomena discovered by Choptuik, which we have reproduced 
here for arbitrary dimension, is that the perfect fluid goes through a similar 
process at the threshold of formation of the black hole with its radius 
scaling as 
\begin{eqnarray}
r_{\rm BH} &\sim& \ell_0 \left|p-p^*\right|^{1/\lambda_{\rm BH}},
\label{radius}
\end{eqnarray}
with $\ell_0$ some fiducial length scale. The exponent $\lambda_{\rm BH}$ is also a pure 
number which depends on the dimension and the speed of sound in the fluid, 
$\sqrt{k}$. It is difficult to find compelling 
holographic arguments to select a particular value of $k$. Nevertheless, we 
believe that the limit of ``radiation fluid" is the correct one. In this 
case $k = 1 / (d-1)$ and the energy--momentum tensor is traceless. This 
would support our intuition that a conformal symmetry underlies the 
crossover region towards the black hole formation. We find further evidence in 
this direction when we look at Table \ref{tab1} and see that the values for $\lambda$ 
in the radiation fluid case are 2.58, 2.48 and 2.42 in dimensions five, six 
and seven, respectively. These values are very similar to $\lambda_{\rm cr}$ 
in the gauge theory side. 

At the stage we are in our holographic analysis it is not possible to 
precisely determine 
in which dimension the dual gravitational physics must live. From the 
previous arguments it seems it should not be far from dimension four. In a 
previous analysis \cite{us} it was argued that the mapping 
could be to dimension five. 
This was based on the study of spherically symmetric collapse of a massless 
scalar field \cite{choptuikPRL}. There, a relation identical to (\ref{radius})
was found for the size of the tiny black hole formed when imploding an 
initial radial density of static scalar field. It turns out that in this case 
$\lambda_{\rm BH}$ is very similar to $\lambda_{\rm cr}$ in QCD when the 
scalar field propagates in five dimensions. As discussed in the Introduction, 
the collapse of a scalar field 
also manifests discrete 
self--similarity. This means that the metric and field components at the 
critical manifold, generally denoted by $Z_{*} (t,r)$, reproduce themselves 
after a finite echoing period $\Delta$. This means that $Z_{*}(t,r)= Z_{*}(e^\Delta t,e^\Delta r)$,
where in five dimensions $\Delta \simeq 3.44$ \cite{scalar_hD}. However, it turns out that there 
is no analogue of this echoing period in the Yang--Mills theory side. 

As we already have discussed, in the case of a spherical collapse of a perfect fluid 
with equation of state $p = k \rho$, there is also self-similarity but in 
this case it is continuous (CSS). This implies that we can 
write the critical solution as a function of the variable $z$ defined in Eq. (\ref{tau,z}) only, 
{\it i.e.} $Z_*(t,r) = Z_*(z)$.

\begin{figure}[ht]
\centering
\includegraphics[width=9cm,angle=-90]{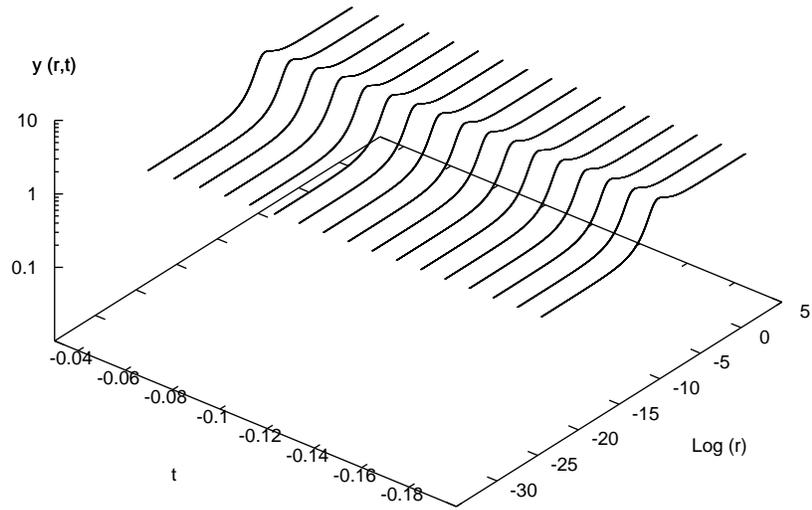}
\includegraphics[width=9cm,angle=-90]{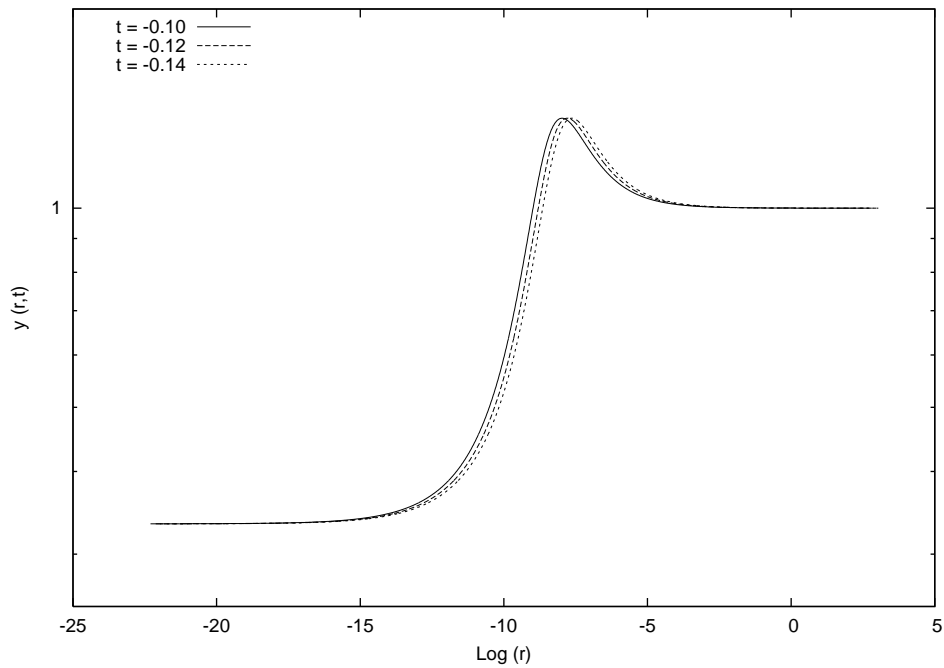}
\caption{A typical solution to gravitational collapse of a perfect fluid with 
continuous self--similarity.}
\label{CSSyrt}
\end{figure}

In Section \ref{sec_back}
we found the critical CSS solution by solving the 
Einstein equations sourced by a perfect fluid. 
As an example in four dimensions, we can look at the profile of the function 
$y(r,t)$ which 
is proportional to the ratio of the mean density inside the sphere of radius 
$r$ to the local density at $r$. In Fig.~\ref{CSSyrt} we show how $y(r,t)$ 
maintains a constant $r$--profile for different values of $t$. This implies 
that the solution is CSS since any continuous change in the time coordinate 
can be compensated by a change in $r$ leaving the solution invariant. The 
mapping of this scaling symmetry in the QCD side corresponds to the dual 
geometrical scaling in Eq.~(\ref{YQlambda}).

It is important to note that in QCD geometric scaling is only present 
in the region very close to the saturation line. In the same way 
the CSS present in the gravitational critical solution breaks down very 
rapidly when we introduce a Liapunov perturbation of the form shown in Eq. (\ref{perturbations}).

\begin{figure}[ht]
\centering
\includegraphics[width=9cm,angle=-90]{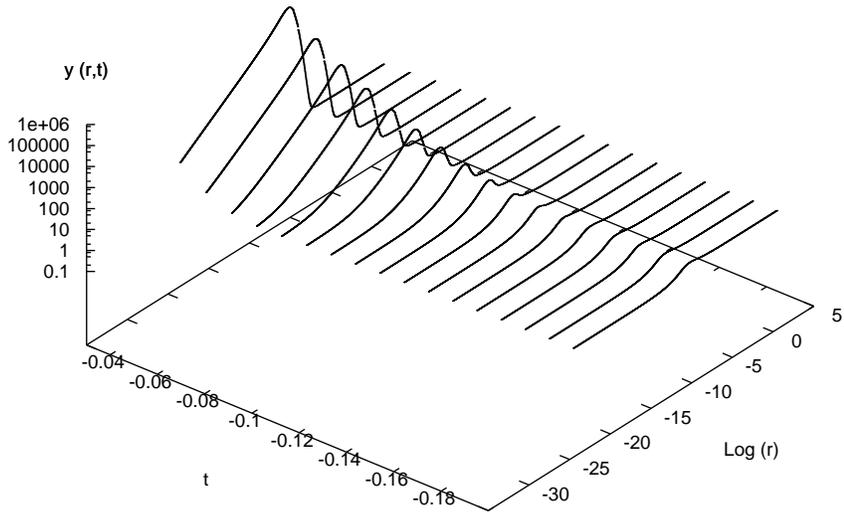}
\includegraphics[width=9cm,angle=-90]{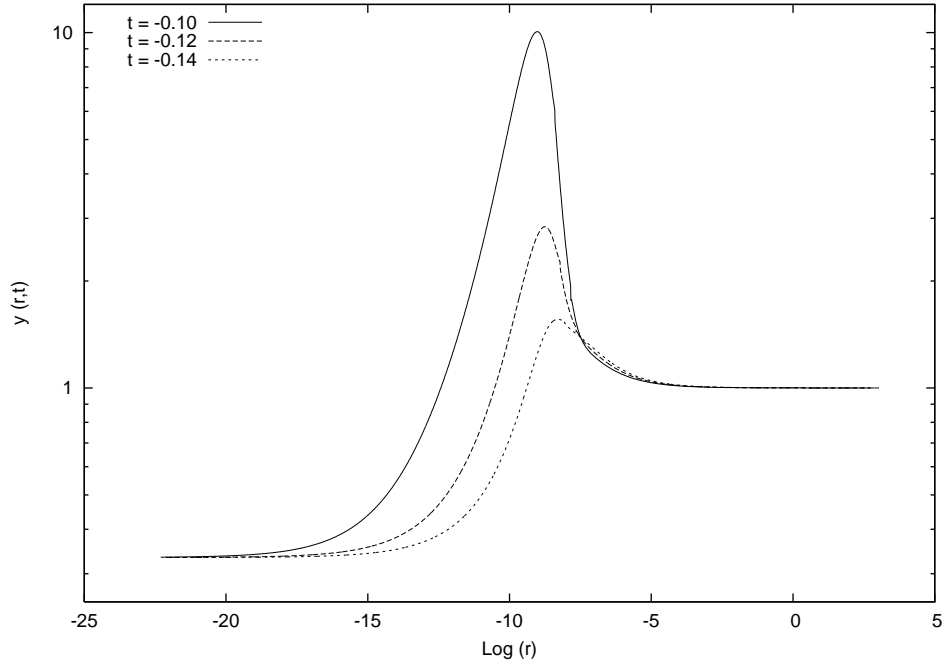}
\caption{Continuous self--similarity is lost by an exponentially growing mode 
in $t$.}
\label{pertCSSyr}
\end{figure}

The Choptuik exponent characterizing the black hole radius of 
Eq.~(\ref{radius}) can be obtained by 
searching for these Liapunov modes of instability of the CSS solution. In 
Fig.~\ref{pertCSSyr} we show how the exponentially growing mode removes 
CSS from the solution to the collapse. The rate of growth of this mode is 
given by a coefficient $\lambda$ which 
coincides with the Choptuik exponent. As explained in previous Sections, we have 
numerically extracted this coefficient in different dimensions. 

To summarize this Section, in QCD there exist two interesting types of 
nonperturbative physics: one is the confinement/deconfinement phase 
transition with a gravity dual \cite{witten_conf} given by a Hawking-Page 
type of phase transitions \cite{hawking_page}. 
The other one, stressed in this article, is the phenomena of parton 
saturation or dilute/dense system of partons transition which we suggest to 
have its gravity dual in the critical formation of tiny black holes. 
Whereas in the confinement/deconfinement phase transition in Hawking-Page 
representation we have an entropic balance to characterize black hole 
nucleation, in the parton saturation case we have an infrared--ultraviolet 
energy competition which we map into a potential/kinetic energy competition in the 
gravity dual. 

\clearpage

\clearpage

\clearpage

\section{Unanswered questions and work in progress}
\label{sec_final}

We are aware of the fact that we are exploring unchartered territory.
On the gravitational side we are looking at the formation of small
black holes, with the corresponding high curvatures and hence
the need to introduce corrections in $\alpha'$ in the Einstein
equations of motion.  We try  to 
relate holographically this region with a yet not fully understood
region in QCD or Yang-Mills theory: the realm  of the hard Pomeron.
Some rudiments of the would-be correspondence have been spelt out
in the previous Section.  We list here a number of open questions,
general philosophy, and work in progress.

So far we have considered the collapse of  either scalar fields
or perfect fluids.  The next step is to consider gravitational
collapse in the case of a full type-IIB string  theory.  We are
currently working on this problem, and  there are some  interesting
provisos worth discussing.  First the beta-function equations
in type-IIB receive corrections starting at four-loop order in $\alpha'$, hence
if we do not consider curvatures close to the string length
our arguments should be reliable at least qualitatively.
Within the spirit of Choptuik computations, since we are
considering the formation of very small black holes, the
cosmological constant should  not be relevant in the computation
of the Choptuik exponents.   What is not clear however, is
whether  we are forming small five-dimesional  AdS black holes
\cite{hawking_page}
or a small ten-dimensional Schwarzschild 
black holes embedded in AdS$_5\,\,\times\,\,$S$^5$, a possibility
shown in \cite{horowitz_hubeny}.
The Choptuik exponent is likely to depend on the various
scenarios. This is currently being  investigated, and
we hope to report our results in the future \cite{ours_future}.

In all our computations, supersymmetry does not
seem to play any r\^ole.  It is important to stress that
the viability of the proposed holographic relation lies in
the fact that
we are computing critical exponents, normally protected
by some form of universality.  This  is the only
stability protection we can advocate against all
corrections imaginable. In addition, on the gauge theory side
we have an effective large-$N$ expansion generated
by the BFKL analysis.  Their effective $N_{\rm eff}$ is 
proportional to $\overline{\alpha}_s \log s$.  To what 
extent this leads to a systematic ``large-$N_{c}$" expansion
remains to be seen, but should such an expansion
exist it would also provide some substance to our
musings.

The region we are trying to explore, at least
on the gauge theory side is accessible to experiment.
How the BFKL evolution is unitarized remains to be 
understood.  Our  claim is that under
a holographic map this should give precious
clues on how gravitational collapse becomes
``unitarized" before forming a singularity.
On the QCD side the BFKL evolution is thought to
lead to the color glass condensate \cite{CGC} (for a review see \cite{CGC_review})
or  some similar scenario in which we go from a single reggeized particle
to a many-body state. This mechanism is somewhat reminiscent of
the Klein paradox. 

In string theory \cite{string_bh}, it is possible to give a 
heuristic description of how, as we approach the formation of
the horizon in gravitational collapse, 
the description in terms of a ``single" string 
should evolve into a multi-string system, thus
diluting the formation of a singularity.  This
analysis is based on entropy arguments, and for
the time being we do not know how to holographically
match these arguments with an entropic argument
in the CGC.  The formation of the latter, as 
well as the scaling arguments in gravitational
collapse are deeply dynamical phenomena, far
from any  notion of equilibrium. At this moment it
does not seem easy to make a connection between
both arguments.  We believe, however, that the
formation of the CGC (or whatever unitarizes the
BFKL evolution) will, through a holographic
map, provide a description of what
happens in the final stages of gravitational
collapse. If this picture is right, 
the exploration of the Regge region in
future colliders is likely to give us 
a completely new panorama of what
happens when black holes form.

\section*{Acknowledgments}

We have benefited from discussions and criticisms from many colleagues.  
We would like to thank J. L. F. Barb\'on, J. Bartels, D. Berenstein, M. Choptuik,
M. Ciafaloni, L. Cornalba, J. Edelstein, R. Emparan, 
D. Gross, H. Itoyama, B.  Kol, A. Kumar, G. Kunstatter, 
K. Kunze, L. Lipatov, H. Liu, 
M. Mari\~no, J. M. Mart\'{i}n-Garc\'{\i}a, D. Mateos, T. Ort\'{\i}n, F. Schaposnik, 
M. M. Sheikh-Jabbari, and S. Wadia 
for comments and discussion which
helped sharpen our ideas and presentation. 
The work of C.G. has been partially supported by the Spanish
DGI contract FPA2003-02877 and the CAM grant HEPHACOS
P-ESP-00346. A.T. thanks the Marie Curie
and the Freydoon Mansouri foundations for support, and the
CERN Theory Group for hospitality.
M.A.V.-M. acknowledges partial support from the Spanish
Government Grants PA2005-04823, FIS2006-05319 and Spanish Consolider-Ingenio
2010 Programme CPAN (CSD2007-00042), and thanks the CERN
Theory Group for hospitality.  

\renewcommand{\thesection}{ }
\section{Appendix A. The Einstein equations}
\renewcommand{\thesection}{A}
\label{appA}

In this Appendix we detail the derivation of the Einstein equations for the line element
\begin{eqnarray}
ds^{2}=-\alpha(t,r)^{2}dt^{2}+a(t,r)^{2}dr^{2}+R(t,r)^{2}d\Omega_{\kappa,D-2}^{2}
\end{eqnarray}
coupled to a perfect-fluid energy-momentum tensor (\ref{em_tensor}).  
Here $d\Omega_{\kappa,D-1}^{2}$ is the line element for a 
$(D-2)$-dimensional space of constant curvature $\kappa=0,\pm 1$. 
The case with $\kappa=1$ corresponds
to the spherically symmetric case studied in this paper.

In order to write the geometric side of the Einstein equations it is very convenient to introduce a system of
tetrads
\begin{eqnarray}
e^{0}=\alpha(t,r)dt, \hspace*{0.5cm} e^{1}=a(t,r)dr, \hspace*{0.5cm} e^{i}=R(t,r)\theta^{i},
\end{eqnarray}
where $\theta^{i}$ is a system of tetrads for the transverse space satisfying $d\Omega_{\kappa,D-2}^{2}=
\delta_{ij}\theta^{i}\theta^{j}$.

The Riemannnian spin connection can be computed from the 
torsion-free conditions $de^{a}+\omega^{a}_{\,\,\,\,b}\wedge e^{b}=0$, 
with the result
\begin{eqnarray}
\omega^{0}_{\,\,\,\,1}={\alpha_{,r}\over a\alpha}\,e^{0}+{a_{,t}\over a \alpha}\,e^{1} \hspace*{1cm}
\omega^{0}_{\,\,\,\,i}={R_{,t}\over \alpha R}\,e^{i}, \hspace*{1cm} \omega^{1}_{\,\,\,\,i}=-{R_{,r}\over a R}\,e^{i}, \hspace*{1cm}
\omega^{i}_{\,\,\,\,j}=\sigma^{i}_{\,\,\,\,j},
\end{eqnarray}
with $\sigma^{i}_{\,\,\,\,j}$ the spin connection for the constant curvature transverse space, satisfying 
$d\theta^{i}+\sigma^{i}_{\,\,\,\,j}\wedge \theta^{j}=0$. The nonvanishing components of curvature form 
$\Omega^{a}_{\,\,\,\,b}=d\omega^{a}_{\,\,\,\,b}+\omega^{a}_{\,\,\,\,c}\wedge \omega^{c}_{\,\,\,\,b}$ are 
given by
\begin{eqnarray}
\Omega^{0}_{\,\,\,\,1} &=& {1\over \alpha\,a}\left[\left({a_{,t}\over \alpha}\right)_{,t}-\left({\alpha_{,r}\over a}\right)_{\,r}\right]
e^{0}\wedge e^{1}, \nonumber \\
\Omega^{0}_{\,\,\,\,i} &=& \left[{1\over \alpha \,R}\left({R_{,t}\over \alpha}\right)_{,t}-{R_{,r}\alpha_{,r}\over \alpha\,a^{2}\,R}\right]
e^{0}\wedge e^{i}+\left[{1\over a\, R}\left({R_{,t}\over \alpha}\right)_{,r}-{R_{,r}a_{,t}\over \alpha\,a^{2}\,R}\right]
e^{1}\wedge e^{i}, \\
\Omega^{1}_{\,\,\,\,i} &=& \left[-{1\over \alpha\,R}\left({R_{,r}\over a}\right)_{,t}-{R_{,t}\alpha_{,r}\over \alpha^{2}\,a\,R}\right]
e^{0}\wedge e^{i}+\left[-{1\over a\,R}\left({R_{,r}\over a}\right)_{,r}-{R_{,t}a_{,t}\over \alpha^{2}\,a\,R}\right]
e^{1}\wedge e^{i}. \nonumber  \\
\Omega^{i}_{\,\,\,\,j} &=& \left[\left({R_{,t}\over \alpha\,R}\right)^{2}-\left({R_{\,r}\over a\,R}\right)^{2}\right]e^{i}\wedge e^{j}+
\mathcal{R}^{i}_{\,\,\,\,j} ,\nonumber 
\end{eqnarray}
where $\mathcal{R}^{i}_{\,\,\,\,j}$ is the curvature form of the constant curvature space
\begin{eqnarray}
\mathcal{R}^{i}_{\,\,\,\,j}={\kappa\over 2}\left(\delta_{k}^{i}\delta_{j\ell}-\delta^{i}_{\ell}\delta_{jk}\right)\theta^{k}\wedge \theta^{\ell}.
\end{eqnarray}

The components of the Ricci tensor in the tetrad frame $R_{ab}$ can be read from this expressions using
\begin{eqnarray}
R_{ab}=\Omega^{c}_{\,\,\,\,acb}, \hspace*{1cm} \mbox{where} \hspace*{1cm} \Omega^{a}_{\,\,\,\,b}={1\over 2}
\Omega^{a}_{\,\,\,\,bcd}\,e^{c}\wedge e^{d}.
\end{eqnarray}
Then the Einstein tensor in the orthonormal frame 
$G_{ab}=R_{	ab}-{1\over 2}\eta_{ab}R$, with $R=\eta^{ab}R_{ab}$, 
is given by
\begin{eqnarray}
G_{00}&=& (D-2)\left[-{1\over a R}\left({R_{,r}\over a}\right)_{,r}+{a_{,t}R_{,t}\over \alpha^{2} a R}\right] \nonumber \\
& & \,\,+\,\,{(D-2)(D-3)\over 2}\left[\left({R_{,t}\over \alpha R}\right)^{2}-\left({R_{,r}\over a R}\right)^{2}+
{\kappa\over R^{2}}\right] ,\nonumber \\
G_{11}&=& (D-2)\left[-{1\over \alpha R}\left({R_{,t}\over \alpha}\right)_{,t}+{R_{,r}\alpha_{,r}\over \alpha a^{2} R}\right]
\nonumber\\
& & \,\,-\,\,{(D-2)(D-3)\over 2}\left[\left({R_{,t}\over \alpha R}\right)^{2}-\left({R_{,r}\over a R}\right)^{2}+
{\kappa\over R^{2}}\right]  ,\\
G_{01}&=& (D-2)\left[-{1\over a R}\left({R_{,t}\over \alpha}\right)_{,r}+{a_{,t}R_{,r}\over \alpha a^{2} R}\right],
\nonumber \\
G_{ij}&=& \delta_{ij}\left\{(D-3)\left[{1\over a R}\left({R_{,r}\over a}\right)_{,r}+{\alpha_{,r}R_{,r}\over \alpha a^{2} R}\right]
-(D-3)\left[-{1\over \alpha R}\left({R_{,t}\over \alpha}\right)_{,t}-{a_{,t}R_{,t}\over \alpha^{2} a R}\right]\right. \nonumber  \\
& & \,\,+\left.\,\,\left[{1\over \alpha a}\left({a_{,t}\alpha}\right)_{,t}+{1\over \alpha a}\left({\alpha_{\,r}\over a}\right)_{\,r}\right]
-{(D-3)(D-4)\over 2}\left[\left({R_{,t}\over \alpha R}\right)^{2}-\left({R_{,r}\over a R}\right)^{2}+
{\kappa\over R^{2}}\right]\right\}. \nonumber 
\end{eqnarray}

The next step is to compute the right-hand side of the Einstein equations. Again we work in the tetrad frame where the 
four-velocity is given by
\begin{eqnarray}
u^{a}\equiv e^{a}_{\,\,\,\,\mu}u^{\mu}=(1,\vec{0}),
\end{eqnarray}
so the energy-momentum tensor has the simple form $T_{ab}={\rm diag}\,(\rho,p,\ldots,p)$. With this we can write the
full set of Einstein equations $G_{ab}=8\pi G_{N}T_{ab}$. For the sake of simplicity, however, it is convenient to replace the 
more complicated $ij$ components by the Bianchi identities
(\ref{bi}). For the 00, 01 and 11 component of the Einstein equations one arrives after simple algebraic manipulations
to Eqs. (\ref{ee}).

\renewcommand{\thesection}{ }
\section{Appendix B. Calculation of the coefficients in Eq. (\ref{series_sp})   }
\renewcommand{\thesection}{B}
\label{appB}
\setcounter{equation}{0}

In this Appendix we outline the procedure to compute the second coefficients $\mathcal{C}_{\eta}^{(1)}(y_{\rm sp})$ 
of the expansion 
of $(\log\eta)'$ as a series expansion in powers of $\log(z/z_{\rm sp})$. Taking a derivative with respect to 
$\log{z}$ in the last equation in (\ref{ee_ss}) and defining the right-hand side of the expression using the 
l'H\^opital rule we arrive at
\begin{eqnarray}
2(V_{z}^{2})'_{\rm sp}(\log\eta)_{\rm sp}''=\mathcal{K}''_{\rm sp}-(\log\eta)'(V_{z}^{2})'',
\label{equationlogprime2}
\end{eqnarray}
where by $\mathcal{K}$ we denote the numerator of the right-hand side of the last equation in (\ref{ee_ss})
\begin{eqnarray}
\mathcal{K}\equiv {(1+k)^{2}\over d-2}\eta^{k-1\over k+1}S^{4-2d}-(d-2)(y-1)V_{z}^{2}-2k.
\end{eqnarray}
At the sonic point this satisfies
\begin{eqnarray}
\mathcal{K}''_{\rm sp}&=& \Big\{\mathcal{K}'_{\rm sp}+(d-2)\Big[ky'_{\rm sp}+(y_{\rm sp}-1)(V_{z}^{2})'_{\rm sp}\Big]\Big\}
\left[{k-1\over k+1}(\log\eta)'_{\rm sp}+{y_{\rm sp}-1\over k+1}(4-2d)\right] \nonumber \\[0.2cm]
&+& k\Big[2+(d-2)(y_{\rm sp}-1)\Big]\left[{k-1\over k+1}(\log\eta)''_{\rm sp}+(4-2d){y'_{\rm sp}\over k+1}\right]
\label{kappabig}\\[0.2cm]
&-& (d-2)\Big[ky_{\rm sp}''+2y'_{\rm sp}(V_{z})'_{\rm sp}+(y-1)(V_{z}^{2})_{\rm sp}''\Big].
\nonumber 
\end{eqnarray}

As we have already argued, all derivatives at the sonic point should be a function of just one parameter
that we can take to be $y_{\rm sp}$. Indeed, all the derivatives appearing in 
Eq. (\ref{kappabig}) can be written in terms of $y_{\rm sp}$ and $(\log\eta)''_{\rm sp}$ 
by using
\begin{eqnarray}
\mathcal{K}'_{\rm sp}&=& {k\over k+1}\Big[2+(d-2)(y_{\rm sp}-1)\Big]\Big[(k-1)(\log\eta)'_{\rm sp}+(y_{\rm sp}-1)(4-2d)\Big]
\nonumber \\[0.2cm] 
& & -\,\,\, (d-2)\Big[ky_{\rm sp}'+(y_{\rm sp}-1)(V_{z}^{2})'_{\rm sp}\Big],
\end{eqnarray}
together with
\begin{eqnarray}
y'_{\rm sp} &=& \left({1-y_{\rm sp}\over 1+k}\right)\Big[k(d-3)+(d-1)y_{\rm sp}\Big]-y_{\rm sp}(\log\eta)'_{\rm sp},
\nonumber \\[0.2cm]
y''_{\rm sp} &=& {y'_{\rm sp}\over k+1}\Big[(d-1)(1-2y_{\rm sp})-k(d-3)\Big]-y'_{\rm sp}(\log \eta)'_{\rm sp}
-y_{\rm sp}(\log\eta)''_{\rm sp} 
\end{eqnarray}
and
\begin{eqnarray}
(V_{z}^{2})'_{\rm sp} &=& {2k\over k+1}\Big\{(1-k)\Big[1-(\log\eta)_{\rm sp}'\Big]+(2-d)(y_{\rm sp}-1)\Big\},
\\[0.2cm]
(V_{z}^{2})''_{\rm sp} &=& {2\over k+1}(V_{z}^{2})'_{\rm sp}\Big\{(1-k)\Big[1-(\log\eta)'_{\rm sp}\Big]+(2-d)(y_{\rm sp}-1)\Big\}
\nonumber \\
& & -\,\,\,{2k\over k+1}\Big[(1-k)(\log\eta)''_{\rm sp}+(d-2)y_{\rm sp}'\Big].
\nonumber 
\end{eqnarray}
In addition, Eq. (\ref{solutionquadratic}) with the choice of the minus branch is also needed. 

Plugging all these expressions into Eq. (\ref{equationlogprime2}) we arrive at a linear equation for $(\log\eta)''_{\rm sp}$ that
determines also $\mathcal{C}^{(1)}_{\eta}(y_{\rm sp})$.
The fact that this equation is linear is of conceptual importance. It means that, fixed the ambiguity in the choice of
the branch determining $(\log\eta)'_{\rm sp}$ there is no further ambiguity in the calculation of $(\log\eta)''_{\rm sp}$. 
In fact, this extends to the calculation of all higher derivatives. The algebraic equations giving the values of the
$n$-th derivative at
the sonic point are always linear.

\renewcommand{\thesection}{ }
\section{Appendix C. Some details of the calculation of perturbations}
\renewcommand{\thesection}{C}
\label{appC}
\setcounter{equation}{0}

In this Appendix we give the expressions for the derivatives 
of the coefficients needed to solve the perturbation
equations around the singular points. In order to evaluate 
(\ref{coefficients_pert}) the following derivatives 
are required
\begin{eqnarray}
\mathcal{Q}' &=& -(V_{z}^{2}-k)(2y+k-1)y'-(V^{2}_{z})'(y-1)(y+k), \nonumber \\
\mathcal{D}_{M}' &=& k(V_{z}^{2})'-2k\left[V_{z}^{2}(1-y)+1+{y\over k}\right]y'  \nonumber \\
& & \,\,+\,\,{\lambda(k+1)\over (d-3)}\Big\{(V_{z}^{2})'y(1-y)+\left[V_{z}^{2}(1-2y)+k+2y\right]y'\Big\} \nonumber \\
& & \,\,-\,\, {(k+1)^{3}y\,\eta^{k-1\over k+1}S^{4-2d}\over (d-2)(d-3)}\left[{2k\over k+1}(\log{\eta})'-2(d-2)(\log{S})'\right], \\
\mathcal{D}_{S}'&=& [1-(d-2)k]\Big[(V_{z}^{2})'(y-1)^{2}+2(y-1)V_{z}^{2}y'\Big] \nonumber \\
& & \,\,- \,\, 2k (d-1)(y+k)y'+\lambda(k+1)\Big[(V_{z}^{2})'(y-1)+(V_{z}^{2}+k)y'\Big] \nonumber \\
& & \,\,+\,\, {(k+1)^{3}y\,\eta^{k-1\over k+1}S^{4-2d}\over (d-2)}\left[{2k\over k+1}(\log{\eta})'-2(d-2)(\log{S})'\right]. \nonumber
\end{eqnarray}
In addition, the derivatives of $V_{z}^{2}$, $\log{y}$ and $\log{\eta}$ are also needed. For the first two we have
\begin{eqnarray}
(V_{z}^{2})' &=& 2V_{z}^{2}\left\{\left({k-1\over k+1}\right)\Big[(\log{\eta})'-1\Big]-(d-2){y-1\over k+1}\right\}, \nonumber \\
(\log{y})' &=& {k(d-3)\over k+1}\left({1\over y}-1\right)-(\log{\eta})'-(d-1)\left({y-1\over k+1}\right),
\end{eqnarray}
whereas $(\log{\eta})'$ can be written as
\begin{eqnarray}
(\log{\eta})' = -\mathcal{T}-\sqrt{\mathcal{T}^{2}-\mathcal{S}},
\end{eqnarray}
where $\mathcal{T}$ and $\mathcal{S}$ are given respectively by
\begin{eqnarray}
4(k-1)\mathcal{T} &=&- {(k+1)^{2}(k-1)\over d-2}(2D)^{2k\over k+1}\left({z^{2}\over \eta}\right)^{k-1\over k+1}
\nonumber \\
& & \,\,+\,\, 4d-6-2k(d-1)+(d-2)(k-5)y,\nonumber \\
{2(k-1)\over y-1}\mathcal{S} &=& 2(k+1)^{2}(2D)^{2k\over k+1}\left({z^{2}\over\eta}\right)^{k-1\over k+1} \\
&  & \,\,-\,\, (d-2)\Big[(d-1)(k-2)+(3d-5)y\Big] .\nonumber 
\end{eqnarray}

\end{document}